\documentclass[english, 12pt]{article}
\usepackage[bookmarks=true, colorlinks=true, linkcolor=blue, citecolor=blue, breaklinks=true]{hyperref}
\usepackage{natbib}
\usepackage[T1]{fontenc}
\usepackage[utf8]{inputenc}  
\usepackage[active]{srcltx}
\usepackage{booktabs}
\usepackage{babel}
\usepackage{float}
\usepackage{amsthm}
\usepackage{amsmath}
\usepackage{amssymb}
\usepackage{graphicx, graphics}

\usepackage{setspace}
\usepackage{esint}
\usepackage{algorithm}
\usepackage{comment}
\usepackage{siunitx}
\usepackage[flushleft]{threeparttable}
\usepackage{multirow}
\usepackage[noend]{algpseudocode}
\usepackage{subcaption}
\usepackage[normalem]{ulem}
\usepackage{placeins}
\usepackage[usenames]{color}
\definecolor{mygreen}{rgb}{0,.5,0}

\usepackage{suffix}
\usepackage{mathtools}

\newcommand{\be}{\begin{equation}}
	\newcommand{\ee}{\end{equation}}
\newcommand{\bee}{\begin{equation*}}
	\newcommand{\eee}{\end{equation*}}

\theoremstyle{plain}  

\usepackage{fullpage}
\usepackage{pdfsync}
\usepackage[title,titletoc]{appendix}

\providecommand{\theoremname}{Theorem}
\providecommand{\corollaryname}{Corollary}

\begin{document}
	
	\title{Deep Learning for Dynamic Programming with Recursive Utility\thanks{The research is supported by the National Natural Science Foundation of China (Grant No. 72573011).}}
	
	\author{Xianhua Peng\thanks{HSBC Business School, Peking University, University Town, Nanshan District, Shenzhen, 518055, China. Email: xianhuapeng@pku.edu.cn.}
		\and  Wu Guo\thanks{HSBC Business School, Peking University, University Town, Nanshan District, Shenzhen, 518055, China. Email: wuguo@stu.pku.edu.cn.}
	}

	\date{This version: \today}
	
	\maketitle
	
	\baselineskip 18pt
	
	\begin{abstract}
		We propose the first deep learning algorithm, the Certainty Equivalent Learning (CEL) algorithm, for solving high-dimensional discrete-time dynamic programming problems with recursive utility. Dynamic programming with recursive utility is numerically challenging because the recursive utility does not have an explicit representation and the Bellman equation contains a certainty equivalent that is difficult to evaluate. The CEL algorithm learns this certainty-equivalent value directly with neural networks and jointly approximates value functions, policy functions, and certainty-equivalent functions. The CEL algorithm is mesh-free and simulation-based, allowing high-dimensional state and control spaces, and does not rely on Euler equations, first-order conditions, or differentiability of the state transition function. The CEL algorithm also works for dynamic programming problems with expected utility as expected utility is a special case of recursive utility. We apply the CEL to discounted linear exponential quadratic Gaussian control, small-noise robust control, Epstein--Zin DSGE, and multivariate strategic asset allocation problems. Compared with closed-form and VFI-based  benchmarks, the CEL delivers accurate value and policy approximations, remains effective in high-dimensional problems, achieves accuracy comparable to VFI in the small-noise robust-control case, and produces out-of-sample Bellman errors and Euler or first-order residuals that are in the range from $\text{1.0e-4}$ to $\text{1.0e-3}$ for most problems.

        \emph{Keywords}: recursive utility, deep learning, dynamic programming, neural network, dynamic models,  portfolio choice, strategic asset allocation
        
        \emph{JEL classification}: C12, C14, C22, C52, C61, C63, C65, C68, C88, E32, E37, G11, G12
        
	\end{abstract}

	\section{Introduction}
	\label{sec:introduction}
	
	Dynamic programming problems are central to economics and finance, with applications in stochastic growth, real business cycle, asset pricing, etc. A large literature has formulated and developed numerical methods for such problems; see, for example, \citet{stokey1989recursive}, \citet{Hansen-Sargent-2013}, \citet{Ljungqvist-Sargent-2018_updated}, and \citet{miao2020economic}. 
	Traditional grid-based numerical methods suffer from the curse of dimensionality, i.e., the number of grid points grows exponentially with the number of state variables. 
    As a result, grid-based numerical methods are limited to low- or moderate-dimensional problems.
	
    A growing literature has developed deep learning methods for high-dimensional dynamic programming problems with expected utility, using neural networks to approximate value functions, policy functions, and first-order optimality conditions. These methods show that deep learning can be effective for high-dimensional dynamic programming problems with expected utility. 
	
	However, many economic and financial applications require preference specifications beyond standard expected utility. Recursive preferences, such as Epstein--Zin and risk-sensitive preferences, provide a flexible framework because they separate risk aversion from intertemporal substitution and allow agents to respond to long-run risk, model uncertainty, and stochastic volatility. Expected utility is nested in this recursive formulation as a special case. When the transformation is linear and the intertemporal aggregator is additive, the certainty-equivalent term reduces to the usual conditional expectation, and the Bellman equation collapses to the standard expected-utility Bellman equation.

    Recursive utility has therefore been widely used in asset pricing, long-horizon portfolio choice, robust control, climate policy, and other dynamic decision problems under uncertainty.
	
    Despite their theoretical and practical appeal, discrete-time dynamic programming problems with recursive utility are difficult to solve numerically. First, recursive utility generally has no explicit expression in terms of future states and controls. Unlike expected utility, where the value function can be explicitly represented as the sum of a discounted stream of future utilities, recursive utility is defined implicitly as the fixed point of the Bellman equation. Second, the Bellman equation contains a certainty-equivalent term, which is a nonlinear transformation of the conditional expectation of a nonlinear function of future utility. The certainty equivalent is a high-dimensional nonlinear function of the state and control and is difficult to calculate.   
    Third, because of the nonlinear function in the certainty equivalent, the estimation of the certainty equivalent is more delicate than that in expected-utility models. With expected utility, conditional expectation of the future value can be estimated by average of samples of future values without bias. With recursive utility, however, nonlinear function of the sample average of future values is not an unbiased estimate of the certainty equivalent, which is the nonlinear function of conditional expectation. 
    Fourth, the certainty-equivalent value is a function of state and control. During policy improvement, the algorithm must evaluate certainty-equivalent values at candidate controls that are not optimal instead of only at optimal controls. Fifth, certainty-equivalent learning and policy improvement are tightly coupled. The certainty-equivalent target depends on the future value approximation, the policy update depends on the learned certainty-equivalent value, and the updated policy changes the distribution of future simulated states. These feedback loops can amplify approximation errors across iterations and motivate learning the state-control certainty-equivalent value as a separate object. Section~\ref{subsec:background_challenges} discusses these challenges in more detail.

    These computational difficulties are compounded by a general limitation of gradient-based deep learning methods. Neural-network training typically involves nonconvex optimization, so the computed solution is not guaranteed to be globally optimal. This makes independent accuracy evaluation especially important. For recursive utility, this issue is particularly relevant because the learned solution must jointly satisfy the Bellman equation, the nonlinear certainty-equivalent relation, and the implied policy optimality conditions. Moreover, to the best of our knowledge, existing deep-learning approaches to recursive utility have not systematically validated learned solutions against closed-form benchmark solutions. This motivates our use of closed-form benchmarks, VFI comparisons, and out-of-sample Bellman errors and Euler or first-order residuals in the numerical evaluation.
	
	This paper proposes the Certainty Equivalent Learning (CEL) algorithm, the first deep learning algorithm for solving high-dimensional discrete-time dynamic programming problems with recursive utility. The key idea is to learn the certainty-equivalent value directly, rather than repeatedly evaluating nested nonlinear expectations during each update. CEL jointly approximates value functions, policy functions, and certainty-equivalent functions using neural networks. This provides a simulation-based and mesh-free approach for nonlinear, stochastic, and high-dimensional decision problems with recursive utility.
	
    This paper makes six main methodological contributions. First, we develop CEL as a deep learning framework that treats the certainty-equivalent value as a separate learnable object.
    Instead of repeatedly evaluating nested nonlinear expectations during each update, CEL learns the state-control certainty-equivalent function that enters the Bellman equation. This directly addresses the main computational difficulty created by recursive preferences, namely that the future value enters the current value through a nonlinear certainty-equivalent operator.
	
	Second, CEL only requires forward simulation of the discrete-time state transition and does not use Euler equations or first-order conditions as training objectives. Unlike Euler-equation or first-order-condition-based methods, which typically require differentiability of the transition law whenever derivatives of future states enter the optimality conditions, CEL does not backpropagate through the state transition function. The law of motion may be nonlinear, non-Gaussian, non-smooth, high-dimensional, or available only as a simulator. Therefore, CEL can be applied to general discrete-time Markov state evolution.
	
	Third, the paper introduces a modular family of neural architectures for recursive utility. The baseline three-network architecture approximates the value function, the policy function, and the certainty-equivalent function. The four-network architecture further decomposes the certainty-equivalent term into a standard conditional expectation component and a nonlinear risk-adjustment component. The two-network architecture removes the explicit value network and represents the current value implicitly through the policy and certainty-equivalent networks. These architectures allow CEL to adapt the number of learned objects to the numerical difficulty of the model.
	
    Fourth, CEL is mesh-free and simulation-based. The algorithm does not rely on tensor-product grids, interpolation over state-control grids, perturbation around a steady state, problem-specific projection bases, Euler equations, or first-order conditions. As a result, CEL is suitable for high-dimensional discrete-time dynamic programming problems with recursive utility, nonlinear policy functions, high-dimensional state and control variables, and complex stochastic dynamics.

    Fifth, the paper evaluates the learned solutions using out-of-sample residual diagnostics computed from the model's Bellman equation and optimality conditions. Following the convention of using residual-based accuracy measure in numerical dynamic programming, especially the Euler equation residuals \citep{santos2000accuracy}, we assess the final learned value and policy functions on independently simulated test states. Specifically, we report Bellman errors and Euler or first-order residuals obtained by substituting the learned policy, value, and certainty-equivalent functions into the original Bellman equation and optimality conditions. The conditional expectations in these diagnostics are recomputed using new Monte Carlo samples rather than the training samples. These accuracy measures are therefore out-of-sample diagnostics instead of in-sample ones, and they provide separate evidence on value-function consistency and policy optimality.

    Sixth, the CEL algorithm can also be applied for dynamic programming problems with expected utility, as expected utility is a special case of recursive utility when the nonlinear function in the certainty equivalent reduces to linear and the time aggregator function is simple summation. 
    
	We further provide numerical evidence across several benchmark problems, including Gaussian linear-quadratic control, small-noise robust control, Epstein--Zin DSGE models with stochastic volatility, and multivariate strategic asset allocation. These applications cover different recursive-utility specifications, model uncertainty, stochastic volatility, and high-dimensional state and control spaces. The numerical results show that CEL can learn stable value and policy approximations, accurately handle certainty-equivalent terms, and provide reliable solutions in models where traditional grid-based methods are computationally expensive or infeasible. In particular, the high-dimensional Gaussian benchmark and the multivariate portfolio-choice application show that the method can scale to state and control dimensions that are difficult for conventional grid-based approaches. In the final evaluation, the reported out-of-sample Bellman errors and Euler or first-order residuals further show that the learned solutions remain consistent with the original Bellman equation and optimality conditions outside the training sample.
   
    To the best of our knowledge, \citet{friedl2023deep} is the only paper that uses deep learning to solve a discrete-time dynamic programming problem with recursive utility. They study an integrated assessment model with Epstein--Zin utility, which has a 12-dimensional state with a 3-dimensional shock and a 3-dimensional control. 
    Their approach builds on deep equilibrium networks \citep{azinovic2022deep}, using neural networks to approximate policy functions, Lagrange multipliers, and the value function and solving the model by minimizing residuals from a model-specific system of equations.
    
    CEL differs from this approach in several important respects. First, CEL introduces a certainty-equivalent network that directly learns the certainty-equivalent value which is a nonlinear function of the state and control. In contrast, \citet{friedl2023deep} use a 3-dimensional Gauss--Hermite quadrature to calculate the certainty equivalent by integrating against the density of the 3-dimensional shock \citep{scheidegger2026deep}. This quadrature approach becomes infeasible or inaccurate for shocks with 4 or higher dimension because of curse of dimensionality. 
    Second, CEL does not use Euler equations, first-order conditions, or model-specific residuals as training objectives. Instead, CEL learns the policy by directly maximizing the Bellman objective using the learned certainty-equivalent function. In contrast, \citet{friedl2023deep} learn the policy indirectly by minimizing residuals from a model-specific system of equations, so the policy update depends on the derived system of first-order and other model-specific equations. 
    Third, CEL uses a modular architecture that separates value approximation, policy learning, and certainty-equivalent learning, and updates these objects through an alternating learning procedure. In contrast, \citet{friedl2023deep} use a deep equilibrium network to approximate several model-specific objects jointly, including policies, multipliers, and the value function.
    Fourth, CEL updates the value function in a separate step by fitting the value network to the Bellman target constructed from the learned policy and certainty-equivalent function. In contrast, \citet{friedl2023deep} include the value function as one output of their deep equilibrium network, together with choice variables and Lagrange multipliers, and discipline it through a value function residual within the joint residual loss rather than through a separate Bellman-consistency objective. Fifth, CEL evaluates the learned value and policy functions using out-of-sample Bellman errors and Euler or first-order residuals across benchmark applications. In contrast, \citet{friedl2023deep} do not report residual-based accuracy errors or compare the learned solution with closed-form or high-accuracy numerical benchmark solutions. Sixth, because CEL does not require multiplier networks, it reduces the number of model-specific objects that must be approximated, whereas \citet{friedl2023deep} explicitly approximate Lagrange multipliers as part of the model-specific system. Seventh, CEL incorporates target networks and delayed policy updates, which substantially improve the stability of the learning process by stabilizing target values and avoiding overly frequent policy updates. In contrast, \citet{friedl2023deep} optimize all these model-specific objects jointly through a single residual loss without these separate stabilization devices. Finally, CEL uses the state transition rule only as a simulator for next-period states and does not require differentiability of the transition function. In contrast, the residual-system construction in \citet{friedl2023deep} requires the transition equations to be differentiable with respect to both state variables and policy variables in order to derive and evaluate the relevant first-order and other model-specific equations.
	
	The rest of the paper is organized as follows. Subsection \ref{subsec:literature_review} reviews related literature. Section \ref{sec:model_setting} presents the general recursive utility problem and the Bellman equation. Section \ref{sec:algorithm} introduces the CEL algorithm, including the neural-network architectures and training objectives. Section \ref{sec:results} presents the numerical results for discounted linear exponential quadratic Gaussian control, small-noise robust control, Epstein--Zin DSGE, and multivariate strategic asset allocation, and evaluates the accuracy of the learned solutions using closed-form benchmarks, VFI comparisons, out-of-sample Bellman errors, and Euler or first-order residuals. Section \ref{sec:conclusion} concludes.

	\subsection{Literature Review}
	\label{subsec:literature_review}
	
    This subsection reviews several related strands of literature: recursive utility theory and applications, dynamic programming foundations for recursive preferences, traditional numerical methods for dynamic programming with recursive utility, and machine-learning methods for dynamic programming and equilibrium computation.
	
	The theoretical foundation of recursive utility goes back to \citet{krepsPorteus1978}, \citet{epstein1989substitution}, and \citet{Weil1990}. Recursive preferences have become important in macroeconomics and finance because they disentangle risk aversion from the elasticity of intertemporal substitution. They have been widely used to study the equity premium puzzle \citep{WEIL1989}, long-run risk asset pricing \citep{bansal2004risks}, portfolio choice \citep[see, e.g.,][]{DuffieEpstein,schroder1999optimal}, fiscal policy \citep{kaplan2014model}, robust control \citep[see, e.g.,][]{hansen1995discounted,hansen2008robustness,Hansen-Sargent-2013}, and long-horizon policy evaluation. Recursive utility has also been applied to strategic asset allocation with time-varying investment opportunities \citep[see, e.g.,][]{campbell1999consumption,campbell2003multivariate}, integrated assessment models with climate risk \citep[see, e.g.,][]{Cai-Lontzek-2019,zhao2023SCC}, and consumption-investment problems in incomplete markets \citep[see, e.g.,][]{kraft2013consumption,kraft2017optimal,xing2017consumption,matoussi2018convex,feng2021optimal,feng2024consumption}. These applications show that recursive preferences are useful for studying dynamic decisions in which long-run risk, uncertainty, and intertemporal substitution play distinct roles.
	
	Recent theoretical work has studied the existence, uniqueness, and dynamic programming foundations of recursive utility. \citet{hansen2012recursive} analyze recursive utility in Markov environments with stochastic growth. \citet{christensen2022existence}, \citet{pohl2024existence}, and \citet{stachurski2024asset} provide existence and uniqueness results for recursive utility and asset-pricing models with unbounded states, stochastic volatility, rare disasters, ambiguity, and preference shocks. \citet{stachurski2021dynamic}, \citet{ma2021dynamic}, \citet{ma2022unbounded}, and \citet{Jaskiewicz2024recursive} further develop dynamic programming methods for recursive preferences. These papers clarify when recursive utility is well defined and when Bellman equations are valid. Our paper builds on this theoretical foundation but focuses on a computational method for solving such models.
	
	Traditional numerical methods provide important tools for dynamic programming and recursive utility models, but in practice they are usually limited to low-dimensional or low- to moderate-dimensional problems. Standard approaches include value function iteration, policy iteration, projection methods, perturbation methods, and Euler-equation methods; see \citet{Judd-1998}, \citet{Miranda-Fackler-2002}, and \citet{carroll2006,Carroll2011,Carroll2022_updated} for general treatments. Accuracy evaluation in this literature often relies on residual-based diagnostics. \citet{santos2000accuracy} provides theoretical foundations for using Euler equation residuals to assess numerical solutions by relating the size of Euler residuals to the approximation errors of policy and value functions. In recursive-utility settings, value function iteration and policy iteration have been used by \citet{ma2021dynamic}, \citet{ma2022unbounded}, \citet{stachurski2021dynamic}, \citet{coeurdacier2020financial}, \citet{Cai-Lontzek-2019}, and \citet{zhao2023SCC}. Projection-based methods have been applied to portfolio-choice and climate-economy models with recursive preferences \citep[see, e.g.,][]{garlappi2010solving,cai2013social,cai2020numericalsolutiondynamicportfolio,dixon2023time}. Perturbation and local approximation methods are common in DSGE models with recursive preferences \citep[see, e.g.,][]{tallarini2000risk,andreasen2012effects}, while global solution approaches have also been developed for macro-financial applications \citep[see, e.g.,][]{cao2023global,cao2023uncovering}. These methods often rely on grids, basis functions, local approximations, or problem-specific structures, and their computational cost grows rapidly with the dimension of the state and control spaces. As a result, they become difficult to apply to high-dimensional recursive utility models.
	
There is a growing literature that develops machine learning methods for dynamic programming problems with expected utility. These studies use neural networks, Gaussian processes, and simulation-based methods to approximate value functions and policy functions, and to train or evaluate these approximations using model-specific conditions, including Bellman equations, optimality conditions, feasibility restrictions, and other consistency conditions, mainly in expected-utility models.

\citet{HanE2016} develop deep learning methods for solving high-dimensional discrete-time finite-horizon dynamic programming problems with expected utility. For representative-agent infinite-horizon models, \citet{Renner-Scheidegger-2018} and \citet{scheidegger-Bilionis-2019} use Gaussian process regression to approximate value functions, \citet{Lepetyuk2020} use neural networks to learn policy functions, and \citet{valaitis2024machine} use neural networks to approximate expectation terms in Euler equations. \citet{maliar2021deep} propose a general deep learning framework based on lifetime utility maximization, Euler-equation error minimization, and Bellman-equation error minimization. Related work studies finite-horizon and overlapping-generations models using deep learning to approximate value and policy functions and to minimize residuals from model-specific systems of equations \citep[see, e.g.,][]{duarte2021simple,azinovic2022deep,Azinovic2023}. \citet{Pascal2024_published} and \citet{Skavysh2023} further develop Monte Carlo and Bellman-error-based approaches for dynamic economic models.
	
	Machine learning has also been applied to heterogeneous-agent and continuous-time models. \citet{maliar2021deep} demonstrate deep learning methods in both representative-agent and heterogeneous-agent environments. \citet{Han2021} combine supervised and unsupervised learning using neural networks for value and policy approximation. \citet{HallHoffarth2023} approximate heterogeneous-agent New Keynesian models with neural networks trained on first-order-condition errors and hard output constraints. In continuous-time heterogeneous-agent settings, \citet{huang2023probabilistic} and \citet{huang2023breaking} use deep learning methods based on probabilistic formulations and forward-backward stochastic differential equations. Related continuous-time developments in applied mathematics use deep neural networks to solve high-dimensional PDEs, BSDEs, HJB equations, and stochastic control problems \citep[see, e.g.,][]{EHan2017_published,BeckEJ2019_published,Hure-Pham-Bachouch-Lang-2021,Bachouch_2021,Reppen2023}. In continuous-time finance, \citet{duarte2024machine} solve equilibrium models with Epstein--Zin preferences by minimizing Hamilton--Jacobi--Bellman residuals. Their framework is closely tied to continuous-time diffusion or jump-diffusion representations, whereas CEL is designed for discrete-time Bellman equations and only requires forward simulation of next-period states.
	
	A related data-driven and simulation-based literature uses approximate dynamic programming and reinforcement learning to learn value functions, policies, or decision rules from observed or simulated transition data. General references include \citet{Sutton-Barto-1998}, \citet{Powell-2011}, and \citet{Bertsekas-2012}. Applications in economics and finance include \citet{feng2023optimal}, \citet{aboussalah2022value}, and \citet{jiang2025high}. These methods provide flexible tools for learning from transition data, but most of them are designed for expected-utility or time-additive objective functions rather than recursive utility.
	
	Relative to these strands of literature, this paper extends deep learning methods from expected-utility dynamic programming to recursive utility from a certainty-equivalent learning perspective. The key distinction is that CEL directly learns the certainty-equivalent value in the Bellman equation and uses this learned object for policy improvement. This design makes the method mesh-free, simulation-based, and applicable to recursive utility models with nonlinear certainty-equivalent terms, high-dimensional state variables, and general discrete-time stochastic dynamics.

	\section{The Dynamic Programming Problem with Recursive Utility}
	\label{sec:model_setting}
	
	We consider a discrete-time infinite-horizon dynamic programming problem with recursive utility. Time is indexed by $t=0,1,2,\ldots$. At time period $t$, the agent observes a state vector $s_t\in\mathbb{R}^{n_s}$ and chooses a control vector $c_t\in\mathbb{R}^{n_c}$. We write the policy function as
	
	\begin{equation}
		c_t = c(s_t).
	\end{equation}
	The control must satisfy the feasibility constraint
	\begin{equation}
		c_t \in \mathcal{A}(s_t),
		\label{eq:c_t_constraint}
	\end{equation}
	where $\mathcal{A}(s_t)$ denotes the admissible set of controls conditional on the current state. The state evolves according to
	\begin{equation}
		s_{t+1}
		=
		\psi(s_t,c_t,z_{t+1}),
		\label{eq:state_evo_inf}
	\end{equation}
	where $\psi(\cdot)$ is the transition function and $z_{t+1}$ is a random shock. Conditional on $(s_t,c_t)$, the next state is independent of past states, so the state process is Markovian. Additional history dependence can be incorporated by augmenting the state vector.
	
	We adopt a general recursive utility specification and use $V(s_t)$ to denote the value function at state $s_t$. The key object is the certainty-equivalent value conditional on the current state and control. We define it as
	\begin{equation}
		V_c(s_t,c_t)
		\equiv
		f^{-1}
		\left(
		\mathbb{E}
		\left[
		f\left(V(s_{t+1})\right)
		\mid s_t,c_t
		\right]
		\right),
		\label{eq:ce_operator_general_inf}
	\end{equation}
	where $f$ captures the agent's attitude toward risk and $f^{-1}$ is well defined on the relevant range. Thus, $V_c(s_t,c_t)$ is a state-control certainty-equivalent value: for each current state $s_t$ and candidate control $c_t$, it summarizes the nonlinear conditional certainty-equivalent value generated by the next-period value function.
	
	Recursive utility is then defined by
	\begin{equation}
		V(s_t)
		=
		w\left(
		s_t,
		c_t,
		V_c(s_t,c_t)
		\right),
		\qquad t=0,1,2,\ldots,
		\label{eq:recursive_utility_general_inf}
	\end{equation}
	where $w$ is the time aggregator. This notation keeps the state $s_t$, the control $c_t$, the value function $V(s_t)$, the certainty-equivalent value $V_c(s_t,c_t)$, and the aggregator $w$ consistent with the Bellman equation below.
	
	The decision problem is to choose an admissible Markov policy to maximize the initial value,
	\begin{equation}
		\label{eq:recursive_utility_objective}
		\begin{aligned}
			\max_{\{c_t\}_{t\ge 0}}\quad
			& V(s_0) \\
			\text{subject to}\quad
			& c_t\in\mathcal{A}(s_t),
			\qquad t=0,1,2,\ldots, \\
			& s_{t+1}=\psi(s_t,c_t,z_{t+1}),
			\qquad t=0,1,2,\ldots .
		\end{aligned}
	\end{equation}
	The associated Bellman equation is
	\begin{equation}
		V(s_t)
		=
		\max_{c_t\in\mathcal{A}(s_t)}
		w\left(
		s_t,
		c_t,
		V_c(s_t,c_t)
		\right),
		\label{equ:gen_bellman_equ_inf}
	\end{equation}
	where \(V_c(s_t,c_t)\) is the certainty-equivalent value defined in \eqref{eq:ce_operator_general_inf}, and the next-period state is generated by \eqref{eq:state_evo_inf}.
	
	Equivalently, using generic state variables $s$ and $s'$, the Bellman equation can be written as
	\begin{equation}
		V(s)
		=
		\max_{c\in\mathcal{A}(s)}
		w\left(
		s,
		c,
		V_c(s,c)
		\right),
		\label{equ:gen_bellman_equ_inf2}
	\end{equation}
	where
	\begin{equation}
		V_c(s,c)
		\equiv
		f^{-1}
		\left(
		\mathbb{E}
		\left[
		f\left(V(s')\right)
		\mid s,c
		\right]
		\right),
		\qquad
		s'=\psi(s,c,z').
		\label{eq:generic_vc_definition}
	\end{equation}
	Here $s'$ denotes the next-period state generated from the current state $s$, the control $c$, and the shock realization $z'$. Solving the problem amounts to finding a value function, a policy function, and a certainty-equivalent function that jointly satisfy the Bellman equation.
	
	The notation nests several standard specifications in recursive utility and robust control. Risk-sensitive and robust-control formulations commonly use exponential certainty equivalents, as in the risk-sensitive and robust-control literatures \citep[see, e.g.,][]{hansen1995discounted,hansen2008robustness,Hansen-Sargent-2013,tallarini2000risk}. Kreps--Porteus and Epstein--Zin preferences separate risk aversion from intertemporal substitution and generate power certainty equivalents of continuation utility \citep[see, e.g.,][]{krepsPorteus1978,epstein1989substitution,WEIL1989,Weil1990,DuffieEpstein}. Two important examples can be written as special cases of the same notation. In risk-sensitive control, one commonly uses
	\begin{equation}
		f(X)=\exp\left(-\frac{\sigma}{2}X\right),
		\qquad
		V_c(s_t,c_t)
		=
		-\frac{2}{\sigma}
		\log
		\mathbb{E}
		\left[
		\exp\left(
		-\frac{\sigma}{2}V(s_{t+1})
		\right)
		\mid s_t,c_t
		\right],
		\label{eq:risk_sensitive_ce_example}
	\end{equation}
	so that the certainty-equivalent value is an exponential certainty equivalent. In Epstein--Zin preferences, for positive certainty-equivalent values,
	\begin{equation}
		f(X)=X^{1-\gamma},
		\qquad
		V_c(s_t,c_t)
		=
		\left(
		\mathbb{E}
		\left[
		V(s_{t+1})^{1-\gamma}
		\mid s_t,c_t
		\right]
		\right)^{\frac{1}{1-\gamma}},
		\label{eq:ez_ce_example}
	\end{equation}
	and a standard aggregator can be written as
	\begin{equation}
		V(s_t)
		=
		\left[
		(1-\beta)c_t^{1-1/\psi}
		+
		\beta V_c(s_t,c_t)^{1-1/\psi}
		\right]^{\frac{1}{1-1/\psi}}.
		\label{eq:ez_aggregator_example}
	\end{equation}
	These examples illustrate that the same Bellman structure covers both exponential risk adjustments and Epstein--Zin certainty-equivalent values.
	
	The key difficulty is the nonlinear certainty-equivalent value \(V_c(s,c)\). The conditional expectation is taken after applying the nonlinear transformation \(f\) to the future value function, and the inverse transformation is then applied to the expectation. This nested nonlinear state-control object is the central quantity that CEL is designed to approximate.

	\section{The Certainty Equivalent Learning (CEL) Algorithm}
	\label{sec:algorithm}
	
	This section is organized around the main components of CEL. Section~\ref{subsec:background_challenges} explains why nonlinear certainty-equivalent values create a distinct numerical challenge. Section~\ref{subsec:network_framework} introduces the neural-network representations of the value, policy, and certainty-equivalent value. Section~\ref{subsec:alternating_optimization} describes the alternating update strategy. The remaining subsections discuss stabilizing enhancements and then summarize the full training procedure.

    \subsection{Challenges}
    \label{subsec:background_challenges}
    
    Recursive utility changes the computational nature of dynamic programming in a fundamental way. The discussion in the Introduction identifies five related challenges. This subsection explains these challenges in more detail and shows how they motivate the CEL architecture.
    
    The first challenge is that recursive utility generally has no explicit representation in terms of future states and controls. With expected utility, once a policy $c_t=c(s_t)$ is fixed, the value function can often be represented as the sum of a discounted stream of future utilities generated by that policy. In many benchmark models, this structure either yields an explicit expansion in terms of future states and controls or reduces the problem to a standard conditional-mean regression. With recursive utility, this simplification is generally unavailable. The value function is defined implicitly as the fixed point of the Bellman equation, and the current value depends recursively on the future value function. As a result, the Bellman equation cannot usually be unfolded into a simple discounted sum of future utilities.
    
    The second challenge is that the Bellman equation contains a nonlinear certainty-equivalent term $V_c(s_t, c_t)=
    f^{-1}
    \left(
    \mathbb{E}_t
    \left[
    f\left(V(s_{t+1})\right)
    \mid s_t,c_t\right]\right)$,
    where $f$ is a nonlinear function. This term is a nonlinear transformation of the conditional expectation of a nonlinear function of future utility. It is also a function of both the current state and the current control. In high-dimensional models, directly evaluating this state-control certainty-equivalent value is difficult because it requires approximating a nonlinear conditional object over a high-dimensional state-control space. This expression also clarifies how expected utility arises as a special case. If $f$ is linear and the aggregator over time is additive, the certainty-equivalent term reduces to the ordinary conditional expectation of future value. The numerical difficulty studied here arises when the certainty equivalent is no longer a simple conditional mean.

    The third challenge is that finite-sample average is an biased estimate of the certainty equivalent. With expected utility, the future-value term is linear in the conditional expectation, so a sample average of simulated future values is an unbiased estimator of the conditional expectation. With recursive utility, however, the nonlinear transformation makes the plug-in estimator biased in general. Let
    $s_{t+1}^{j}, j=1,\ldots, G$ be i.i.d. samples of $s_{t+1}$. Define
    \[
    M(s_t,c_t)
    =
    \mathbb{E}_t
    \left[
    f\left(V(s_{t+1})\right)
    \mid s_t,c_t
    \right],
    \qquad
    \widehat M
    =
    \frac{1}{G}
    \sum_{j=1}^{G}
    f\left(V(s_{t+1}^{j})\right).
    \]
    Although $\widehat M$ is an unbiased estimator of $M(s_t,c_t)$ conditional on $(s_t,c_t)$, the plug-in estimator $f^{-1}(\widehat M)$ is generally not unbiased for $f^{-1}(M(s_t,c_t))=V_c(s_t, c_t)$, because $f^{-1}$ is nonlinear. More explicitly,
    \[
    \mathbb{E}_t
    \left[
    f^{-1}\left(\widehat M\right)
    \mid s_t,c_t
    \right]
    \neq
    V_c(s_t, c_t)
    \]
    in general. By a local expansion, the bias is approximately $\left(f^{-1}\right)''(M(s_t,c_t))\operatorname{Var}_t(\widehat M)/2$. Hence the bias can become more important when the certainty-equivalent transformation is strongly curved or when the conditional distribution of future values is dispersed, as in high-risk-aversion, risk-sensitive, or robust-control specifications. This is why a naive Monte Carlo treatment of the certainty-equivalent term may be unstable under recursive utility, and this motivates learning the certainty-equivalent value directly. In the expected-utility special case, this finite-sample bias disappears. When $f^{-1}$ is linear, applying $f^{-1}$ to the sample mean preserves unbiasedness, so the plug-in estimator remains unbiased for the continuation term.

    The fourth challenge is that, during policy improvement, the certainty-equivalent value must be evaluated at candidate controls, including controls that are not optimal, rather than only at the optimal control. The relevant object is not merely a state function; it is a state-control function. To compare alternative controls at the same current state, the algorithm needs the certainty-equivalent value associated with each candidate control. Therefore, a value network evaluated only at future states does not by itself provide all the information needed for the maximization step. This motivates learning the certainty-equivalent value as a separate state-control function, which maps each state-control pair into the corresponding certainty-equivalent value. This is the role of the certainty-equivalent network in the three-network architecture.
    
    The fifth challenge is the tight coupling between certainty-equivalent learning and policy improvement. The certainty-equivalent target depends on the future value approximation, while the policy network is updated using the learned certainty-equivalent value. The updated policy then changes the distribution of future simulated states. As a result, errors in the value approximation can affect certainty-equivalent learning, errors in the certainty-equivalent approximation can affect policy improvement, and the policy update can change the future states on which the next certainty-equivalent targets are constructed. These feedback loops can amplify approximation errors across iterations, especially in models with stochastic volatility, rare events, strong risk sensitivity, or high curvature in preferences. CEL therefore separates the learning problem into functional blocks. The value network approximates the value function, the policy network represents the policy function, and the certainty-equivalent network approximates the nonlinear certainty-equivalent term. In the four-network architecture, the certainty-equivalent value is further decomposed into an expected continuation network and a nonlinear difference network. These architectures should be understood as different decompositions of the Bellman equation under recursive utility, rather than as a literal count of how many neural-network modules are used in a particular implementation.

	\subsection{Neural Network Framework}
	\label{subsec:network_framework}
	
	CEL represents the objects in the Bellman equation with neural networks and updates them sequentially. Throughout this section, $s_t$ denotes the current state, $c_t$ denotes the current control, $w$ denotes the time aggregator, and $f^{-1}(\mathbb{E}_t[f(\cdot)])$ denotes the certainty-equivalent transformation. We use the following names consistently. The \emph{value network} approximates the value function, the \emph{policy network} represents the policy function, the \emph{certainty-equivalent network} directly approximates the nonlinear certainty-equivalent value, the \emph{expected continuation network} approximates the ordinary conditional expectation of next-period value, and the \emph{nonlinear difference network} approximates the gap between the expected-value component and the certainty-equivalent value. We consider three related architectures. 
    The terms three-network, four-network, and two-network refer to different functional decompositions of the Bellman equation under recursive utility rather than to the literal number of subnetworks used in every application.
    For example, when a model has several control components, such as consumption and labor or consumption and portfolio weights, these components may be represented by separate policy subnetworks but are treated as one policy-network block in the CEL architecture. The three-network architecture is the baseline specification, the four-network architecture further decomposes the certainty equivalent into an expected continuation component and a nonlinear difference component, and the two-network architecture provides a more compact representation by removing
	the explicit value network and defining the current value implicitly through the policy and certainty-equivalent networks.
	
	\subsubsection{Three-network architecture}
	
	The baseline architecture uses one value-network block, one policy-network block, and one certainty-equivalent-network block:
	\begin{align}
		V(s_t) &\equiv V(s_t;\phi),
		\label{nn:three_value}
		\\
		c_t &\equiv c(s_t;\theta),
		\label{nn:three_policy}
		\\
		V_c(s_t,c_t;\xi)
		&\equiv
		f^{-1}
		\left(
		\mathbb{E}_t
		\left[
		f\left(V(s_{t+1};\phi)\right)
		\mid s_t,c_t
		\right]
		\right).
		\label{nn:three_ce}
	\end{align}
	The value network $V(s_t;\phi)$ approximates the value function, the policy network $c(s_t;\theta)$ approximates the policy function, and the certainty-equivalent network $V_c(s_t,c_t;\xi)$ approximates the nonlinear certainty-equivalent value used by the Bellman operator. 
	
	\subsubsection{Four-network architecture}
	
	The four-network architecture separates the certainty-equivalent value into an expected continuation component and a nonlinear difference component:
	\begin{align}
		V(s_t) & \equiv V(s_t; \phi), \label{nn:vf_four} \\
		c_t & \equiv c(s_t; \theta), \label{nn:control_four} \\
		\mathbb{E}_t\left[V(s_{t+1})\mid s_t,c_t\right]
		&\equiv V_e(s_t,c_t;\rho), \label{nn:expectation} \\
		f^{-1}\left(
		\mathbb{E}_t\left[f\left(V(s_{t+1})\right)\mid s_t,c_t\right]
		\right)
		&\equiv V_e(s_t,c_t;\rho)-D(s_t,c_t;\nu).
		\label{nn:difference}
	\end{align}
	Equivalently,
	\begin{equation}
		V_c(s_t,c_t)
		=
		V_e(s_t,c_t;\rho)-D(s_t,c_t;\nu).
		\label{eq:vc_ve_d_relation}
	\end{equation}
	Here $V_e$ is the expected continuation network, which learns the ordinary conditional expectation of the future value, while $D$ is the nonlinear difference network, which learns the Jensen-type gap between the expected-value component and the certainty equivalent. This decomposition is useful when the nonlinear transformation is highly curved.
	
	The sign of $D$ follows from Jensen's inequality and the monotonicity of $f^{-1}$. Let $X_{t+1}=V(s_{t+1})$. If $f$ is increasing and concave, Jensen's inequality implies
	\begin{equation}
		f\left(\mathbb E_t[X_{t+1}]\right)
		\geq
		\mathbb E_t[f(X_{t+1})].
	\end{equation}
	Since $f^{-1}$ is increasing, applying $f^{-1}$ to both sides gives
	\begin{equation}
		\mathbb E_t[X_{t+1}]
		\geq
		f^{-1}\left(\mathbb E_t[f(X_{t+1})]\right).
	\end{equation}
	Therefore,
	\begin{equation}
		V_e(s_t,c_t)\geq V_c(s_t,c_t),
		\qquad
		D(s_t,c_t)\geq 0.
		\label{eq:jensen_positive_d}
	\end{equation}
	The same conclusion holds when $f$ is decreasing and convex, because applying the decreasing inverse $f^{-1}$ reverses the Jensen inequality. In this case, the four-network implementation can impose non-negativity on $D$, for example by using a positive output activation.
	
	Conversely, if $f$ is increasing and convex, or decreasing and concave, the Jensen inequality is reversed after accounting for the monotonicity of $f^{-1}$. Hence,
	\begin{equation}
		V_e(s_t,c_t)\leq V_c(s_t,c_t),
		\qquad
		D(s_t,c_t)\leq 0.
		\label{eq:jensen_negative_d}
	\end{equation}

	\subsubsection{Two-network architecture}
	
	The two-network architecture removes the explicit value network and keeps only the policy-network block and the certainty-equivalent-network block. The two networks are defined by
	\begin{align}
		c_t &\equiv c(s_t;\theta),
		\label{nn:two_policy}
		\\
		V_c(s_t,c_t;\xi)
		&\equiv
		f^{-1}
		\left(
		\mathbb{E}_t
		\left[
		f\left(V(s_{t+1})\right)
		\mid s_t,c_t
		\right]
		\right).
		\label{nn:two_ce}
	\end{align}
	Here $V_c(s_t,c_t;\xi)$ is the certainty-equivalent network, and the right-hand side uses the value function $V$ to describe the population target rather than to define another network. Since there is no independent value network in this architecture, the current value used during training is represented implicitly by
	\begin{equation}
		\widehat V(s_t;\theta,\xi)
		\equiv
		w\left(
		s_t,
		c(s_t;\theta),
		V_c(s_t,c(s_t;\theta);\xi)
		\right).
		\label{nn:two_implicit_value}
	\end{equation}
	This specification reduces the number of networks and forces the value used in policy improvement to be consistent with the learned certainty-equivalent function. It is most useful when the certainty-equivalent network is accurate enough to support bootstrapping without an independently fitted value network.

	\subsection{Alternating Optimization Strategy}
	\label{subsec:alternating_optimization}
	
	CEL updates the neural networks by alternating among certainty-equivalent estimation, policy improvement, and value-function fitting. The precise set of updates depends on whether the implementation uses the three-network, four-network, or two-network architecture.
	
	\medskip
	\noindent\textbf{Three-network architecture.}
	Given samples $\{(s_t^i,c_t^i,s_{t+1}^{i,j})\}_{i,j}$, the certainty-equivalent network is trained by
	\begin{equation}
		\min_{\xi \in \Theta^\xi}
		\frac{1}{N}\sum_{i=1}^{N}
		\left(
		f\left(V_c(s_t^i,c_t^i;\xi)\right)
		-
		\frac{1}{G}\sum_{j=1}^{G}
		f\left(V(s_{t+1}^{i,j};\phi)\right)
		\right)^2.
		\label{eq:update_inf_xi}
	\end{equation}
	Given the learned certainty-equivalent function, the policy network is updated by
	\begin{equation}
		\max_{\theta \in \Theta^\theta}
		\frac{1}{N}\sum_{i=1}^{N}
		w\left(s_t^i,c(s_t^i;\theta),V_c(s_t^i,c(s_t^i;\theta);\xi)\right),
		\label{eq:update_inf_theta}
	\end{equation}
	and the value network is fitted to the Bellman target:
	\begin{equation}
		\min_{\phi \in \Theta^\phi}
		\frac{1}{N}\sum_{i=1}^{N}
		\left(
		V(s_t^i;\phi)
		-
		w\left(s_t^i,c(s_t^i;\theta),V_c(s_t^i,c(s_t^i;\theta);\xi)\right)
		\right)^2.
		\label{eq:update_inf_phi}
	\end{equation}
	
	\medskip
	\noindent\textbf{Four-network architecture.}
	The four-network version first fits the expected continuation network,
	\begin{equation}
		\min_{\rho \in \Theta^\rho}
		\frac{1}{N}\sum_{i=1}^{N}
		\left(
		V_e(s_t^i,c_t^i;\rho)
		-
		\frac{1}{G}\sum_{j=1}^{G}V(s_{t+1}^{i,j};\phi)
		\right)^2,
		\label{eq:update_ve_rho}
	\end{equation}
	and then fits the nonlinear difference network,
	\begin{equation}
		\min_{\nu \in \Theta^\nu}
		\frac{1}{N}\sum_{i=1}^{N}
		\left(
		f\left(V_e(s_t^i,c_t^i;\rho)-D(s_t^i,c_t^i;\nu)\right)
		-
		\frac{1}{G}\sum_{j=1}^{G}f\left(V(s_{t+1}^{i,j};\phi)\right)
		\right)^2.
		\label{eq:update_d_nu}
	\end{equation}
	The policy and value updates are
	\begin{equation}
		\max_{\theta \in \Theta^\theta}
		\frac{1}{N}\sum_{i=1}^{N}
		w\left(s_t^i,c(s_t^i;\theta),V_e(s_t^i,c(s_t^i;\theta);\rho)-D(s_t^i,c(s_t^i;\theta);\nu)\right),
		\label{eq:update_theta_fournet}
	\end{equation}
	\begin{equation}
		\min_{\phi \in \Theta^\phi}
		\frac{1}{N}\sum_{i=1}^{N}
		\left(
		V(s_t^i;\phi)-w\left(s_t^i,c(s_t^i;\theta),V_e(s_t^i,c(s_t^i;\theta);\rho)-D(s_t^i,c(s_t^i;\theta);\nu)\right)
		\right)^2.
		\label{eq:update_phi_fournet}
	\end{equation}
	
	\medskip
	\noindent\textbf{Two-network architecture.}
	For the two-network version, there is no separate value update. The implicit value is
	\begin{equation}
		\widehat V(s;\theta,\xi)
		=
		w\left(s,c(s;\theta),V_c(s,c(s;\theta);\xi)\right).
		\label{eq:implicit_value_vc_alt}
	\end{equation}
	Thus, the certainty-equivalent network is trained by replacing the explicit next-period value in \eqref{eq:update_inf_xi} with the implicit value \(\widehat V(s_{t+1};\theta,\xi)\). In other words, the update uses the policy network, the certainty-equivalent network, and the implicit value representation in \eqref{nn:two_policy}--\eqref{nn:two_implicit_value}. The policy update remains the same as \eqref{eq:update_inf_theta}, but the current value is always evaluated through the implicit composition in \eqref{eq:implicit_value_vc_alt}. This compact variant is therefore a two-network counterpart to the three-network CEL scheme.
	
	\subsection{Algorithmic Enhancements for Stability and Efficiency}
	
	To enhance the numerical robustness and convergence properties of CEL, we introduce several methodological refinements. These improvements address numerical challenges in deep learning for dynamic programming, including instability arising from bootstrapped targets, insufficient exploration in high-dimensional control spaces, and value-estimation errors caused by overly frequent policy updates. The proposed modifications consist of three components: exploratory control perturbation during path simulation, target networks for stabilized value updates, and delayed policy updates for more stable policy improvement.
	
	The subsequent discussion applies to the three-network, four-network, and two-network variants. In the four-network case, the certainty-equivalent value is decomposed as $V_c=V_e-D$; in the three-network case, the same object is learned directly as $V_c$; and in the two-network case, the current value is represented implicitly through $w(s_t,c(s_t;\theta),V_c(s_t,c(s_t;\theta);\xi))$. Hence the stabilization devices below can be adapted to all three architectural variants with only minor notational changes.

	\subsubsection{Exploratory Control Perturbation During Path Simulation}
	
	Accurate policy estimation requires not only precise function approximation but also sufficient exploration of the control space. In high-dimensional decision environments, naively following the policy network's output may lead to premature convergence to local optima or under-exploration of feasible states. To mitigate this issue, we implement a randomized perturbation mechanism that injects structured exploration into the simulation process. This procedure is provided in \ref{alg:exploration}.
	
	The simulation process is designed to cover the relevant regions of the state space. Each simulation epoch begins by sampling initial states $S_0$ from the estimated stationary distribution of the controlled process, ensuring that our policy evaluation starts from states that are representative of the long-run behavior of the system. This approach provides a more stable foundation for evaluating policy performance compared to fixed or arbitrary initial conditions. For state variables and environmental parameters not explicitly governed by the policy network, we employ a conservative initialization strategy, drawing their values from a predefined uniform distribution over their plausible ranges. This ensures that our exploration covers a diverse set of scenarios while maintaining physiological and economic plausibility.
	
	At each iteration, when we update the network $V_e(s_t, c_t; \rho)$ and $D(s_t, c_t; \nu)$, we simulate the Gaussian path exploration process in reinforcement learning, with a specific focus on adding exploratory noise to the policy actions. The simulation starts from these properly initialized states and generates $N$ trajectories of length $T$. At each time step $t$, it first generates the base action using the policy network $c(s_t; \theta)$ based on the current state. Subsequently it adds a uniformly distributed random perturbation to the action space. This design maintains the policy's dominance while encouraging the system to explore unknown state regions through targeted exploration. Finally, the state for the next time step is updated according to the transition equation, completing a single iteration. This entire process repeats for $T$ steps, ultimately outputting the complete trajectories containing both states and actions. This procedure is provided in \ref{alg:exploration}.
	
	Since the input to the neural network $V_e(s_t, c_t; \rho)$ and $D(s_t, c_t; \nu)$ includes not only the state $s_t$ but also the action $c_t$, this simulation process updates the values not just for the state-action pairs $(s_t, a_t)$ on the optimal path, but for all pairs within its neighborhood.
	
	\vspace{1em}
	\noindent \textbf{Case 1: Discounted Linear Exponential Quadratic Gaussian Control.}  
	In this benchmark case, the control network outputs unbounded real-valued actions (e.g., via linear or ReLU activations), and exploration is enhanced through Gaussian-distributed noise injection. Specifically, for a subset of simulated paths, we perturb the network-generated action $c_t$ as:
	\[
	\tilde{c}_t = c(s_t; \theta) + \zeta \cdot \eta_t, \quad \eta_t \sim \mathcal{N}(0, I),
	\]
	where $\zeta$ is a scaling hyperparameter and $I$ denotes the identity matrix. To maintain numerical stability, the perturbed values are clipped to a prescribed feasible range when necessary, ensuring compatibility with the underlying physical or economic constraints.
	
	\noindent \textbf{Case 2: High-Dimensional Control with Low-Discrepancy Sequences.}  
	For problems where the control space is high-dimensional and standard Gaussian perturbations may lead to inefficient exploration, we employ a quasi-Monte Carlo approach using Sobol sequences. This strategy generates deterministic, low-discrepancy perturbations that provide more uniform coverage of the control space. The perturbation is defined as:
	\[
	\tilde{c}_t = c(s_t; \theta) + \zeta \cdot (2q_t - 1),
	\]
	where $q_t \in [0,1]^{n_c}$ is a vector drawn from a $n_c$-dimensional Sobol sequence, and $(2q_t - 1)$ linearly transforms the sequence to the interval $(-1,1)^{n_c}$. This approach yields a systematic exploration pattern that covers the control space more evenly than random Gaussian noise, particularly in high-dimensional settings, while maintaining the perturbation magnitude controlled by the scaling hyperparameter $\zeta$.
	
	\vspace{1em}
	\begin{algorithm}[]
		\caption{Exploratory Control Perturbation During Path Simulation}
		\label{alg:exploration}
		\begin{algorithmic}[1]
			\State \textbf{Input:} initial state $s_0$, policy network $c(\cdot; \theta)$, transition rule $\psi$, shock distribution $P_z$, simulation horizon $T$, perturbation $\text{Method} \in \{\text{Gaussian}, \text{Sobol}\}$, scale factor $\zeta$, control dimension $n_c$
			\State \textbf{Optional:} Pre-initialized Sobol sequence generator $\mathcal{Q}$ for $\text{Method} = \text{Sobol}$
			\State \textbf{Output:} sequence of perturbed controls $\{\tilde{c}_t\}_{t=0}^{T-1}$, state trajectory $\{s_t\}_{t=0}^{T}$
			\State Initialize $s \gets s_0$
			\For{$t = 0$ to $T-1$}
			\If{$\text{Method} = \text{Gaussian}$}
			\State Sample noise vector: $\eta_t \sim \mathcal{N}(\mathbf{0}, I_{n_c})$
			\State Apply additive perturbation: $\tilde{c}_t \gets c(s; \theta) + \zeta \cdot \eta_t$
			\ElsIf{$\text{Method} = \text{Sobol}$}
			\State Generate low-discrepancy point: $q_t \gets \text{next}(\mathcal{Q})$
			\State Scale point to $(-1,1)^{n_c}$: $\xi_t \gets 2q_t - 1$
			\State Apply deterministic perturbation: $\tilde{c}_t \gets c(s; \theta) + \zeta \cdot \xi_t$
			\EndIf
			\State \textbf{Enforce feasibility:} Clip $\tilde{c}_t$ to the admissible control range $[c_{\text{min}}, c_{\text{max}}]$
			\State Sample shock $z_{t+1}\sim P_z$
			\State \textbf{Step dynamics:} Update state $s \gets \psi(s, \tilde{c}_t,z_{t+1})$
			\EndFor
			\State \Return $\{\tilde{c}_t\}_{t=0}^{T-1}, \{s_t\}_{t=0}^{T}$
		\end{algorithmic}
	\end{algorithm}

	\subsubsection{Target Networks for Stabilized Value Updates}
	
	A fundamental challenge in solving dynamic programming problems with recursive utility using deep learning stems from the inherent instability of bootstrapping in value function estimation. Recursive utility specifications, which nest nonlinear transformations of future utilities within current decisions, are particularly susceptible to this issue. When the same network parameters are used to both generate and fit temporal-difference targets, approximation errors can propagate and amplify through successive updates, leading to training divergence or oscillatory behavior.
	
	To address this instability, we implement a target network architecture following the methodology pioneered by \citet{lillicrap2015continuous}. This mechanism maintains separate, slowly-evolving copies of key networks that are used exclusively for constructing bootstrapped targets. The targets are thereby decoupled from the rapidly changing online networks, introducing a form of hysteresis that improves learning stability. The parameters of these target networks are updated via Polyak averaging, ensuring a smooth and controlled evolution of the learning targets. This procedure is provided in \ref{alg:target_soft_update}.
	
	\vspace{0.5em}
	\noindent \textbf{Three-Network Architecture.}
	In the three-network framework, we maintain a target network for the certainty-equivalent network. The parameters of this target network, denoted $\bar{\xi}$, are updated via soft averaging according to:
	\begin{align}
		\bar{\xi}^{(k)} = \tau \xi^{(k)} + (1 - \tau)\bar{\xi}^{(k-1)},
	\end{align}
	where $\tau \ll 1$ is a small update rate that controls the degree of smoothing. This target network, $\bar{V}_c(s_t, c_t; \bar{\xi})$, is used to provide fixed targets for the value function update, leading to the stabilized objective:
	\begin{equation}
		\min_{\phi} \frac{1}{N}\sum_{i=1}^{N} \left( V(s_t^i; \phi) - w\left( s_t^i, {c}(s_t^i; {\theta}),  \bar{V}_c(s_t^i, {c}(s_t^i; {\theta}); \bar{\xi}) \right) \right)^2. \label{eq:phi_target_3net}
	\end{equation}
	Here, the certainty equivalent $\bar{V}_c$ is evaluated using target parameters, effectively freezing the target during the value function optimization.
	
	The same target-network construction applies to the two-network architecture, where the target certainty-equivalent network is used to construct the bootstrapped implicit value, while no separate value-network target is needed.

	\vspace{0.5em}
	\noindent \textbf{Four-Network Architecture.}
	The extension to the four-network architecture necessitates maintaining target parameters for both the expected continuation network and nonlinear difference network. The update rules are given by:
	\begin{align}
		\bar{\rho}^{(k)} &= \tau \rho^{(k)} + (1 - \tau)\bar{\rho}^{(k-1)}, \\
		\bar{\nu}^{(k)} &= \tau \nu^{(k)} + (1 - \tau)\bar{\nu}^{(k-1)}.
	\end{align}
	The target certainty-equivalent value is then constructed as the combination:
	\begin{equation}
		\bar{V}_c(s_t, c_t) := \bar{V}_e(s_t, c_t; \bar{\rho}) - \bar{D}(s_t, c_t; \bar{\nu}).
	\end{equation}
	This formulation yields the corresponding stabilized objective functions for policy and value learning:
	\begin{align}
		\min_{\phi} \quad & \frac{1}{N}\sum_{i=1}^{N} \left( V(s_t^i; \phi) - w\left( s_t^i, {c}(s_t^i; {\theta}), \bar{V}_e(s_t^i, c_t^i; \bar{\rho}) - \bar{D}(s_t^i, c_t^i; \bar{\nu}) \right) \right)^2. \label{eq:phi_target_4net}
	\end{align}

	\vspace{0.5em}
	\noindent \textbf{Unified Update Procedure.}
	The update of all target network parameters follows the same soft averaging scheme. For any online parameter set $\omega$ and its target copy $\bar{\omega}$, the update is
	\begin{equation}
		\bar{\omega}^{(k)} \leftarrow \tau \omega^{(k)} + (1 - \tau) \bar{\omega}^{(k-1)},
		\qquad
		\omega\in\{\xi,\rho,\nu,\theta\},
		\label{eq:target_soft_update_all}
	\end{equation}
	where the relevant subset of parameters depends on the chosen network architecture. This procedure is summarized in Algorithm \ref{alg:target_soft_update}, which delineates the specific parameters updated in each architectural variant.
	
	\begin{algorithm}[H]
		\caption{Target Network Soft Update Procedure}
		\label{alg:target_soft_update}
		\begin{algorithmic}[1]
			\State \textbf{Input:} Current online parameters $\theta^{(k)}$, previous target parameters $\bar{\theta}^{(k-1)}$, update coefficient $\tau$
			\If{using three-network architecture}
			\State $\bar{\xi}^{(k)} \gets \tau \xi^{(k)} + (1 - \tau) \bar{\xi}^{(k-1)}$
			\ElsIf{using four-network architecture}
			\State $\bar{\rho}^{(k)} \gets \tau \rho^{(k)} + (1 - \tau) \bar{\rho}^{(k-1)}$
			\State $\bar{\nu}^{(k)} \gets \tau \nu^{(k)} + (1 - \tau) \bar{\nu}^{(k-1)}$
			\EndIf
		\end{algorithmic}
	\end{algorithm}
	
	This target network framework follows a stabilization principle: the quantities used for bootstrapping must evolve on a slower timescale than the parameters being updated. By decoupling the target computation from the immediate gradient updates, the learning process benefits from a stationary objective, which is essential for convergent behavior in approximate value iteration schemes. The modifications outlined in this section---encompassing exploratory perturbation, continuation methods, and target networks---collectively form a robust numerical toolkit for solving recursive utility models with deep learning, addressing distinct sources of instability and inefficiency and improving convergence stability across a broad spectrum of economic environments.

	\subsubsection{Policy Network Delay}
	
	Inspired by the delayed policy update mechanism in the Twin Delayed Deep Deterministic Policy Gradient (TD3) algorithm, we implement a similar delayed update schedule for the policy network $\theta$. Specifically, we update the policy parameters $\theta$ only once every $d$ iterations, where $d$ is a hyperparameter typically set between 1 and 5. During the intervening iterations, we continue to update the value function and certainty-equivalent networks, allowing for more accurate and stable value estimates before each policy improvement step.
	
    This delay mitigates the accumulation of value estimation errors that can arise when the policy network changes too frequently. By updating the policy less often than the value networks, we ensure that the policy gradient steps are taken based on more reliable value estimates, leading to more stable convergence in training.
	
	\subsection{Algorithmic Framework and Training Procedure}
	
	To enhance numerical robustness and accelerate convergence in these challenging nonlinear optimization problems, the Certainty Equivalent Learning (CEL) algorithm incorporates several key methodological innovations: (1) exploratory control perturbation during path simulation for structured exploration; (2) target networks for stable bootstrapping in value function learning; and (3) delayed policy updates to mitigate value estimation errors.
	
	CEL cycles through certainty-equivalent estimation, policy improvement, and, when an explicit value network is used, value-function fitting. In the three-network architecture the certainty-equivalent block is parameterized by $\xi$; in the four-network architecture it is decomposed into $(\rho,\nu)$; and in the two-network architecture the value function is represented implicitly and the explicit value-update step is omitted. The complete procedure alternates between these phases until convergence.
	
	The optimization proceeds through a sequential reinforcement learning procedure, where the CEL parameters are updated in a fixed cyclic order to improve stability and convergence. The enhanced update procedure is as follows:

	\subsubsection{Three-Network Architecture}
	\begin{enumerate}
		\item \textbf{Simulation with exploratory control perturbation:}
		For each optimization iteration $k$, we generate $N$ trajectories of length $T$ using the exploratory control perturbation mechanism described in Algorithm \ref{alg:exploration}. Starting from initial states sampled from the estimated stationary distribution, we simulate:
		\begin{equation}
			\{\tilde{c}_t^{i}, s_t^{i}\}_{t=0}^{T-1} \sim \mathcal{P}(\cdot|s_0^{i}, \theta^{(k-1)}, \text{Method}, \zeta)
		\end{equation}
		for $i = 1, \ldots, N$, where Method $\in$ \{Gaussian, Sobol\}. These perturbed trajectories cover the relevant regions of the state-action space while maintaining the policy's dominance.
		
		\item \textbf{Update the certainty-equivalent network ($\xi$):}
		Given fixed value network parameters ${\phi}^{(k-1)}$, we update the certainty-equivalent network using the stabilized objective:
		\begin{equation}
			\min_{\xi \in \Theta^\xi} \frac{1}{N}\sum_{i=1}^{N} \left( {f\left(V_c(s_t^i, \tilde{c}_t^i; \xi)\right)} - \frac{1}{G} \sum_{j=1}^G {f\left( V(s_{t+1}^{i,j}; \phi^{(k-1)}) \right)} \right)^2.
			\label{eq:update_d_nu_enhanced_weighted_3}
		\end{equation}
		where $\tilde{c}_t^i$ denotes the perturbed control action generated during simulation using Algorithm \ref{alg:exploration}. Note that the control variable $\tilde{c}_t^i$ incorporates exploratory noise, ensuring that the certainty-equivalent network is trained on a diverse set of state-action pairs.
		
		\item \textbf{Update the policy network ($\theta$):}
		Given fixed parameters ${\xi}^{(k)}$, we optimize the policy parameters only if $k \mod d = 0$, where $d$ is the delay hyperparameter. When updating, we maximize:
		\begin{equation}
			\max_{\theta \in \Theta^\theta} \frac{1}{N} \sum_{i=1}^{N} w\left( s_t^i, c(s_t^i; \theta),  V_c(s_t^i, c(s_t^i; \theta); {\xi}^{(k)}) \right).
			\label{eq:update_theta_delayed_3}
		\end{equation}
		This delayed update schedule prevents the accumulation of value estimation errors and promotes more stable convergence.
		
		\item \textbf{Update the value network ($\phi$):}
		Given fixed policy parameters $\theta^{(k)}$
        and certainty equivalent parameters  $\bar{\xi}^{(k-1)}$, we refine the value function by optimizing:
		\begin{equation}
			\min_{\phi \in \Theta^\phi} \frac{1}{N}\sum_{i=1}^{N} \left( {V(s_t^i; \phi)} - w\left( s_t^i, c(s_t^i; {\theta}^{(k)}),  V_c(s_t^i, c(s_t^i; {\theta}^{(k)}),\bar{\xi}^{(k-1)})\right) \right)^2.
			\label{eq:update_phi_target_3}
		\end{equation}
		This objective uses target networks to provide stable learning targets, preventing training divergence.
		
		\item \textbf{Soft update target networks:}
		After updating the online networks, we perform soft updates of the target networks using Polyak averaging:
		\begin{align}
			\bar{\xi}^{(k)} = \tau \xi^{(k)} + (1 - \tau)\bar{\xi}^{(k-1)}
		\end{align}
		where $\tau \ll 1$ is a small update rate controlling the degree of smoothing. This ensures that target parameters evolve slowly, providing stable learning targets.
	\end{enumerate}

	\subsubsection{Four-Network Architecture}
	
	The four-network architecture follows the same simulation, delayed policy update, value-fitting, and target-network logic as the three-network architecture. The only difference is that the certainty-equivalent value is decomposed into an expected continuation component and a nonlinear difference component:
	\begin{equation}
		V_c(s_t,c_t)
		=
		V_e(s_t,c_t;\rho)-D(s_t,c_t;\nu).
		\label{eq:four_network_vc_decomposition_35}
	\end{equation}
	Thus, instead of learning $V_c$ directly, the algorithm learns $V_e$ and $D$ separately.
	
	Given the same perturbed state-control samples $\{(s_t^i,\tilde c_t^i)\}_{i=1}^N$ generated by Algorithm \ref{alg:exploration}, the expected continuation network is trained by
	\begin{equation}
		\min_{\rho \in \Theta^\rho}
		\frac{1}{N}
		\sum_{i=1}^{N}
		\left(
		V_e(s_t^i,\tilde c_t^i;\rho)
		-
		\frac{1}{G}
		\sum_{j=1}^{G}
		V(s_{t+1}^{i,j};\phi^{(k-1)})
		\right)^2 .
		\label{eq:update_rho_enhanced_4}
	\end{equation}
	Conditional on the updated expected continuation network, the nonlinear difference network is trained by
	\begin{equation}
		\min_{\nu \in \Theta^\nu}
		\frac{1}{N}
		\sum_{i=1}^{N}
		\left(
		f\left(
		V_e(s_t^i,\tilde c_t^i;\rho^{(k)})
		-
		D(s_t^i,\tilde c_t^i;\nu)
		\right)
		-
		\frac{1}{G}
		\sum_{j=1}^{G}
		f\left(V(s_{t+1}^{i,j};\phi^{(k-1)})\right)
		\right)^2 .
		\label{eq:update_xi_enhanced_4}
	\end{equation}
	The policy network is updated, subject to the same delayed-update schedule as in the three-network architecture, by maximizing
	\begin{equation}
		\max_{\theta \in \Theta^\theta}
		\frac{1}{N}
		\sum_{i=1}^{N}
		w\left(
		s_t^i,
		c(s_t^i;\theta),
		V_e(s_t^i,c(s_t^i;\theta);\rho^{(k)})
		-
		D(s_t^i,c(s_t^i;\theta);\nu^{(k)})
		\right).
		\label{eq:update_theta_fournet_4}
	\end{equation}
    When an explicit value network is used, it is fitted to the stabilized Bellman target constructed from the target expectation and difference networks:
	\begin{equation}
		\min_{\phi \in \Theta^\phi}
		\frac{1}{N}
		\sum_{i=1}^{N}
		\left(
		V(s_t^i;\phi)
		-
		w\left(
		s_t^i,
		c(s_t^i;\theta^{(k)}),
		V_e(s_t^i,c(s_t^i;\theta^{(k)});\bar\rho^{(k-1)})
		-
		D(s_t^i,c(s_t^i;\theta^{(k)});\bar\nu^{(k-1)})
		\right)
		\right)^2 .
		\label{eq:update_phi_fournet_4}
	\end{equation}
	The target parameters are updated by Polyak averaging,
	\begin{align}
		\bar{\rho}^{(k)}
		&=
		\tau \rho^{(k)}+(1-\tau)\bar{\rho}^{(k-1)}, \\
		\bar{\nu}^{(k)}
		&=
		\tau \nu^{(k)}+(1-\tau)\bar{\nu}^{(k-1)}.
	\end{align}
	Therefore, relative to the three-network architecture, the four-network architecture changes only the representation of the certainty-equivalent value. The common CEL loop---exploratory simulation, certainty-equivalent learning, delayed policy improvement, value fitting, and target-network updating---remains unchanged.
	
	\subsubsection{Two-Network Architecture}
	
	The two-network architecture is a compact version of CEL that removes the explicit value network. It contains only a policy network $c(s_t;\theta)$ and a certainty-equivalent network $V_c(s_t,c_t;\xi)$. The value function is represented implicitly by the recursive aggregator:
	\begin{equation}
		\widehat V(s_t;\theta,\xi)
		=
		w\left(
		s_t,
		c(s_t;\theta),
		V_c\left(s_t,c(s_t;\theta);\xi\right)
		\right).
		\label{eq:two_network_implicit_value}
	\end{equation}
    This architecture therefore replaces the independent value approximation $V(s_t;\phi)$ with the implicit value representation $\widehat V(s_t;\theta,\xi)$.
	
	At iteration $k$, simulated state-control pairs are generated using the same exploratory perturbation procedure as before:
	\begin{equation}
		\{\widetilde c_t^i,s_t^i\}_{t=0}^{T-1}
		\sim
		P\left(\cdot \mid s_0^i,\theta^{(k-1)},\text{Method},\zeta\right),
		\qquad i=1,\ldots,N .
		\label{eq:two_network_simulation}
	\end{equation}
	For each pair $(s_t^i,\widetilde c_t^i)$, the next-period states are generated by
	\begin{equation}
		s_{t+1}^{i,j}
		=
		\psi\left(s_t^i,\widetilde c_t^i,z_{t+1}^{i,j}\right),
		\qquad j=1,\ldots,G .
		\label{eq:two_network_next_state}
	\end{equation}
	Since there is no value network, the next-period value used in the certainty-equivalent target is evaluated through the implicit value function in \eqref{nn:two_policy}--\eqref{nn:two_implicit_value}. To stabilize the bootstrap, the certainty-equivalent component uses the target certainty-equivalent network:
	\begin{equation}
		\widehat V(s_{t+1}^{i,j};\theta^{(k-1)},\bar\xi^{(k-1)})
		=
		w\left(
		s_{t+1}^{i,j},
		c(s_{t+1}^{i,j};\theta^{(k-1)}),
		V_c\left(
		s_{t+1}^{i,j},
		c(s_{t+1}^{i,j};\theta^{(k-1)});
		\bar\xi^{(k-1)}
		\right)
		\right).
		\label{eq:two_network_target_value}
	\end{equation}
	The certainty-equivalent network is then trained by
	\begin{equation}
		\min_{\xi\in\Theta_\xi}
		\frac{1}{N}
		\sum_{i=1}^{N}
		\left[
		f\left(
		V_c(s_t^i,\widetilde c_t^i;\xi)
		\right)
		-
		\frac{1}{G}
		\sum_{j=1}^{G}
		f\left(
		\widehat V(s_{t+1}^{i,j};\theta^{(k-1)},\bar\xi^{(k-1)})
		\right)
		\right]^2 .
		\label{eq:two_network_vc_update}
	\end{equation}
	The policy network is updated, again subject to the delayed-update schedule, by maximizing
	\begin{equation}
		\max_{\theta\in\Theta_\theta}
		\frac{1}{N}
		\sum_{i=1}^{N}
		w\left(
		s_t^i,
		c(s_t^i;\theta),
		V_c\left(s_t^i,c(s_t^i;\theta);\xi^{(k)}\right)
		\right).
		\label{eq:two_network_policy_update}
	\end{equation}
	Finally, the target certainty-equivalent network is updated by
	\begin{equation}
		\bar\xi^{(k)}
		=
		\tau \xi^{(k)}+(1-\tau)\bar\xi^{(k-1)} .
		\label{eq:two_network_target_xi_update}
	\end{equation}
	The two-network architecture therefore omits the value-fitting step entirely. Relative to the three-network architecture, it is more compact and enforces consistency between the policy and value representation, but it can be less flexible because the value function is no longer learned as an independent object.

	\begin{algorithm}[H]
		\caption{CEL Algorithm with Three-Network Architecture}
		\label{alg:3net_enhanced}
		\begin{algorithmic}[1]
			\State \textbf{Initialize:} parameters $(\phi^0, \theta^0, \xi^0)$; target network $\bar{\xi}^0 = \xi^0$
			\State \textbf{Initialize:} exploration parameters: Method $\in$ \{Gaussian, Sobol\}, scale factor $\zeta$, delay parameter $d$
			\State \textbf{Initialize:} training parameters: Polyak coefficient $\tau$
			
			\For{$k = 1$ to $K$}
			\State \textbf{Step 1: Exploratory Simulation with Control Perturbation}
			\State Sample initial states $\{s_0^{(i)}\}_{i=1}^N$ from stationary distribution
			\For{$i = 1$ to $N$}
			\State Generate a perturbed trajectory $(\{\tilde c_t^{(i)}\}_{t=0}^{T-1},\{s_t^{(i)}\}_{t=0}^{T})$ using Algorithm~\ref{alg:exploration} with initial state $s_0^{(i)}$, policy $c(\cdot;\theta^{k-1})$, transition rule $\psi$, and shock distribution $P_z$
			\For{$t = 0$ to $T-1$}
			\State Store transition $(s_t^{(i)}, \tilde{c}_t^{(i)}, s_{t+1}^{(i)})$ in buffer $\mathcal{B}$
			\EndFor
			\EndFor
			
			\State \textbf{Step 2: Update Certainty-Equivalent Network ($\xi$)}
			\For{$k_\xi = 1$ to $K_\xi$}
			\State Sample mini-batch from $\mathcal{B}$, compute loss as in \eqref{eq:update_d_nu_enhanced_weighted_3}
			\State Update the online certainty equivalent parameters by Adam using $\nabla_\xi L_\xi$
			\EndFor
			\State Denote the resulting parameters by $\xi^k$
			
			\State \textbf{Step 3: Update Policy Network ($\theta$) with Delayed Updates}
			\If{$k \mod d = 0$}
			\For{$k_\theta = 1$ to $K_\theta$}
			\State Sample mini-batch from $\mathcal{B}$, compute objective as in \eqref{eq:update_theta_delayed_3}
			\State Update the online policy parameters by Adam using $\nabla_\theta J_\theta$
			\EndFor
			\State Denote the resulting parameters by $\theta^k$
			\Else
			\State Set $\theta^k\gets\theta^{k-1}$
			\EndIf
			
			\State \textbf{Step 4: Update Value Network ($\phi$)}
			\For{$k_\phi = 1$ to $K_\phi$}
			\State Sample mini-batch from $\mathcal{B}$, compute loss as in \eqref{eq:update_phi_target_3}
			\State Update the online value parameters by Adam using $\nabla_\phi L_\phi$
			\EndFor
			\State Denote the resulting parameters by $\phi^k$
			\State \textbf{Step 5: Soft Update Target Networks}
			\State Update target parameters: $\bar{\xi}^k \gets \tau \xi^k + (1 - \tau) \bar{\xi}^{k-1}$	
			\If{convergence criterion satisfied}
			\State \Return $(\phi^k, \theta^k, \xi^k)$
			\EndIf
			\EndFor
		\end{algorithmic}
	\end{algorithm}

	\begin{algorithm}[H]
		\caption{CEL Algorithm with Four-Network Architecture}
		\label{alg:4net_enhanced}
		\begin{algorithmic}[1]
			\State \textbf{Initialize:} parameters $(\phi^0, \theta^0, \rho^0, \nu^0)$; target networks $\bar{\rho}^0 = \rho^0$, $\bar{\nu}^0 = \nu^0$
			\State \textbf{Initialize:} exploration parameters: Method $\in$ \{Gaussian, Sobol\}, scale factor $\zeta$, delay parameter $d$
			\State \textbf{Initialize:} training parameters: Polyak coefficient $\tau$
			
			\For{$k = 1$ to $K$}
			\State \textbf{Step 1: Exploratory Simulation with Control Perturbation}
			\State Sample initial states $\{s_0^{(i)}\}_{i=1}^N$ from stationary distribution
			\For{$i = 1$ to $N$}
			\State Generate a perturbed trajectory $(\{\tilde c_t^{(i)}\}_{t=0}^{T-1},\{s_t^{(i)}\}_{t=0}^{T})$ using Algorithm~\ref{alg:exploration} with initial state $s_0^{(i)}$, policy $c(\cdot;\theta^{k-1})$, transition rule $\psi$, and shock distribution $P_z$
			\For{$t = 0$ to $T-1$}
			\State Store transition $(s_t^{(i)}, \tilde{c}_t^{(i)}, s_{t+1}^{(i)})$ in buffer $\mathcal{B}$
			\EndFor
			\EndFor
			
			\State \textbf{Step 2: Update Expectation Network ($\rho$)}
			\State Using perturbed controls $\tilde{c}_t$ for enhanced exploration
			\For{$k_\rho = 1$ to $K_\rho$}
			\State Sample mini-batch from $\mathcal{B}$, compute loss as in \eqref{eq:update_rho_enhanced_4}
			\State Update the online expectation parameters by Adam using $\nabla_\rho L_\rho$
			\EndFor
			\State Denote the resulting parameters by $\rho^k$
			
			\State \textbf{Step 3: Update Difference Network ($\nu$)}
			\For{$k_\nu = 1$ to $K_\nu$}
			\State Sample mini-batch from $\mathcal{B}$, compute loss as in \eqref{eq:update_xi_enhanced_4}
			\State Update the online difference parameters by Adam using $\nabla_\nu L_\nu$
			\EndFor
			\State Denote the resulting parameters by $\nu^k$
			
			\State \textbf{Step 4: Update Policy Network ($\theta$) with Delayed Updates}
			\If{$k \mod d = 0$}
			\For{$k_\theta = 1$ to $K_\theta$}
			\State Sample mini-batch from $\mathcal{B}$, compute objective as in \eqref{eq:update_theta_fournet_4}
			\State Update the online policy parameters by Adam using $\nabla_\theta J_\theta$
			\EndFor
			\State Denote the resulting parameters by $\theta^k$
			\Else
			\State Set $\theta^k\gets\theta^{k-1}$
			\EndIf
			
			\State \textbf{Step 5: Update Value Network ($\phi$)}
			\For{$k_\phi = 1$ to $K_\phi$}
			\State Sample mini-batch from $\mathcal{B}$, compute loss as in \eqref{eq:update_phi_fournet_4}
			\State Update the online value parameters by Adam using $\nabla_\phi L_\phi$
			\EndFor
			\State Denote the resulting parameters by $\phi^k$
			
			\State \textbf{Step 6: Soft Update Target Networks}
			\State Update target parameters: $\bar{\rho}^k \gets \tau \rho^k + (1 - \tau) \bar{\rho}^{k-1}$
			\State Update target parameters: $\bar{\nu}^k \gets \tau \nu^k + (1 - \tau) \bar{\nu}^{k-1}$
			
			\If{convergence criterion satisfied}
			\State \Return $(\phi^k, \theta^k, \rho^k, \nu^k)$
			\EndIf
			\EndFor
		\end{algorithmic}
	\end{algorithm}

	\begin{algorithm}[!htbp]
		\caption{CEL Algorithm with Two-Network Architecture}
		\label{alg:cel_two_network}
		\begin{algorithmic}[1]
			\State \textbf{Initialize:} online parameters $(\theta^0,\xi^0)$ and target certainty equivalent parameters $\bar{\xi}^0=\xi^0$
			\State \textbf{Initialize:} exploration parameters: Method $\in\{\text{Gaussian},\text{Sobol}\}$, scale factor $\zeta$, delay parameter $d$
			\State \textbf{Initialize:} training parameters: Polyak coefficient $\tau$
			
			\For{$k=1$ to $K$}
			
			\State \textbf{Step 1: Exploratory Simulation with Control Perturbation}
			\State Sample initial states $\{s_0^{(i)}\}_{i=1}^{N}$ from the stationary distribution
			\For{$i=1$ to $N$}
			\State Generate a perturbed trajectory $(\{\tilde c_t^{(i)}\}_{t=0}^{T-1},\{s_t^{(i)}\}_{t=0}^{T})$ using Algorithm~\ref{alg:exploration} with initial state $s_0^{(i)}$, policy $c(\cdot;\theta^{k-1})$, transition rule $\psi$, and shock distribution $P_z$
			\For{$t=0$ to $T-1$}
			\State Store $(s_t^{(i)},\tilde c_t^{(i)},s_{t+1}^{(i)})$ in buffer $\mathcal{B}$
			\EndFor
			\EndFor
			
			\State \textbf{Step 2: Update Certainty-Equivalent Network $(\xi)$}
			\For{$k_\xi=1$ to $K_\xi$}
			\State Sample mini-batch from $\mathcal{B}$
			\State Compute the implicit target value $\widehat V(s';\theta^{k-1},\bar{\xi}^{k-1})$ as in \eqref{eq:two_network_target_value}
			\State Compute the certainty-equivalent loss as in \eqref{eq:two_network_vc_update}
			\State Update the online certainty equivalent parameters by Adam using $\nabla_\xi L_\xi$
			\EndFor
			\State Denote the resulting parameters by $\xi^k$
			
			\State \textbf{Step 3: Update Policy Network $(\theta)$ with Delayed Updates}
			\If{$k \bmod d=0$}
			\For{$k_\theta=1$ to $K_\theta$}
			\State Sample mini-batch from $\mathcal{B}$
			\State Compute the policy objective as in \eqref{eq:two_network_policy_update}
			\State Update the online policy parameters by Adam using $\nabla_\theta J_\theta$
			\EndFor
			\State Denote the resulting parameters by $\theta^k$
			\Else
			\State Set $\theta^k\gets\theta^{k-1}$
			\EndIf
			
			\State \textbf{Step 4: Soft Update Target Certainty-Equivalent Network}
			\State $\bar{\xi}^k \gets \tau \xi^k+(1-\tau)\bar{\xi}^{k-1}$
			
			\If{convergence criterion satisfied}
			\State \Return $(\theta^k,\xi^k)$
			\EndIf
			
			\EndFor
		\end{algorithmic}
	\end{algorithm}

	\section{Numerical Results}
	
	\label{sec:results}
	
	This section evaluates the accuracy and reliability of the value and policy functions learned by the proposed CEL variants. Assessing solution quality in dynamic programming problems requires problem-specific metrics, as the appropriate validation approach depends on the availability of closed-form benchmarks. When an analytical solution exists—common in linear-quadratic or exponential-utility specifications—we directly compare the learned functions against the ground truth, reporting pointwise errors and global deviation measures. When the reference is a numerical benchmark rather than the exact solution, we report differences rather than errors. Such comparisons offer the most transparent evidence of numerical accuracy.
	
	In the more typical case where no closed-form solution is available, we instead measure the extent to which the learned solution satisfies the model's Bellman equation and optimality conditions. We focus on problem-specific diagnostic metrics, including Bellman errors, Euler residuals, FOC residuals, and static-condition errors. The Bellman error quantifies the discrepancy between the two sides of the Bellman equation. Euler and FOC residuals evaluate violations of the model's intertemporal or intratemporal optimality conditions, while static-condition errors provide additional model-specific consistency checks. These measures are zero at the exact solution; hence, small values indicate that the learned policy and value functions approximately satisfy the relevant optimality restrictions.
	
	Computing these diagnostics involves evaluating conditional expectations of future values under the learned policy. We employ a nested simulation approach: for each state encountered along a test trajectory, we simulate multiple independent future paths, average the relevant next-period quantities, and evaluate the right-hand side of the Bellman equation. This simulation-based expectation estimation is flexible and model-agnostic, avoiding strong parametric assumptions about state transitions or noise distributions. By using a sufficiently large number of inner simulations, we obtain precise estimates of the Bellman errors and the relevant optimality-condition residuals, thereby obtaining reliable evidence on the accuracy of our numerical solution.
	
	To further validate the learned value function, we use a nested-simulation diagnostic inspired by \citet{Judd-1998}. The idea is to simulate multiple future states from the current state, compute the simulated certainty-equivalent value, and compare the value-network output with this simulation-based estimate. By calculating the standard error of the simulated values, we construct a confidence interval for the value obtained from the right-hand side of the Bellman equation. If the learned value function lies within this interval, the result provides additional evidence that the computed solution is consistent with the Bellman equation.

	Building on the evaluation framework above, we now present the numerical results of the proposed deep learning algorithm for dynamic programming problems with recursive utility. We consider four main applications that cover linear-quadratic control, robust control with a nonlinear certainty-equivalent term, DSGE models with stochastic volatility, and strategic portfolio choice with predictable returns.
	
	The first application is a Gaussian linear-quadratic control problem with a closed-form benchmark. This experiment is used to evaluate whether the proposed method can accurately recover both the value function and the policy function when the true solution is known. We first study a benchmark case with $n_s=8$ and $n_c=4$, and then consider a high-dimensional case with $n_s=100$ and $n_c=50$ to examine scalability.

	The second application is a small-noise robust control problem with nonlinear recursive utility and model uncertainty. We first consider a general nonlinear case without an analytical solution and compare the learned solution with value function iteration (VFI). We then study a homothetic benchmark case with an analytical solution, which allows us to directly evaluate the approximation errors of the value and policy functions. This application is designed to test whether the proposed method remains accurate and stable when robustness concerns and nonlinear certainty-equivalent values become important.
	
	The third application is a DSGE model with recursive preferences and stochastic volatility. This environment is richer than the previous benchmark problems because it contains a multidimensional state vector, multiple controls, and a nonlinear equilibrium structure. We use this example to examine whether the proposed algorithm can solve macroeconomic dynamic programming problems with recursive preferences in which traditional grid-based methods become computationally expensive and sensitive to interpolation errors.
	
	The fourth application is a multivariate strategic asset allocation problem with predictable returns. In this model, the investor jointly chooses the consumption-wealth ratio and a vector of portfolio weights in response to a multivariate predictive state vector. This problem allows us to evaluate the performance of the proposed framework in a high-dimensional portfolio-choice environment with recursive preferences. We assess the learned normalized value function, consumption policy, portfolio policy, and several internal accuracy measures, including Bellman errors and static-condition errors.

    Unless otherwise stated, the figure legends in this section use the following notation. \textit{closedform} denotes the analytical benchmark. \textit{CEL} denotes the direct deep-learning value or policy network output learned by CEL. 
    \textit{CEL-td} denotes the value obtained from a one-period evaluation of the right-hand side of the Bellman equation, where the certainty-equivalent term is evaluated through Monte Carlo simulation of next-period states and then plugged into the Bellman right-hand side.
    \textit{CEL-vc} denotes the corresponding value obtained from the same one-period evaluation, but with the certainty-equivalent term replaced by the learned certainty-equivalent network.
    Thus, the closeness among \textit{CEL}, \textit{CEL-td}, and \textit{CEL-vc} provides an internal consistency check for the learned solution.
	
	The results are organized as follows. Section~\ref{subsec:gaussian_results} reports the Gaussian linear-quadratic control results. Section~\ref{subsec:smallnoise_results} reports the small-noise robust control results. Section~\ref{subsec:dsge_results} reports the DSGE results with recursive preferences and stochastic volatility. Section~\ref{subsec:campbell_results} reports the multivariate strategic asset allocation results.

	\subsection{Gaussian Control: Closed-Form Benchmark and Scalability}
	\label{subsec:gaussian_results}

	The first example builds on the discounted risk-sensitive linear regulator framework developed by \citet{hansen1995discounted}, which extends the classical linear-quadratic Gaussian (LQG) control problem by incorporating exponential risk adjustments into the cost functional. This model is widely used in engineering and robust control, and provides a tractable benchmark for testing the accuracy of CEL under Gaussian dynamics and convex quadratic costs.
	
	\subsubsection{Problem Setup and Network Representation}
	\label{subsubsec:gaussian_setup}
	
	The system evolves according to a linear stochastic difference equation:
	\begin{equation}
		s_{t+1} = A s_t + B c_t + C w_{t+1}, \quad w_{t+1} \sim \mathcal{N}(0, I),
	\end{equation}
	where $s_t \in \mathbb{R}^{n_s}$ denotes the state vector, $c_t \in \mathbb{R}^{n_c}$ denotes the control (or action) vector, and $\{w_t\}$ is an i.i.d. sequence of standard Gaussian shocks. The initial state $s_0$ is given exogenously.

    The decision maker evaluates outcomes under a risk-sensitive recursive utility specification with an exponential certainty-equivalent term.
	The objective is to minimize the resulting exponential-quadratic cost functional:
	\begin{equation}
		V(s_t) = c_t^\top Q c_t + s_t^\top R s_t 
		- \frac{2\beta}{\sigma} \log \mathbb{E}_t \Bigl[ \exp\Bigl( -\frac{\sigma}{2} V(s_{t+1}) \Bigr) \Bigr],
		\label{eq:leqg-recursive}
	\end{equation}
	where $Q \succeq 0$ and $R \succ 0$ are penalty matrices on controls and states, respectively, $\beta \in (0,1)$ is the discount factor, and $\sigma > 0$ parameterizes risk sensitivity. The specification retains the quadratic structure of the standard LQG problem while introducing recursive risk aversion via the exponential certainty equivalent.
	
	For this benchmark, the state is the vector $s_t\in\mathbb{R}^{n_s}$, the control is $c_t\in\mathbb{R}^{n_c}$, and the recursive utility is the exponential-quadratic specification in \eqref{eq:leqg-recursive}. Here $c_t$ denotes the generic control vector in the benchmark, rather than consumption. This example is a canonical validation case because the infinite-horizon solution is linear and time invariant, so the learned value and policy functions can be compared directly with the closed-form benchmark.
	
	To solve this problem using deep learning, we implement the three-network CEL architecture:
	\begin{align*}
		V(s_t) &\equiv V(s_t;\phi),  \\
		c_t &\equiv c(s_t;\theta),  \\
		f^{-1}\left(
		\mathbb{E}_t\left[f\left(V(s_{t+1})\right)\mid s_t,c_t\right]
		\right)
		&\equiv V_c(s_t,c_t;\xi),
	\end{align*}
	where $\phi$, $\theta$, and $\xi$ are the parameters of the value network, policy network, and certainty-equivalent network, respectively. This matches the structure of the Gaussian benchmark, where the nonlinear risk-sensitive certainty-equivalent value can be learned directly by $V_c$ without introducing a separate nonlinear difference network.
	
	For the benchmark experiment, we set $n_s=8$, $n_c=4$, $\sigma=1$, and $\beta=0.96$.
	
	In this experiment, the system is specified by the transition matrices $A \in \mathbb{R}^{8 \times 8}$ and $B \in \mathbb{R}^{8 \times 4}$, the covariance-related matrix $C \in \mathbb{R}^{8 \times 8}$, the state cost matrix $R \in \mathbb{R}^{8 \times 8}$, and the control cost matrix $Q \in \mathbb{R}^{4 \times 4}$. The full numerical specification is reported in Appendix~\ref{app:high_dim_parameters}.

	\subsubsection{Benchmark Case: $n_s=8$, $n_c=4$}
	\label{subsubsec:gaussian_84}
	
\begin{figure}[!htbp]
		\centering
		\includegraphics[width=0.85\textwidth]{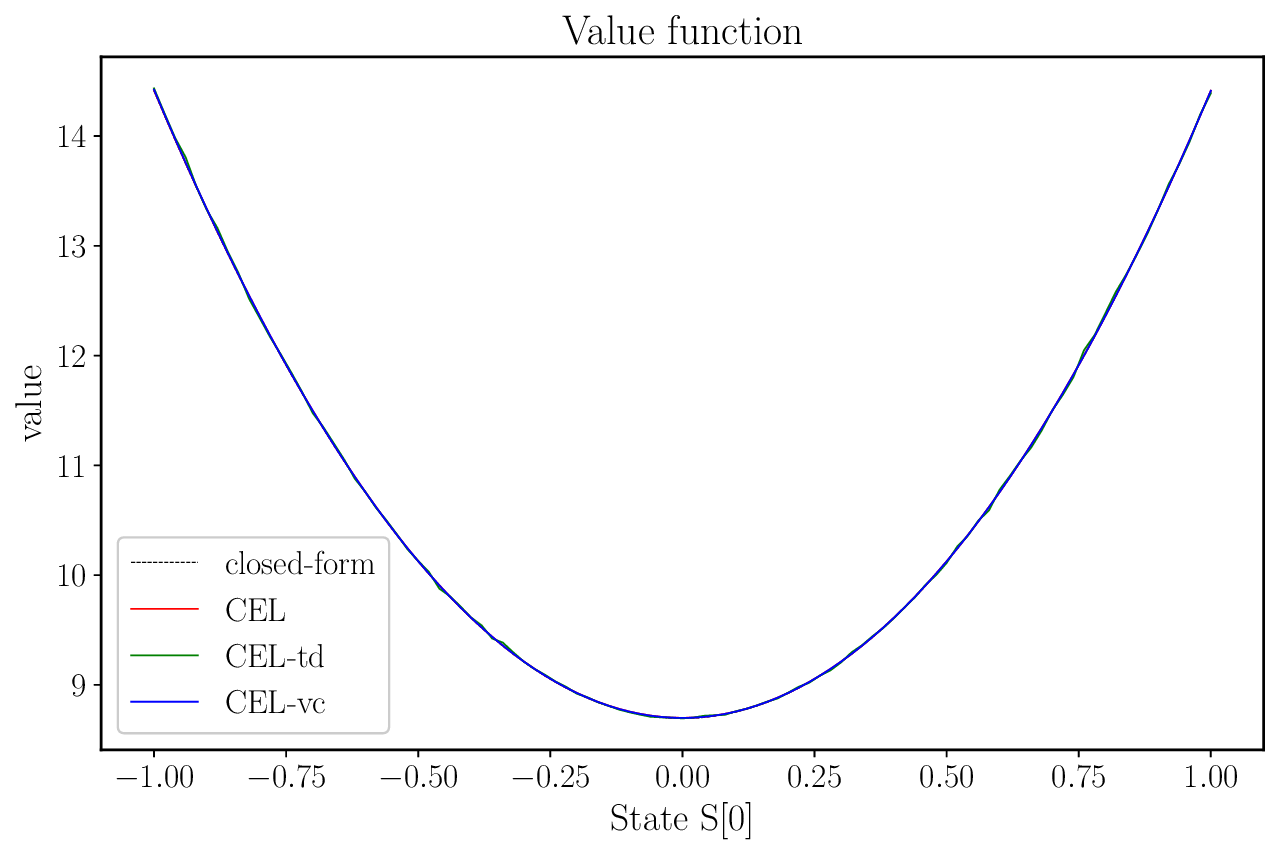}
		\caption{Value function $V(s_t)$ comparison for the Gaussian control problem with $n_s=8$ and $n_c=4$.}
		\label{fig:vf_gauss_84}
	\end{figure}

	\textbf{Value function approximation.} Figure~\ref{fig:vf_gauss_84} reports the learned value function $V(s_t)$ for the Gaussian benchmark case with $n_s=8$ and $n_c=4$. In this experiment, we vary the first component of $s_t$ while fixing all other state variables at their steady-state values. The figure compares the direct value network from the deep learning method with the analytical closed-form benchmark and the one-step certainty-equivalent estimates.
	
	The value-function curves are almost indistinguishable across the evaluation domain. In particular, the CEL value function closely matches the closed-form benchmark, indicating that the value network accurately captures the level, curvature, and scale of the true value function. The certainty-equivalent estimates are also closely aligned with the direct value network, which suggests that the certainty-equivalent components learned by the auxiliary networks are internally consistent with the Bellman equation.
	
\begin{figure}[!htbp]
		\centering
		\includegraphics[width=0.85\textwidth]{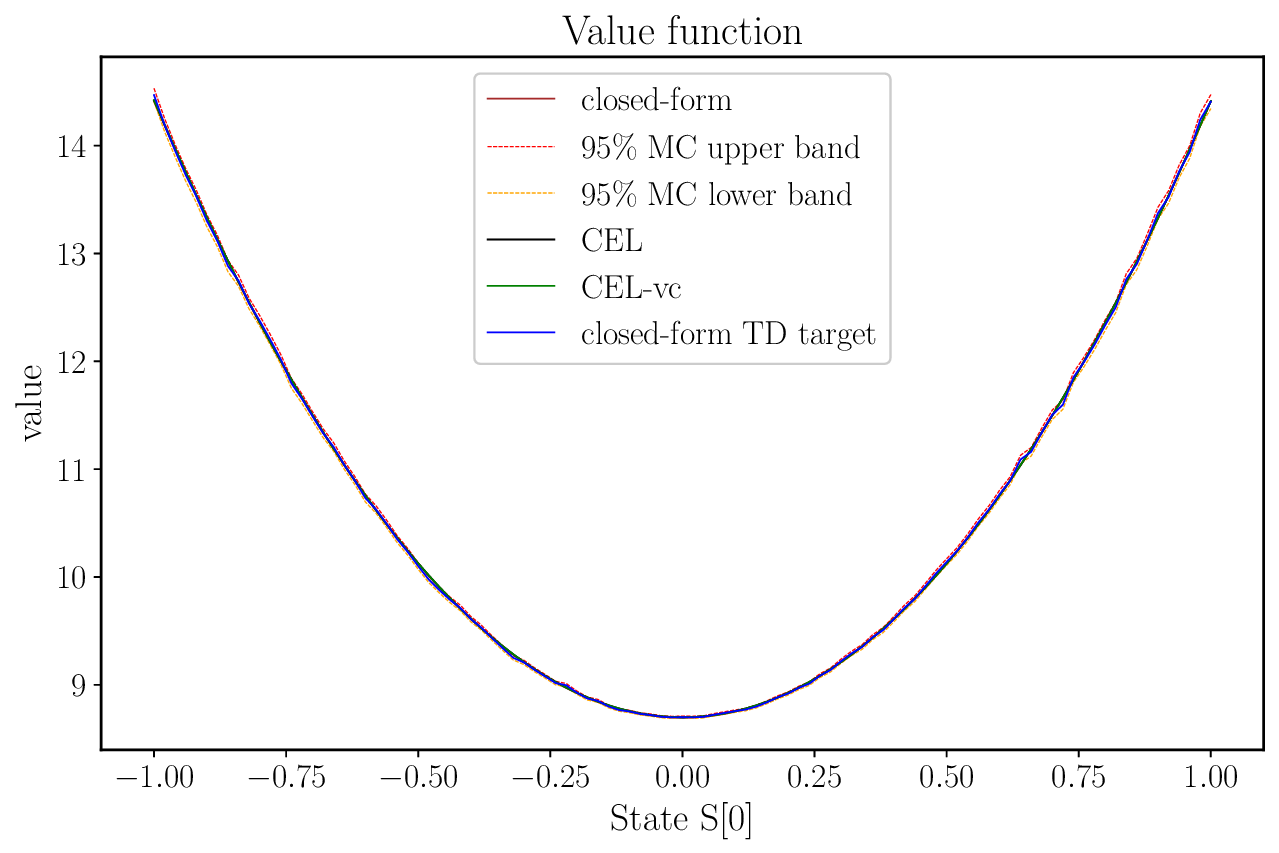}
		\caption{Nested-simulation diagnostic test for the Gaussian benchmark case with $n_s=8$ and $n_c=4$. The dashed lines represent the 95\% confidence bands from the nested-simulation diagnostic.}
		\label{fig:value_multi_84}
	\end{figure}

	To further examine Bellman consistency, we conduct a nested-simulation diagnostic test inspired by \citet{Judd-1998}, and derive the 95\% confidence interval using the delta method. The result is shown in Figure~\ref{fig:value_multi_84}. The learned value function and the certainty-equivalent function both lie within the confidence band over the evaluation range. This provides additional evidence that the deep learning solution satisfies the Bellman equation.
	
\begin{figure}[!htbp]
		\centering
		\includegraphics[width=0.85\textwidth]{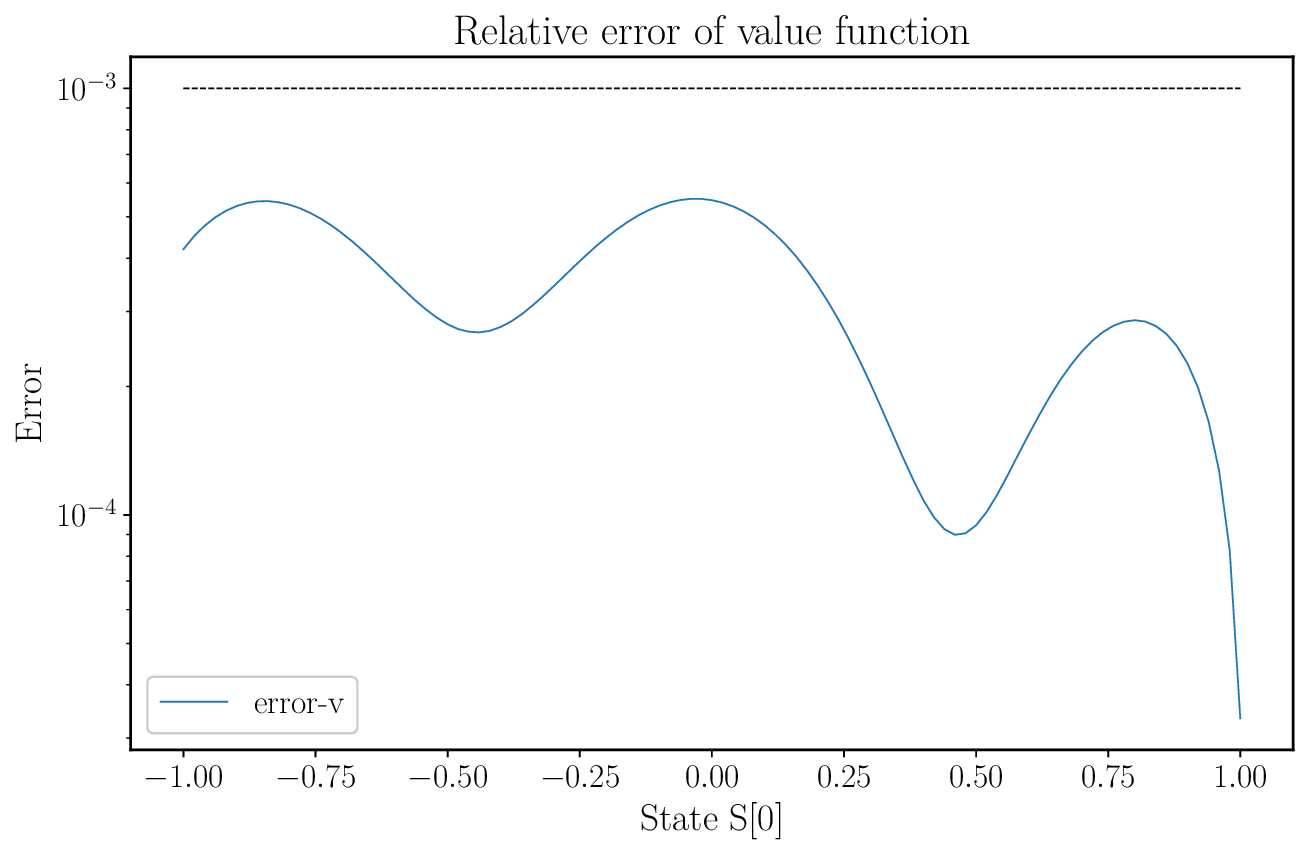}
		\caption{Relative error between the CEL value network and the closed-form benchmark when the first component of $s_t$ varies and all remaining components are fixed at their steady-state values.}
		\label{fig:relative_value_error84}
	\end{figure}

	Figure~\ref{fig:relative_value_error84} reports the relative error between the learned value function and the closed-form benchmark. The relative error remains below $10^{-3}$ across the state range. This confirms that the learned value function is not only visually close to the analytical benchmark, but also quantitatively accurate. Therefore, the proposed method provides an accurate approximation to the value function in the moderate-dimensional Gaussian benchmark case.
	
\begin{figure}[!htbp]
		\centering
		\includegraphics[width=0.85\textwidth]{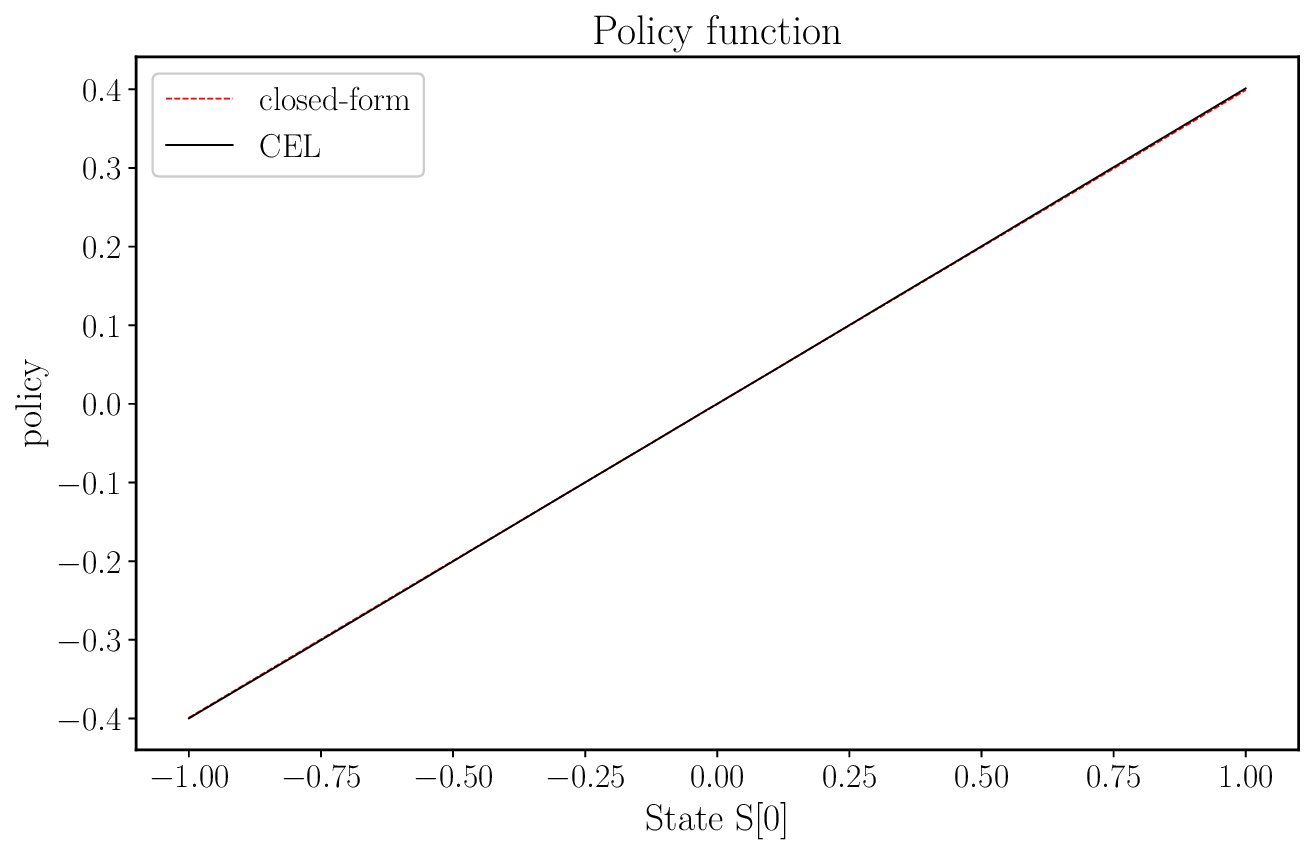}
		\caption{Policy function comparison between the learned CEL policy and the closed-form benchmark when the first component of $s_t$ varies and all remaining components are fixed at their steady-state values.}
		\label{fig:policy84}
	\end{figure}

	\noindent\textbf{Policy function approximation.} The policy function $c(s_t)$ is evaluated by varying the first component of $s_t$ while fixing the remaining state variables at their steady-state values. Figure~\ref{fig:policy84} compares the learned CEL policy with the analytical closed-form benchmark.
	
	The learned policy closely tracks the closed-form benchmark over the entire evaluation range. This indicates that the policy network successfully learns the structure of the optimal control. Therefore, the proposed method is able to recover not only the value function, but also the economically relevant policy function.
	
\begin{figure}[!htbp]
		\centering
		\includegraphics[width=0.85\textwidth]{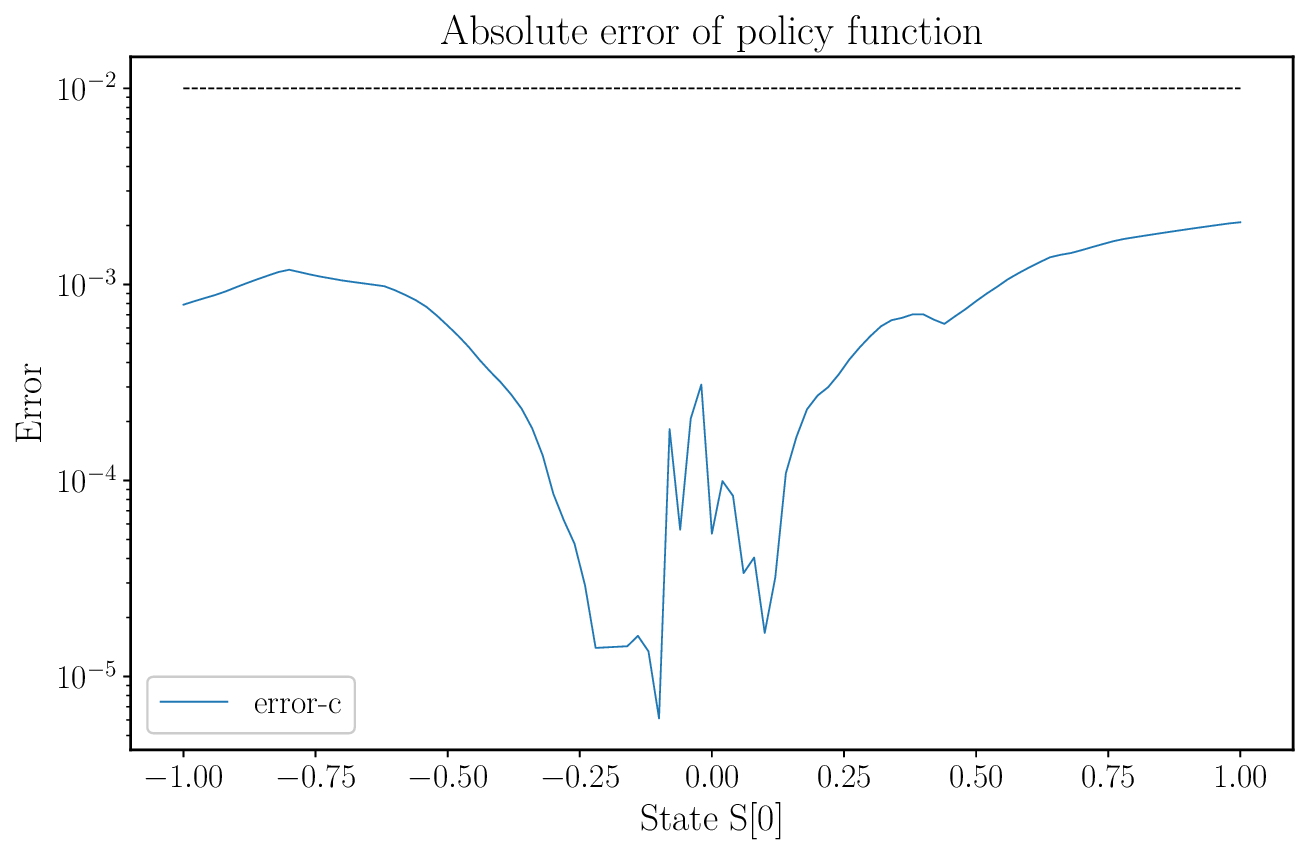}
		\caption{Absolute error between the learned CEL policy and the closed-form benchmark when the first component of $s_t$ varies and all remaining components are fixed at their steady-state values.}
		\label{fig:policy_error84}
	\end{figure}

	Figure~\ref{fig:policy_error84} reports the absolute error between the learned policy and the closed-form benchmark. The policy error remains small across the state range, showing that the policy approximation is quantitatively accurate. 
	
	\FloatBarrier

	\subsubsection{High-Dimensional Case: $n_s=100$, $n_c=50$}
	\label{subsubsec:gaussian_10050}
	
	We next consider a high-dimensional Gaussian control problem with $n_s=100$ and $n_c=50$. This experiment is designed to examine whether the proposed method remains accurate when both the state and control dimensions become large. Such a setting is challenging for traditional grid-based methods because the number of grid points grows exponentially with the dimension. Therefore, this experiment provides a direct test of the scalability of the proposed deep learning algorithm.
	
\begin{figure}[!htbp]
		\centering
		\includegraphics[width=0.85\textwidth]{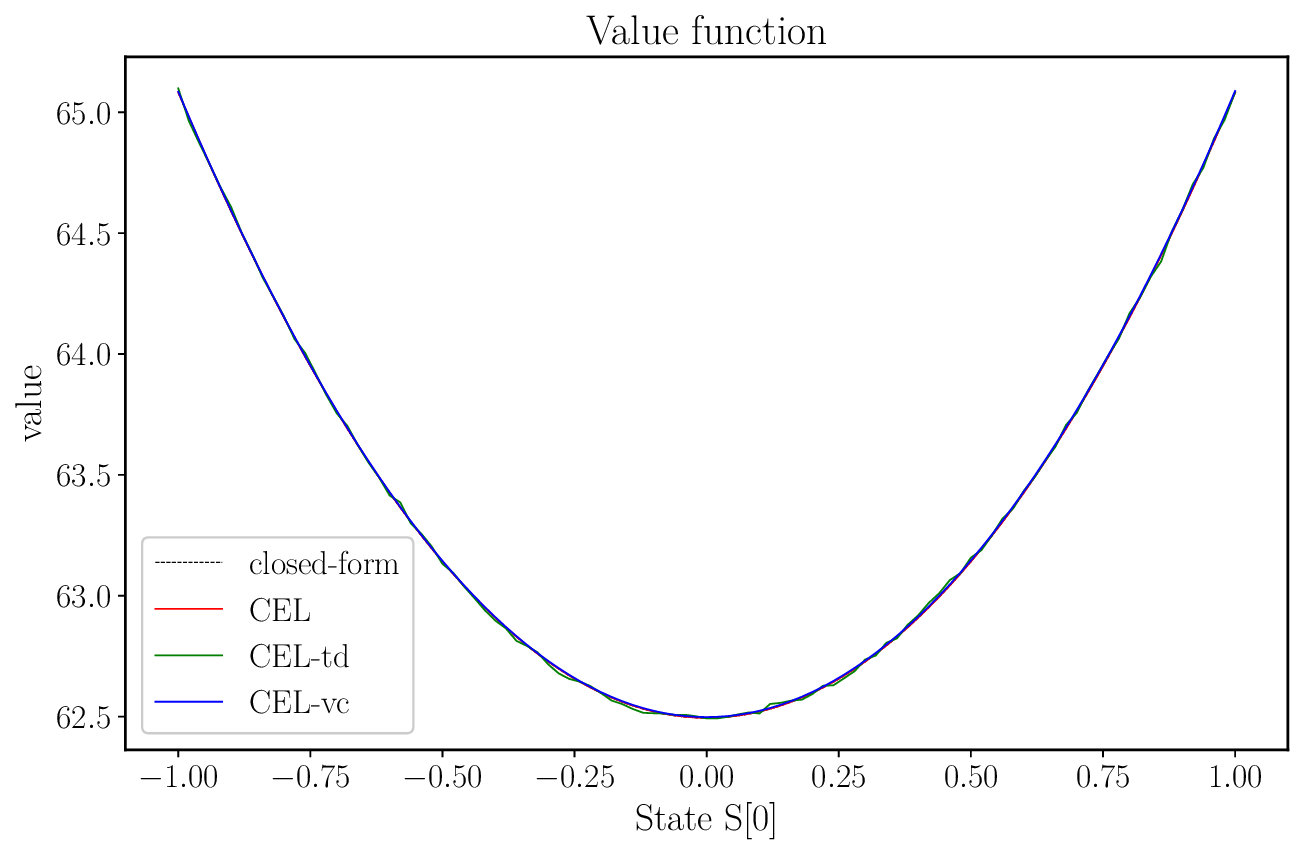}
		\caption{Value function $V(s_t)$ comparison for the Gaussian control problem with $n_s=100$ and $n_c=50$.}
		\label{fig:vf_gauss_10050}
	\end{figure}

	\noindent\textbf{Value function approximation.} 
	Figure~\ref{fig:vf_gauss_10050} reports the learned value function in the high-dimensional case. We vary the first component of $s_t$ while fixing all other state variables at their steady-state values. The figure compares the learned CEL value function with the closed-form benchmark and the one-step certainty-equivalent estimates.
	
	The value-function curves are closely aligned across the full evaluation range. In particular, the learned CEL value function remains almost indistinguishable from the closed-form benchmark. The certainty-equivalent estimates also stay close to the direct value-network output, suggesting that the certainty-equivalent components of the learned solution remain internally consistent even in the high-dimensional environment.
	
\begin{figure}[!htbp]
		\centering
		\includegraphics[width=0.85\textwidth]{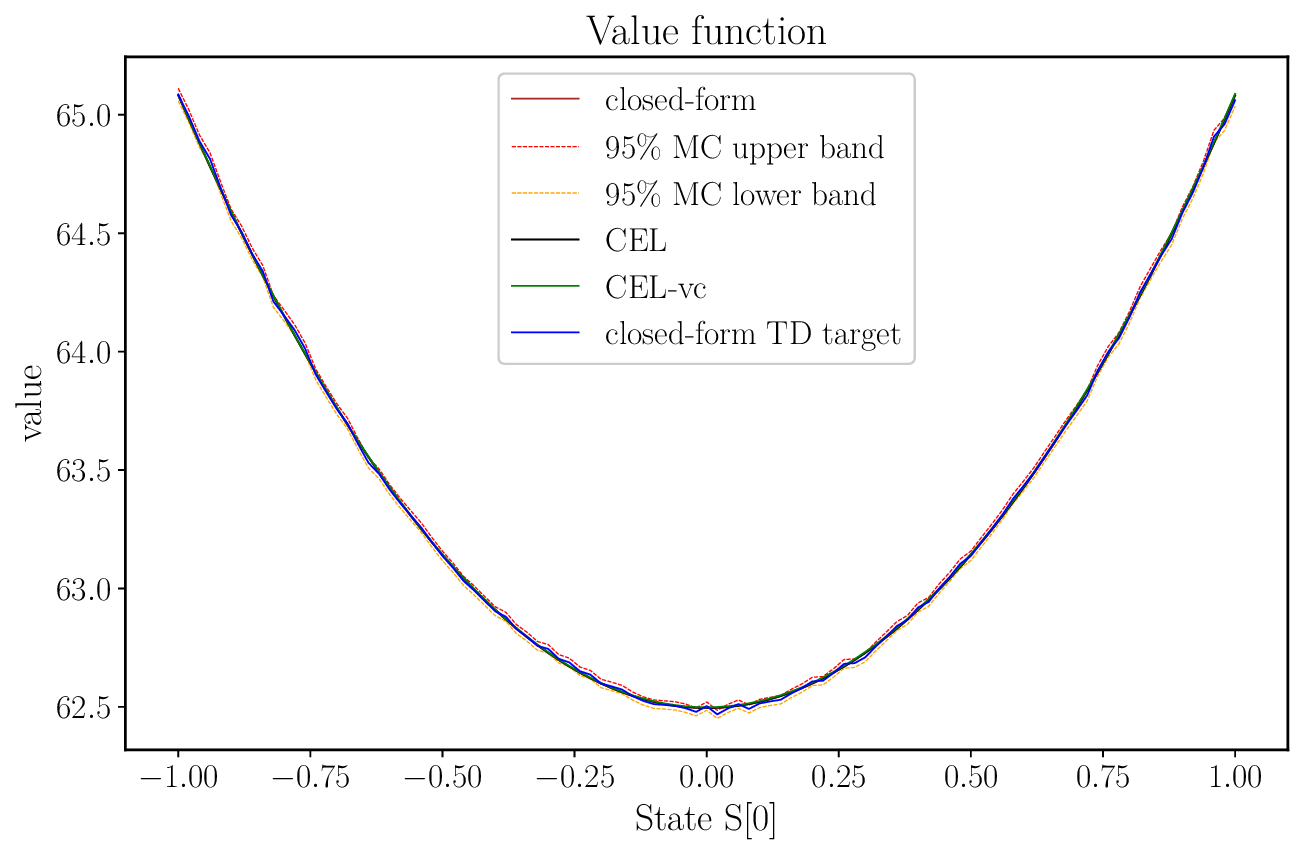}
		\caption{Nested-simulation diagnostic test for the high-dimensional Gaussian control problem with $n_s=100$ and $n_c=50$. The dashed lines represent the 95\% confidence bands from the nested-simulation diagnostic.}
		\label{fig:value_multi_10050}
	\end{figure}

	To further examine whether the learned value function satisfies the Bellman equation, we conduct the same nested-simulation diagnostic test as in the benchmark case. The result is reported in Figure~\ref{fig:value_multi_10050}. The learned value function and certainty-equivalent approximation lie within the 95\% confidence band. This provides additional evidence that the learned solution remains Bellman-consistent when the state dimension increases to 100.
	
\begin{figure}[!htbp]
		\centering
		\includegraphics[width=0.85\textwidth]{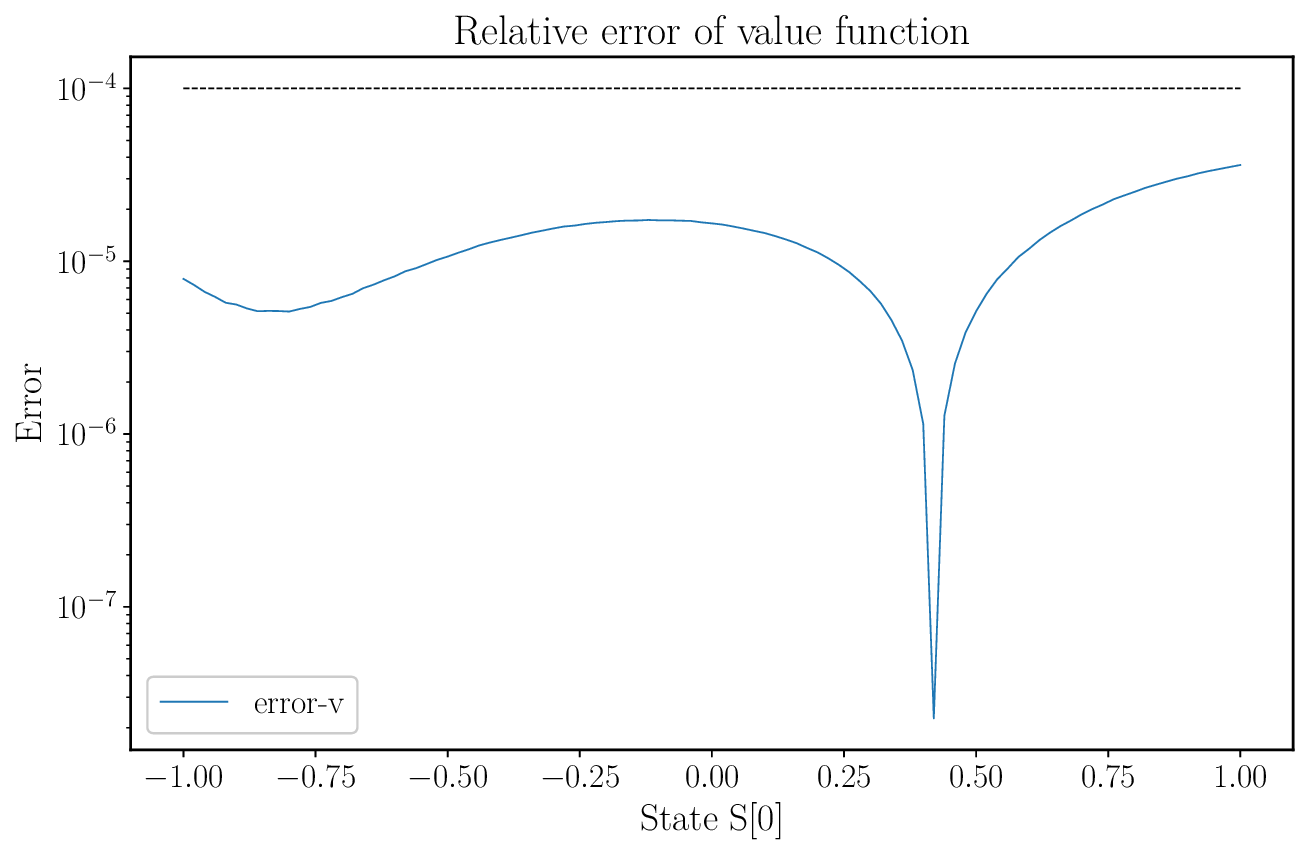}
		\caption{Relative error between the learned CEL value function and the closed-form benchmark when the first component of $s_t$ varies and all remaining components are fixed at their steady-state values.}
		\label{fig:relative_value_error10050}
	\end{figure}

	Figure~\ref{fig:relative_value_error10050} reports the relative error between the learned value function and the closed-form benchmark. The relative error remains below $10^{-4}$ across the evaluation range. This shows that the value approximation is not only visually close to the analytical solution, but also quantitatively accurate in a high-dimensional state space.
	
\begin{figure}[!htbp]
		\centering
		\includegraphics[width=0.85\textwidth]{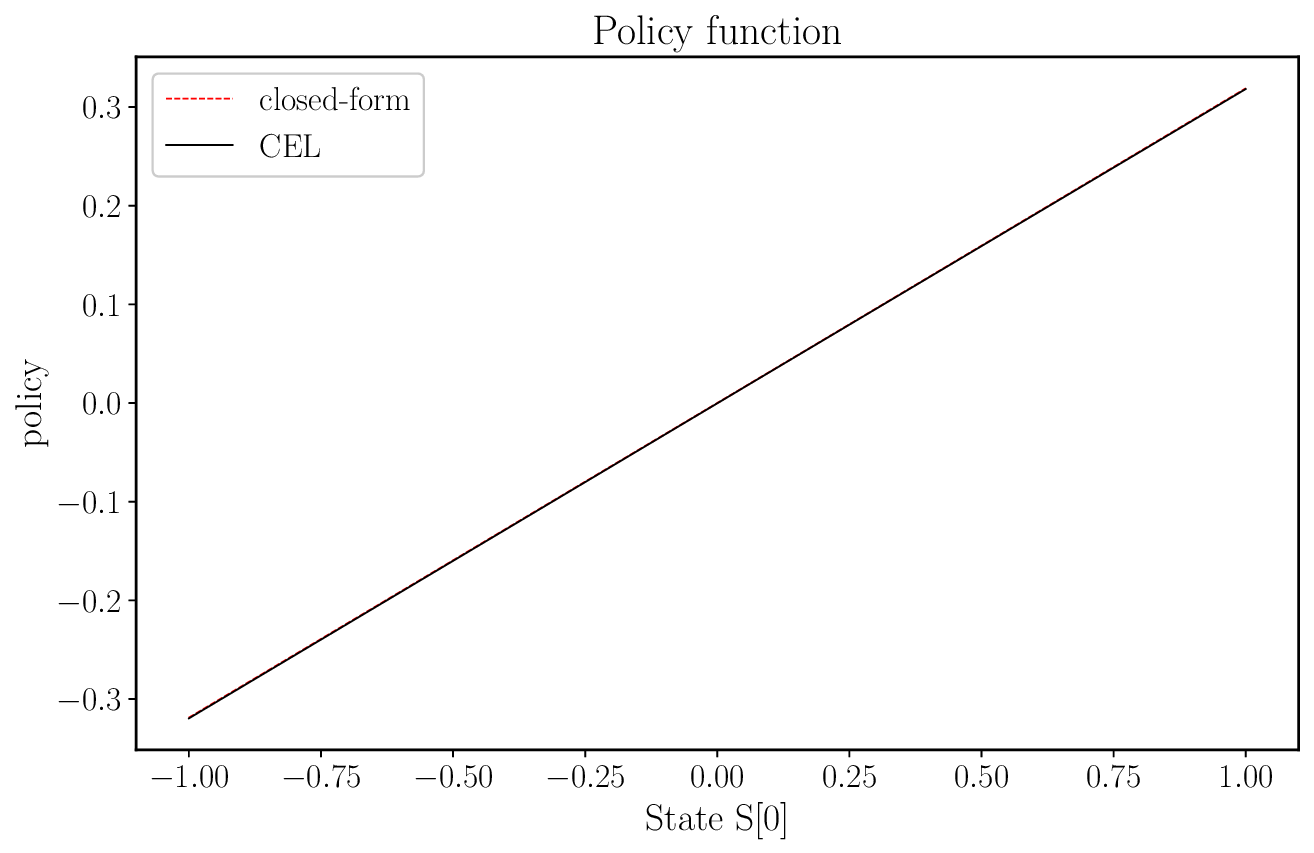}
		\caption{Policy function comparison between the learned CEL policy and the closed-form benchmark when the first component of $s_t$ varies and all remaining components are fixed at their steady-state values.}
		\label{fig:policy10050}
	\end{figure}

	\noindent\textbf{Policy function approximation.} 
	Figure~\ref{fig:policy10050} presents the learned policy function in the high-dimensional case. We again vary the first component of $s_t$ while fixing the remaining state variables at their steady-state values. The learned policy closely follows the closed-form benchmark, indicating that the policy network can still recover the optimal policy function when the control dimension increases to 50.
	
\begin{figure}[!htbp]
		\centering
		\includegraphics[width=0.85\textwidth]{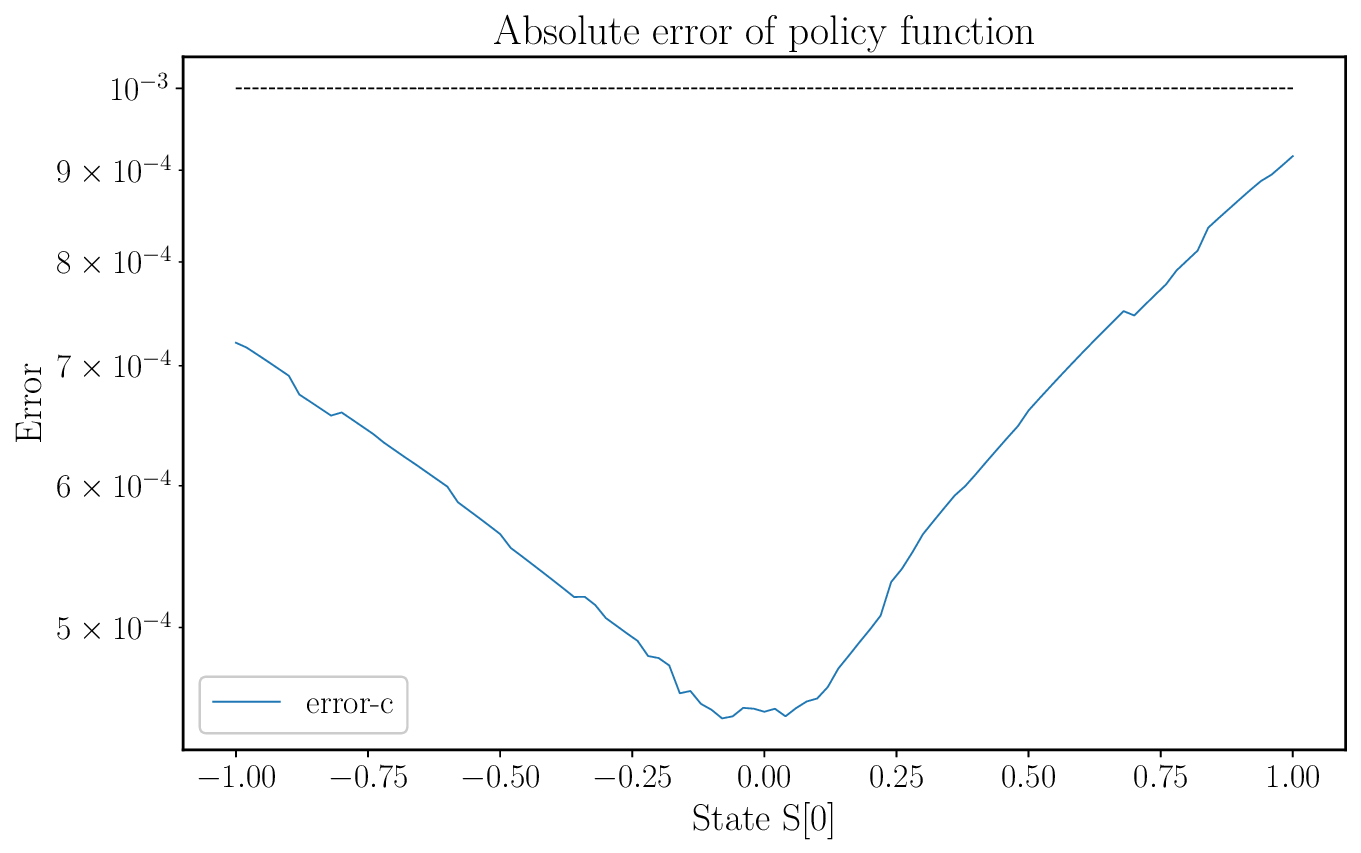}
		\caption{Absolute error between the learned CEL policy and the closed-form benchmark when the first component of $s_t$ varies and all remaining components are fixed at their steady-state values.}
		\label{fig:policy_error10050}
	\end{figure}

	Figure~\ref{fig:policy_error10050} reports the absolute error between the learned policy and the closed-form benchmark. The policy error remains small over the state range, confirming that the learned neural network solution produces accurate control decisions even in the high-dimensional environment.
	
	\FloatBarrier

	\subsection{Small Noise Robust Control: General and Homothetic Cases}
	\label{subsec:smallnoise_results}
	
	Our second example builds on the work of \citet{anderson2012small}, who study robust decision-making under model misspecification and uncertainty. In this framework, a representative agent possesses risk-sensitive preferences as introduced by \citet{hansen1995discounted}. This approach departs from standard expected utility theory by introducing a penalty for potential misperceptions about the stochastic structure of the economy. The penalty is implemented through a log-exponential transformation of future utility, thereby internalizing a form of robustness in intertemporal planning.
	
	\subsubsection{Problem Setup and Network Representation}
	\label{subsubsec:smallnoise_setup}
	
	The agent's problem is characterized by the following recursive utility specification:
	\begin{equation}
		V(K_{t-1}, a_t) = \frac{C_t^{1-\gamma}}{1-\gamma} - \frac{1}{\sigma} \log \mathbb{E}_t \Bigl[ \exp\bigl(-\sigma \beta V(K_t, a_{t+1})\bigr) \Bigr],
		\label{eq:vf_risk_sensitive}
	\end{equation}
	where $\gamma > 0$ denotes the coefficient of relative risk aversion, $\sigma > 0$ parameterizes the sensitivity to model uncertainty (with higher $\sigma$ implying greater concern for robustness), and $\beta \in (0,1)$ is the subjective discount factor. The agent faces a standard Cobb-Douglas resource constraint:
	\begin{equation}
		C_t + K_t = e^{P a_t} K_{t-1}^\alpha + (1 - \delta) K_{t-1},
	\end{equation}
	where $C_t$ is consumption, $K_t$ is the end-of-period capital stock, $\alpha \in (0,1)$ is the capital share, $\delta \in [0,1]$ is the depreciation rate, and $P$ scales the technology shock. The exogenous technology shock $a_t$ follows a linear Gaussian process:
	\begin{equation}
		a_{t+1} = \Omega_0 + \Omega_a a_t + \sqrt{\epsilon} \, \Omega_v w_{t+1}, \quad w_{t+1} \sim \mathcal{N}(0,1),
	\end{equation}
	with persistence parameter $\Omega_a \in (-1,1)$, unconditional mean $\Omega_0 / (1 - \Omega_a)$, and conditional volatility $\sqrt{\epsilon} \, \Omega_v$.
	
	For the deep learning implementation, the state is $s_t=(K_{t-1},a_t)^\top$, and the control is the relative consumption share
	\begin{equation}
		c_t
		=
		\frac{C_t}{e^{P a_t}K_{t-1}^{\alpha}+(1-\delta)K_{t-1}}
		\equiv
		\frac{C_t}{W_t},
	\end{equation}
	where $W_t$ denotes total resources. The constraint $c_t\in(0,1)$ guarantees positive consumption and investment, and the capital law of motion becomes $K_t=(1-c_t)W_t$.

	To capture the nested recursive utility structure, we employ a four-network architecture:
	\begin{align*}
		V(s_t) &\equiv V(s_t;\phi), \\
		c_t &\equiv c(s_t;\theta), \\
		\mathbb{E}\left[V(s_{t+1})\mid s_t,c_t\right]
		&\equiv V_e(s_t,c_t;\rho),\\
		\mathbb{E}\left[V(s_{t+1})\right]
		-
		f^{-1}\left(\mathbb{E}\left[f(V(s_{t+1}))\right]\right)
		&\equiv D(s_t,c_t;\nu),
	\end{align*}
	
	This problem provides a tractable yet rich test case for CEL. In the general nonlinear specification, no analytical solution is available, so we use VFI as a numerical benchmark. In the homothetic benchmark specification, the model admits an analytical solution, which allows us to evaluate the numerical errors of both CEL and VFI directly.

	\subsubsection{General Nonlinear Case without Analytical Solution}
	\label{subsubsec:smallnoise_general}
	
	We first consider the general nonlinear case without an analytical solution. In this setting, value function iteration is used as a numerical benchmark. Since VFI is itself a numerical approximation rather than the exact solution, the comparison below is reported as a relative difference rather than an approximation error. The purpose of this comparison is to examine whether the deep learning solution delivers value and policy functions that are close to the VFI benchmark and whether the residual-based diagnostics are of a similar magnitude.
	
	\noindent\textbf{Results for baseline robustness: $\sigma=1$.}
	
\begin{figure}[!htbp]
		\centering
		\includegraphics[width=0.85\textwidth]{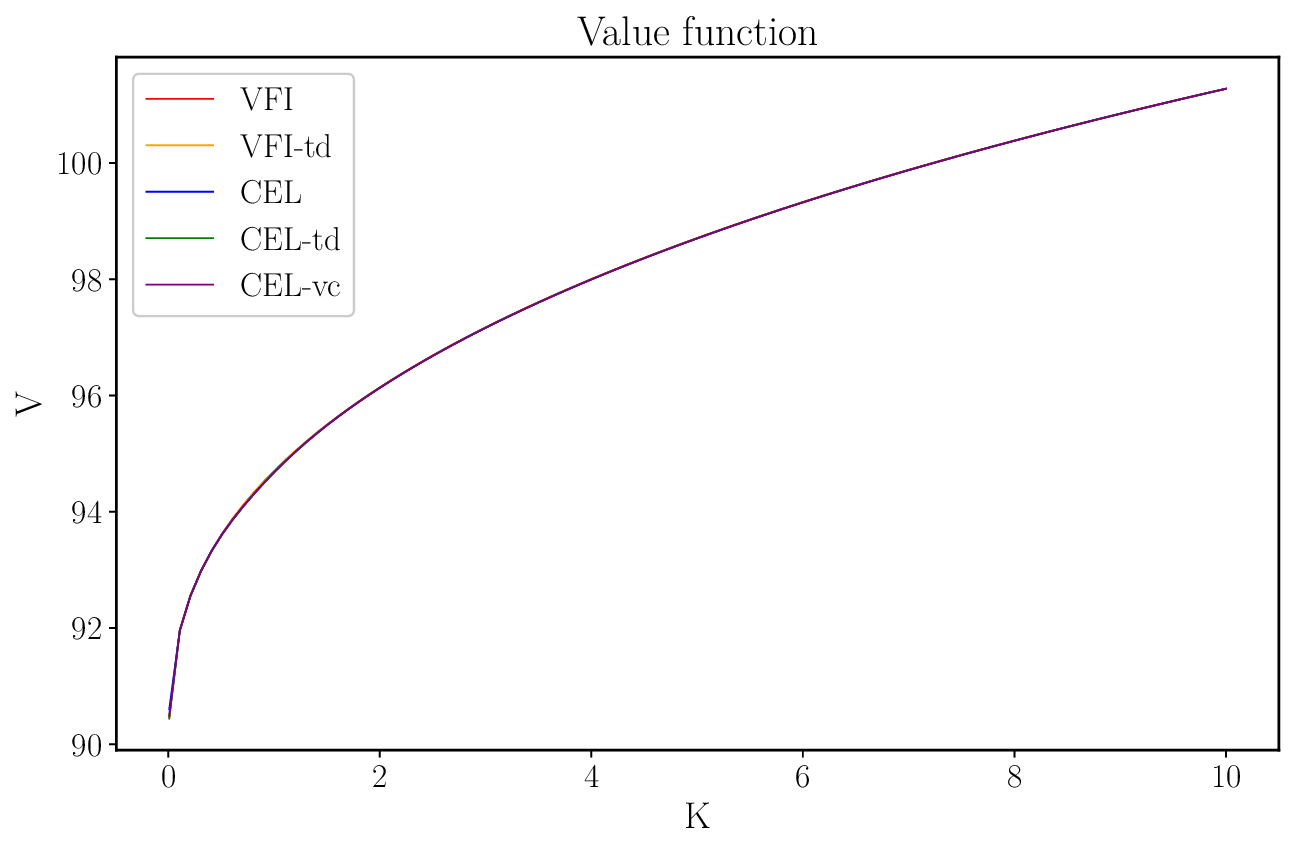}
		\caption{Value function and its one-step expansion from the CEL method and VFI for the general nonlinear small-noise model with $\sigma=1$.}
		\label{fig:vf_small_noise_1}
	\end{figure}

	\noindent\textbf{Value function approximation.}
	The value function $V(s_t)$ is evaluated across varying capital $K_{t-1}$, fixing the technology shock at its steady state $a_{ss}$. Figure~\ref{fig:vf_small_noise_1} compares the learned value function and its one-step expansion from the CEL method with the VFI benchmark.
	
	The learned value function is close to the VFI benchmark over the whole capital range. The one-step expansion is also closely aligned with the direct value network, indicating that the value function approximation is internally consistent. This suggests that the proposed method captures the nonlinear curvature induced by recursive preferences, capital accumulation, and model uncertainty.
	
\begin{figure}[!htbp]
		\centering
		\includegraphics[width=0.85\textwidth]{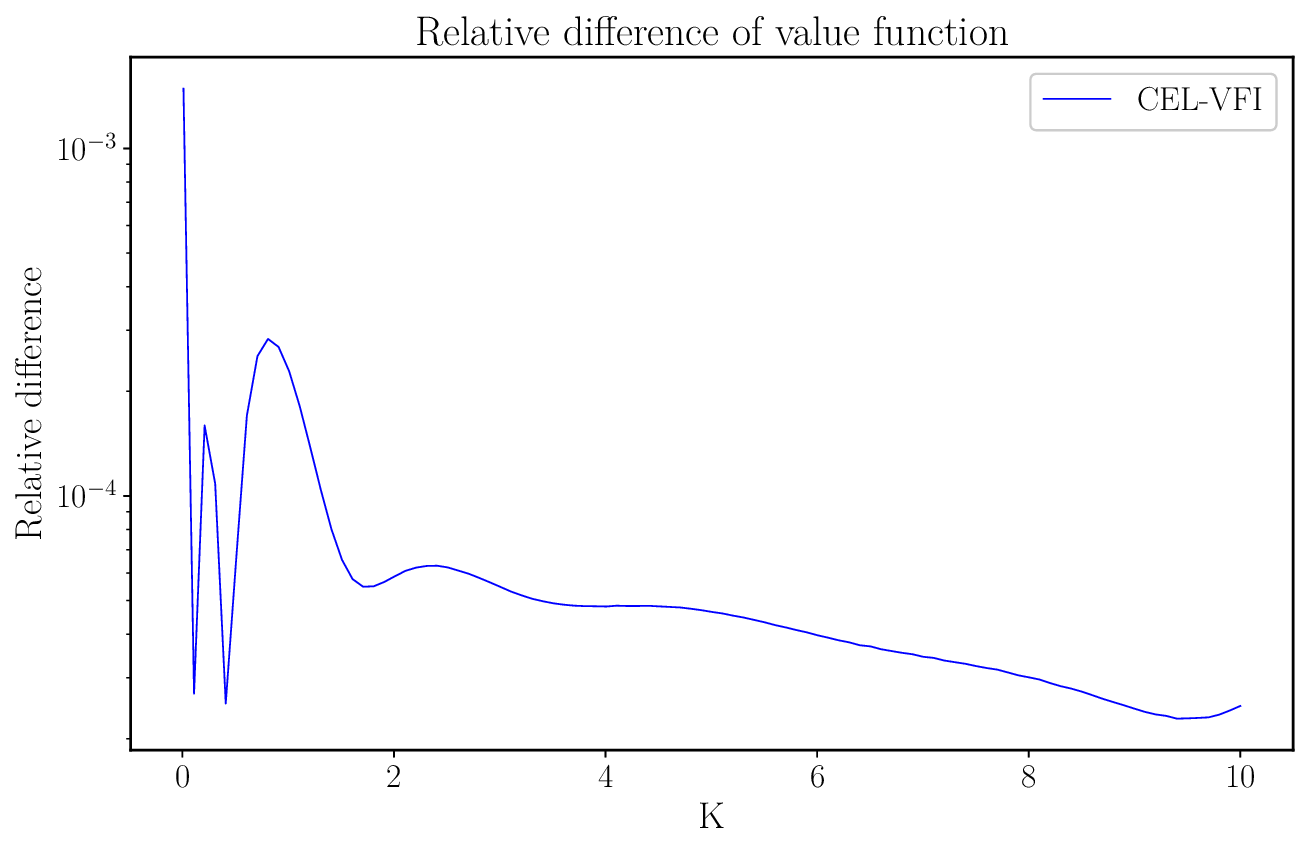}
		\caption{Relative difference between the CEL value network and the VFI benchmark when $K_{t-1}$ varies and $a_t$ is fixed for $\sigma=1$.}
		\label{fig:relative_diff_v_smallnoise_1}
	\end{figure}

	Figure~\ref{fig:relative_diff_v_smallnoise_1} reports the relative difference between the CEL value network and the VFI benchmark. The relative difference remains small across the evaluation range. This shows that the CEL value function is quantitatively close to the VFI benchmark.
	
\begin{figure}[!htbp]
		\centering
		\includegraphics[width=0.85\textwidth]{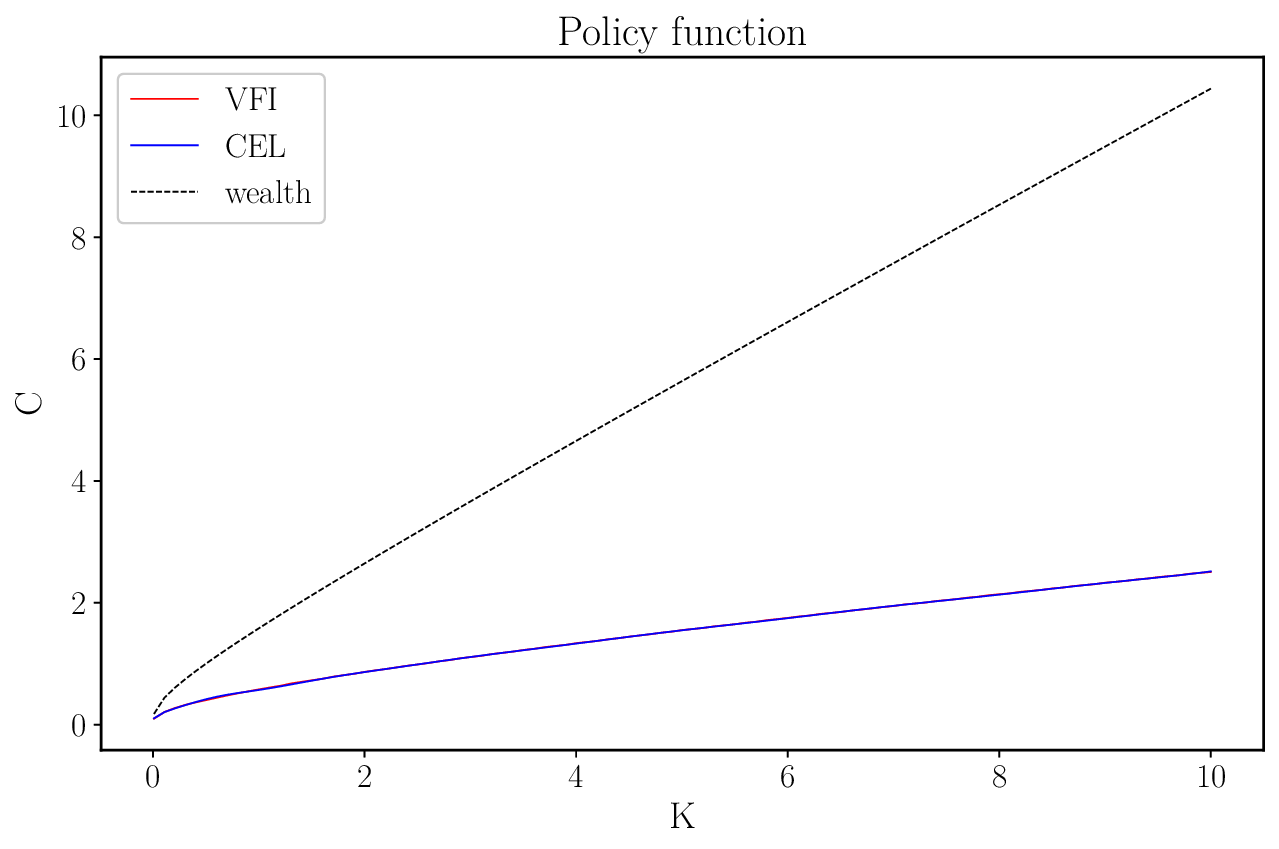}
		\caption{Policy function: consumption when capital $K_{t-1}$ varies for $\sigma=1$.}
		\label{fig:pf_small_noise_1}
	\end{figure}

\begin{figure}[!htbp]
		\centering
		\includegraphics[width=0.85\textwidth]{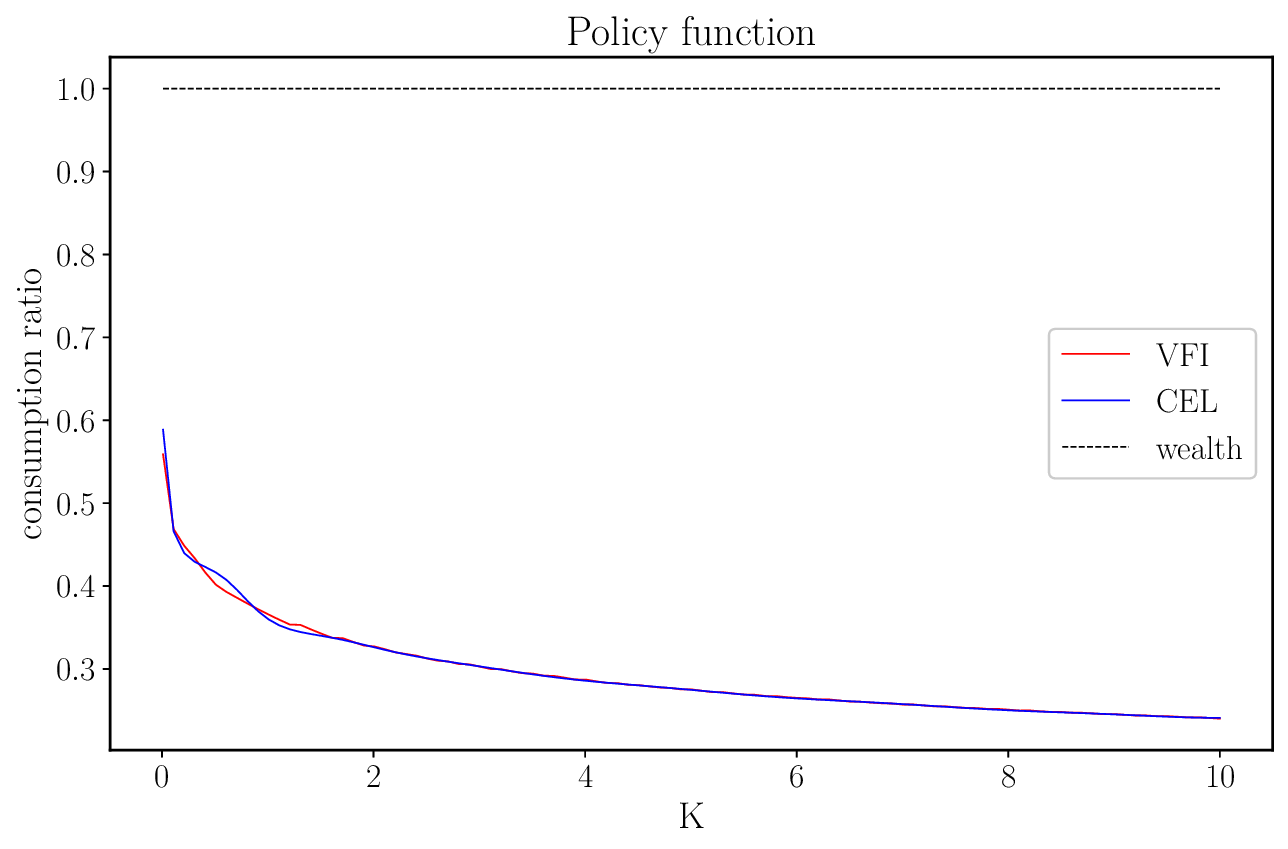}
		\caption{Policy function: consumption-to-wealth ratio when capital $K_{t-1}$ varies for $\sigma=1$.}
		\label{fig:pf_ratio_small_noise_1}
	\end{figure}

	\noindent\textbf{Policy function approximation.} Figure~\ref{fig:pf_small_noise_1} displays the learned consumption policy as capital varies, with the current technology shock fixed at its steady state. Figure~\ref{fig:pf_ratio_small_noise_1} reports the policy function in terms of the consumption-to-wealth ratio.
	
	The level of consumption increases with capital, which is consistent with economic intuition. The consumption-to-wealth ratio provides a normalized view of the consumption-saving decision and shows how the agent adjusts the policy as capital changes. The nonlinear shape of the policy reflects the role of model uncertainty and recursive preferences in the intertemporal trade-off.
	
\begin{figure}[!htbp]
		\centering
		\includegraphics[width=0.85\textwidth]{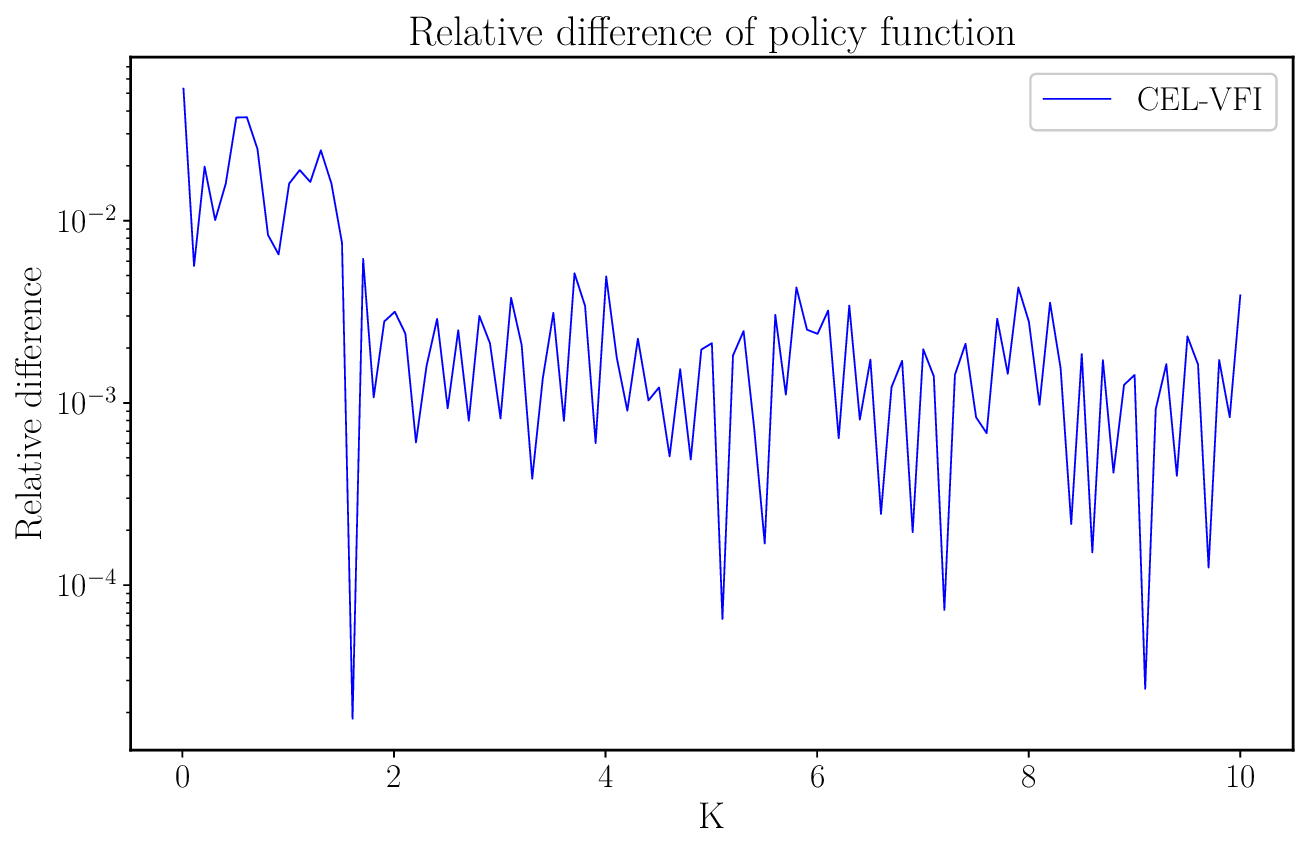}
		\caption{Relative difference between the CEL policy network and the VFI benchmark when $K_{t-1}$ varies and $a_t$ is fixed for $\sigma=1$.}
		\label{fig:relative_diff_c_smallnoise_1}
	\end{figure}

	Figure~\ref{fig:relative_diff_c_smallnoise_1} reports the relative difference between the learned policy and the VFI policy. Since the VFI solution is used as a numerical benchmark, this figure should be interpreted as a comparison between two numerical methods rather than as an error relative to the true solution. The small relative difference indicates that the proposed method approximates not only the value function but also the economically relevant policy function.
	
	\noindent\textbf{Euler residual and relative Bellman error.}
	We further evaluate the learned solution using two diagnostic measures: the Euler residual and the relative Bellman error. The Euler residual evaluates whether the learned consumption-saving policy satisfies the intertemporal optimality condition, while the relative Bellman error evaluates whether the learned value function is consistent with the Bellman equation after normalizing the Bellman discrepancy by the right-hand side of the Bellman equation. The formulas used to compute these diagnostic measures are reported in Appendix~\ref{app:smallnoise_euler_bellman_residual}.
	
\begin{figure}[!htbp]
		\centering
		\includegraphics[width=0.85\textwidth]{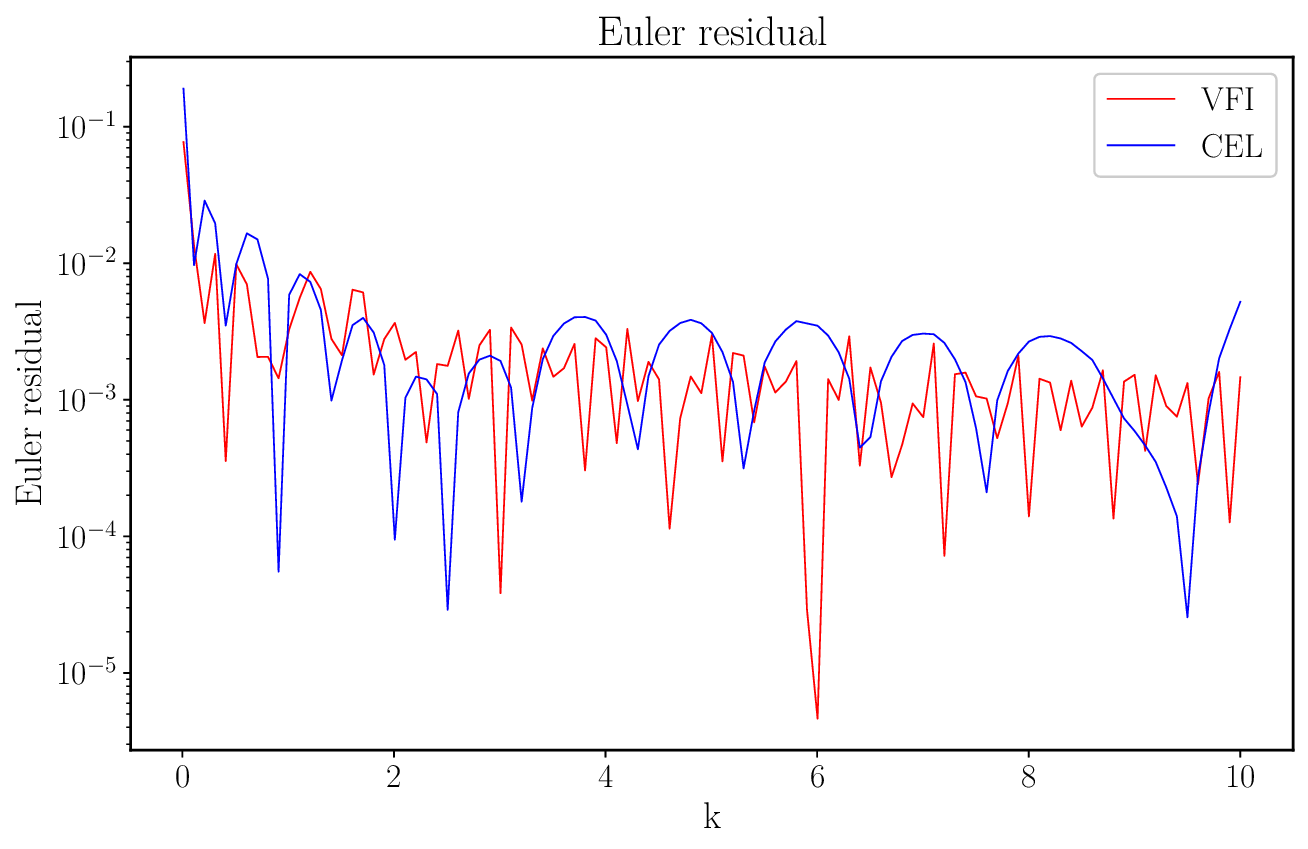}
		\caption{Euler residual for the general nonlinear small-noise model with $\sigma=1$.}
		\label{fig:ee_small_noise_1}
	\end{figure}

	Figure~\ref{fig:ee_small_noise_1} reports the Euler residual. The Euler residual from the CEL method is comparable to that from VFI across the evaluation range. This suggests that the learned policy satisfies the intertemporal optimality condition to a similar degree as the VFI benchmark.
	
\begin{figure}[!htbp]
		\centering
		\includegraphics[width=0.85\textwidth]{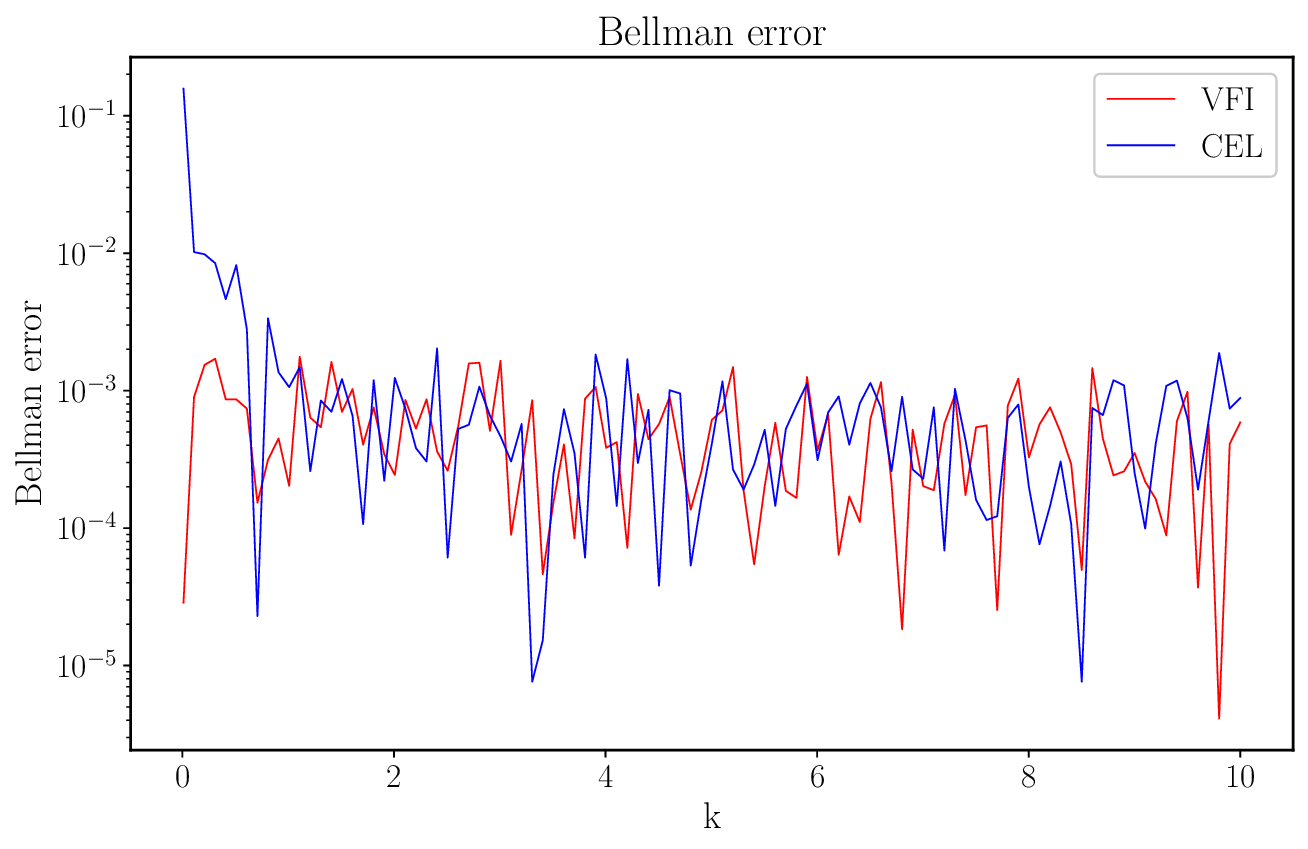}
		\caption{Bellman error for the general nonlinear small-noise model with $\sigma=1$.}
		\label{fig:be_small_noise_1}
	\end{figure}

	Figure~\ref{fig:be_small_noise_1} reports the Bellman error. The Bellman error from the CEL method is also comparable to that obtained from VFI. This indicates that the learned value function is consistent with the Bellman equation at a level similar to the reported grid-based benchmark.
	
	\FloatBarrier

	\noindent\textbf{Results for high robustness: $\sigma=50$.}
	
	We next consider a higher-robustness case with $\sigma=50$. A larger value of $\sigma$ strengthens the recursive exponential adjustment and makes the problem numerically more demanding. This case therefore provides a useful robustness check for the proposed method in the general nonlinear environment.
	
\begin{figure}[!htbp]
		\centering
		\includegraphics[width=0.85\textwidth]{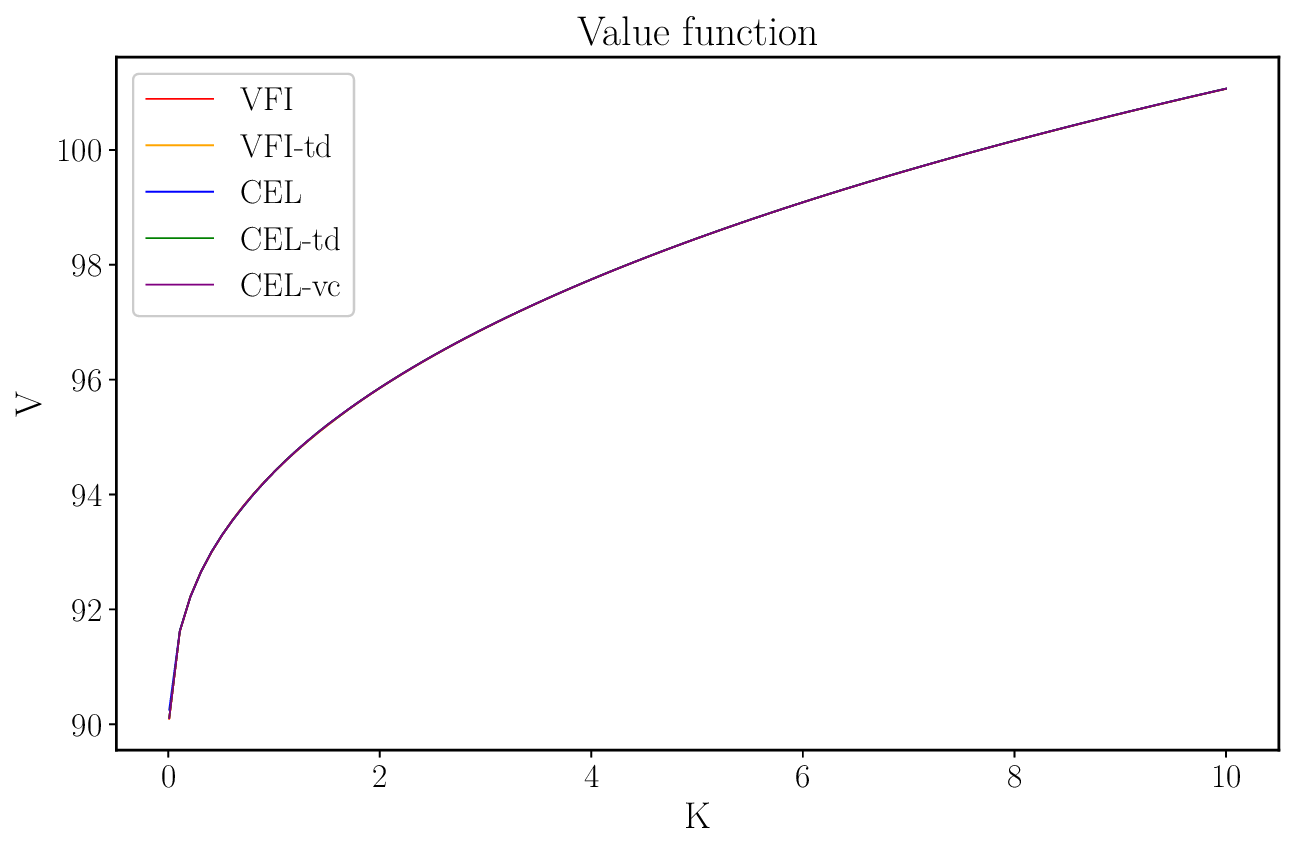}
		\caption{Value function and its one-step expansion from the CEL method and VFI for the general nonlinear small-noise model with $\sigma=50$.}
		\label{fig:vf_small_noise_50}
	\end{figure}

	\noindent\textbf{Value function approximation.}
	Figure~\ref{fig:vf_small_noise_50} compares the learned value function and its one-step expansion with the VFI benchmark for $\sigma=50$.
	
	The learned value function remains close to the VFI benchmark under high robustness. The direct value network and the one-step expansion also remain closely aligned, showing that the recursive approximation remains stable when the robustness parameter increases.
	
\begin{figure}[!htbp]
		\centering
		\includegraphics[width=0.85\textwidth]{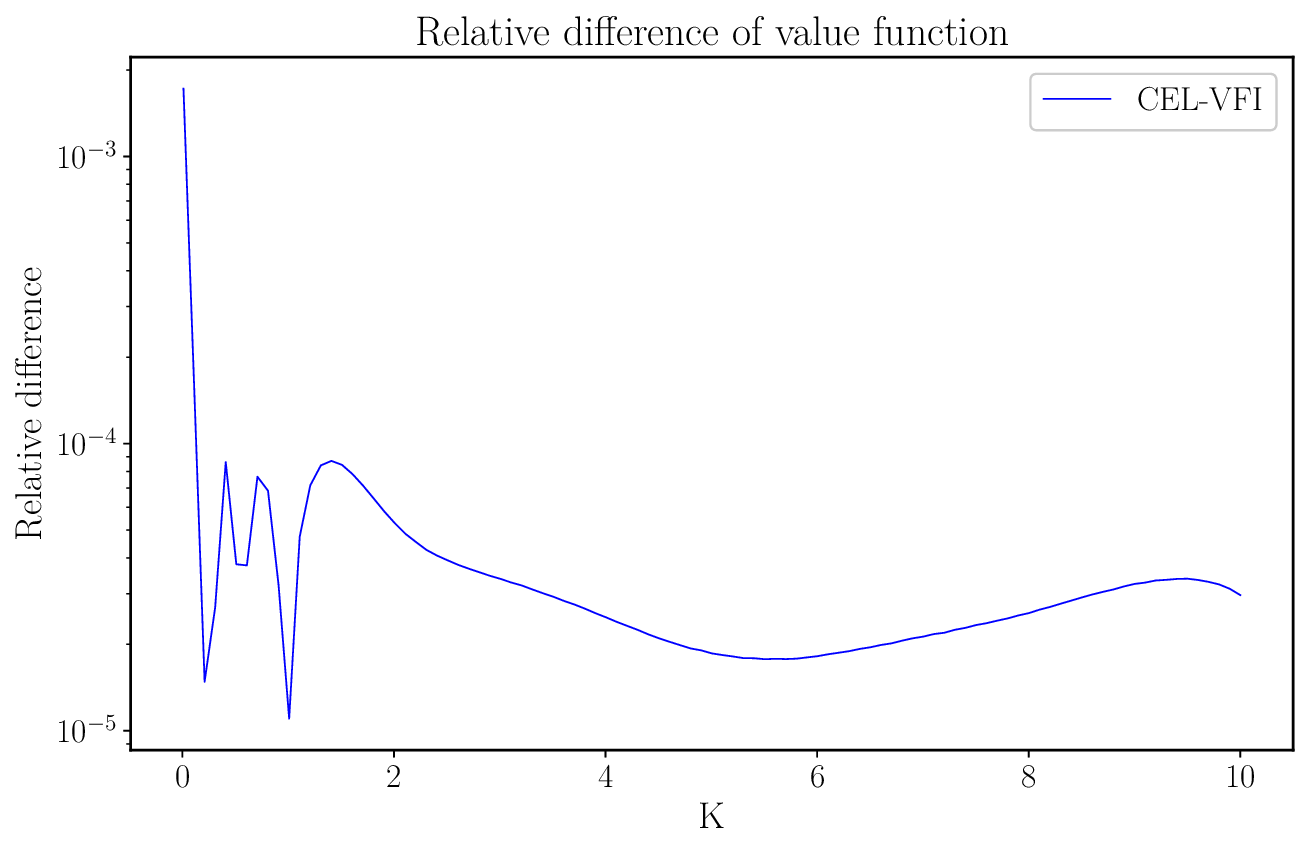}
		\caption{Relative difference between the CEL value network and the VFI benchmark when $K_{t-1}$ varies and $a_t$ is fixed for $\sigma=50$.}
		\label{fig:relative_diff_v_smallnoise_50}
	\end{figure}

	Figure~\ref{fig:relative_diff_v_smallnoise_50} reports the relative difference between the CEL value network and the VFI benchmark. As before, this figure is interpreted as a relative difference between two numerical solutions rather than as an error relative to the exact solution. The small relative difference indicates that the CEL method remains close to the VFI benchmark in the high-robustness case.
	
\begin{figure}[!htbp]
		\centering
		\includegraphics[width=0.85\textwidth]{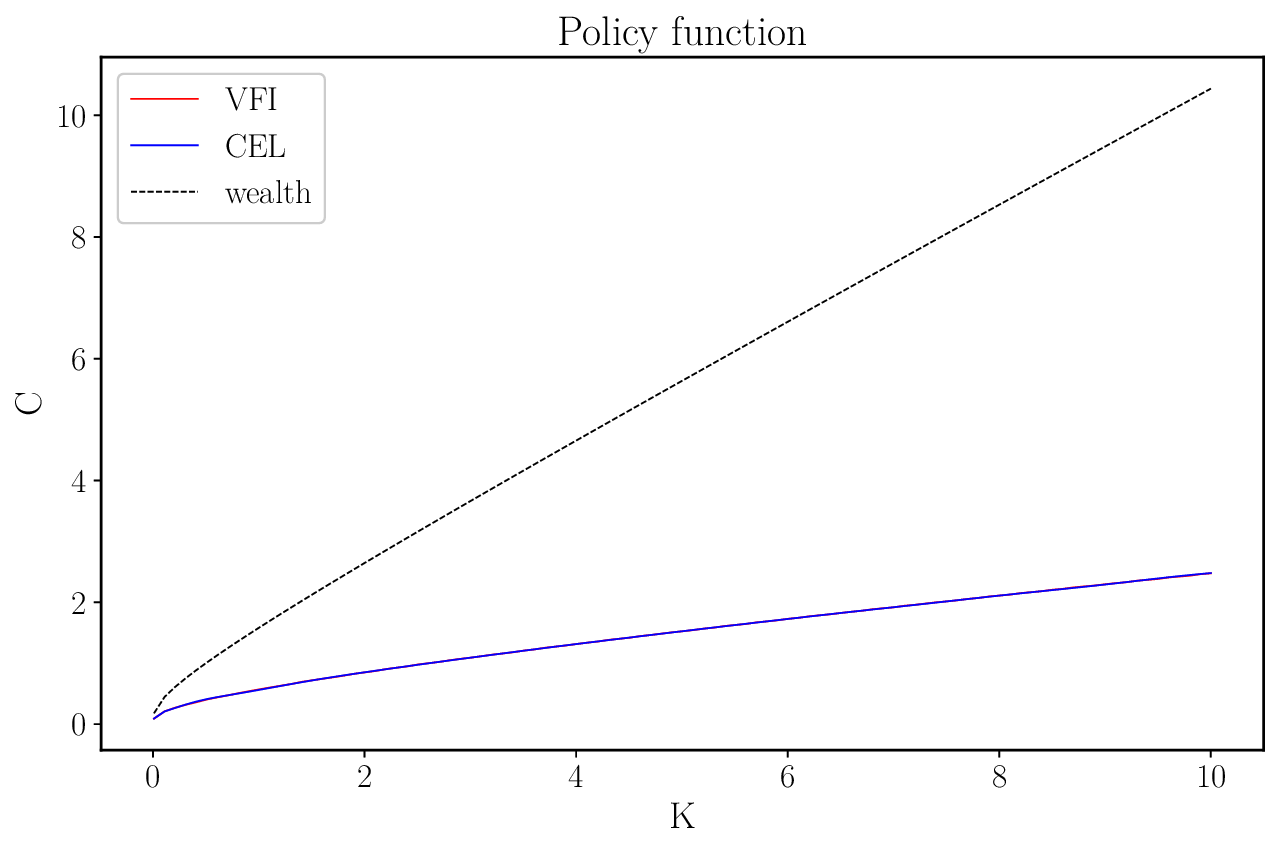}
		\caption{Policy function: consumption when capital $K_{t-1}$ varies for $\sigma=50$.}
		\label{fig:pf_small_noise_50}
	\end{figure}

\begin{figure}[!htbp]
		\centering
		\includegraphics[width=0.85\textwidth]{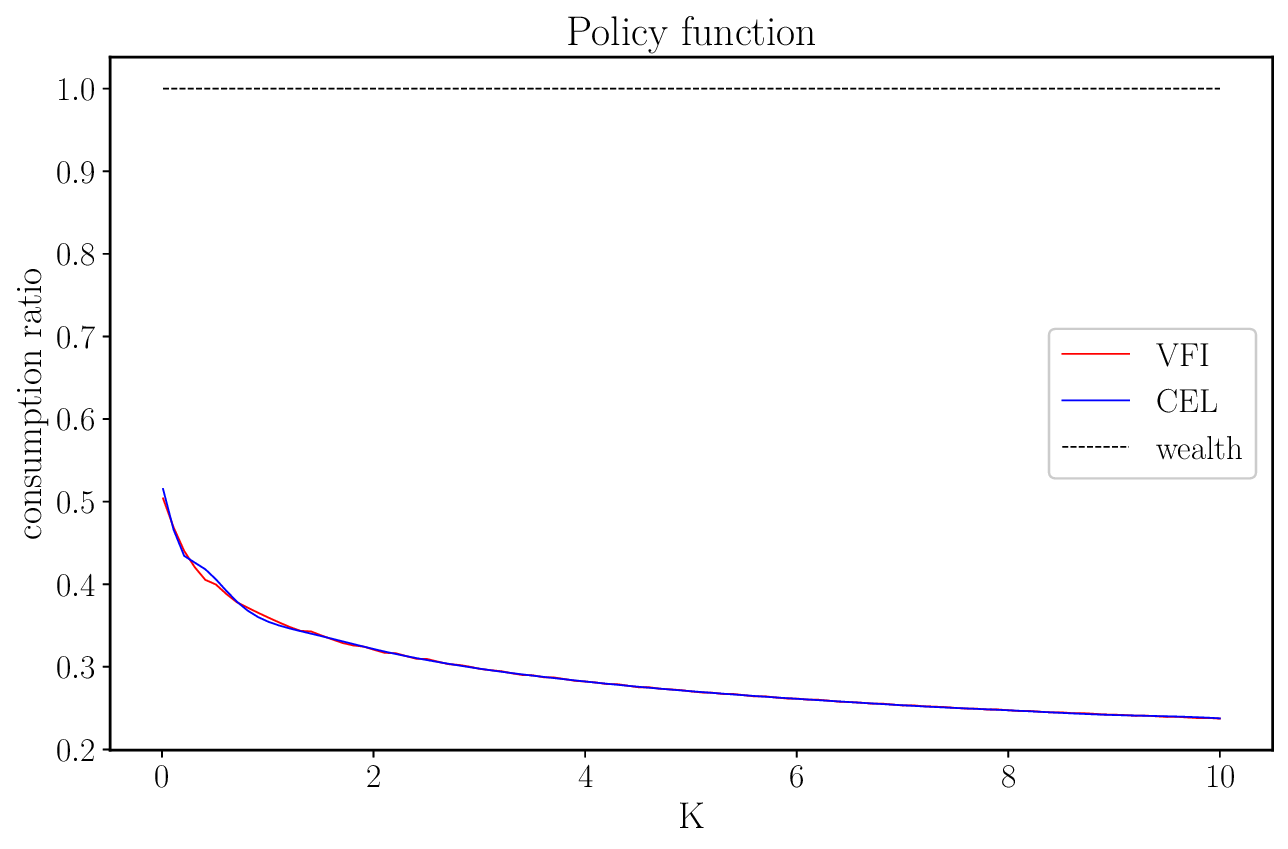}
		\caption{Policy function: consumption-to-wealth ratio when capital $K_{t-1}$ varies for $\sigma=50$.}
		\label{fig:pf_ratio_small_noise_50}
	\end{figure}

	\noindent\textbf{Policy function approximation.}
	Figures~\ref{fig:pf_small_noise_50} and~\ref{fig:pf_ratio_small_noise_50} report the learned policy function and the consumption-to-wealth ratio for $\sigma=50$.
	
	The policy functions remain smooth across the capital range. This suggests that the policy network continues to produce stable consumption decisions even when the recursive utility adjustment becomes stronger.
	
\begin{figure}[!htbp]
		\centering
		\includegraphics[width=0.85\textwidth]{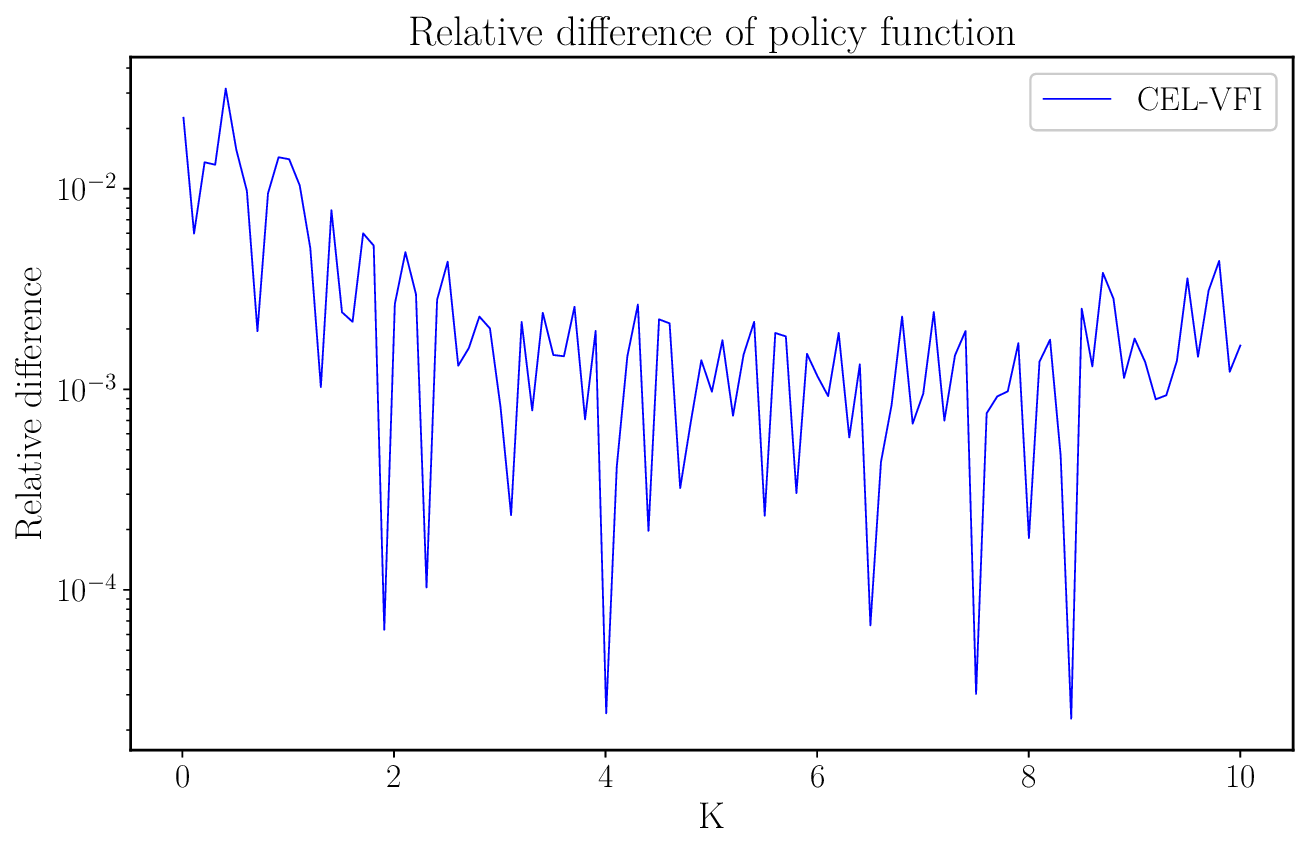}
		\caption{Relative difference between the CEL policy network and the VFI benchmark when $K_{t-1}$ varies and $a_t$ is fixed for $\sigma=50$.}
		\label{fig:relative_diff_c_smallnoise_50}
	\end{figure}

	Figure~\ref{fig:relative_diff_c_smallnoise_50} reports the relative difference between the CEL policy and the VFI policy. The difference remains controlled across the state range, suggesting that the proposed method produces policy functions that are close to the VFI benchmark under high robustness.
	
\begin{figure}[!htbp]
		\centering
		\includegraphics[width=0.85\textwidth]{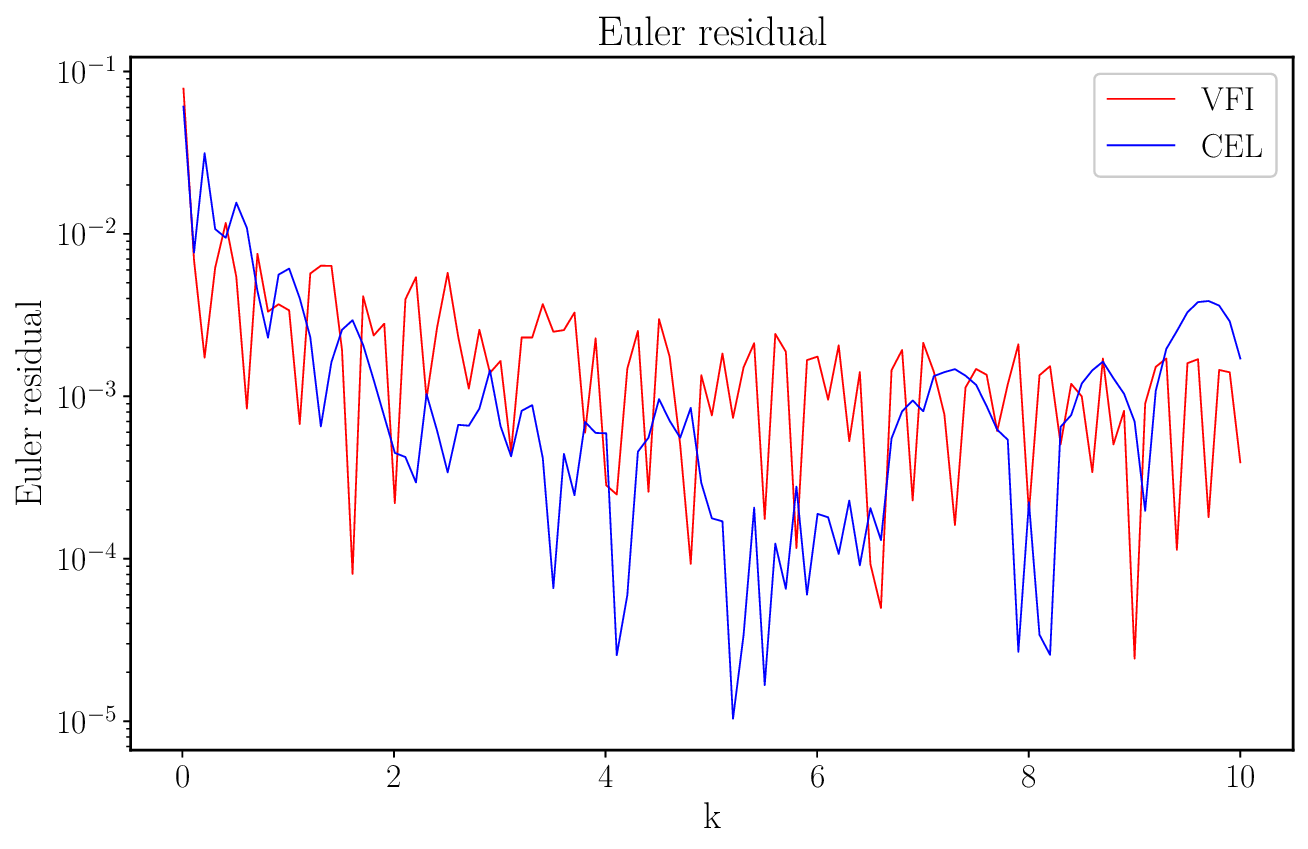}
		\caption{Euler residual for the general nonlinear small-noise model with $\sigma=50$.}
		\label{fig:ee_small_noise_50}
	\end{figure}

	\noindent\textbf{Euler residual and Bellman error.}
	Figure~\ref{fig:ee_small_noise_50} reports the Euler residual for $\sigma=50$. The Euler residual produced by the CEL method is of a similar magnitude to that obtained from VFI. This indicates that the learned policy satisfies the first-order optimality condition at a level comparable to the VFI benchmark.
	
\begin{figure}[!htbp]
		\centering
		\includegraphics[width=0.85\textwidth]{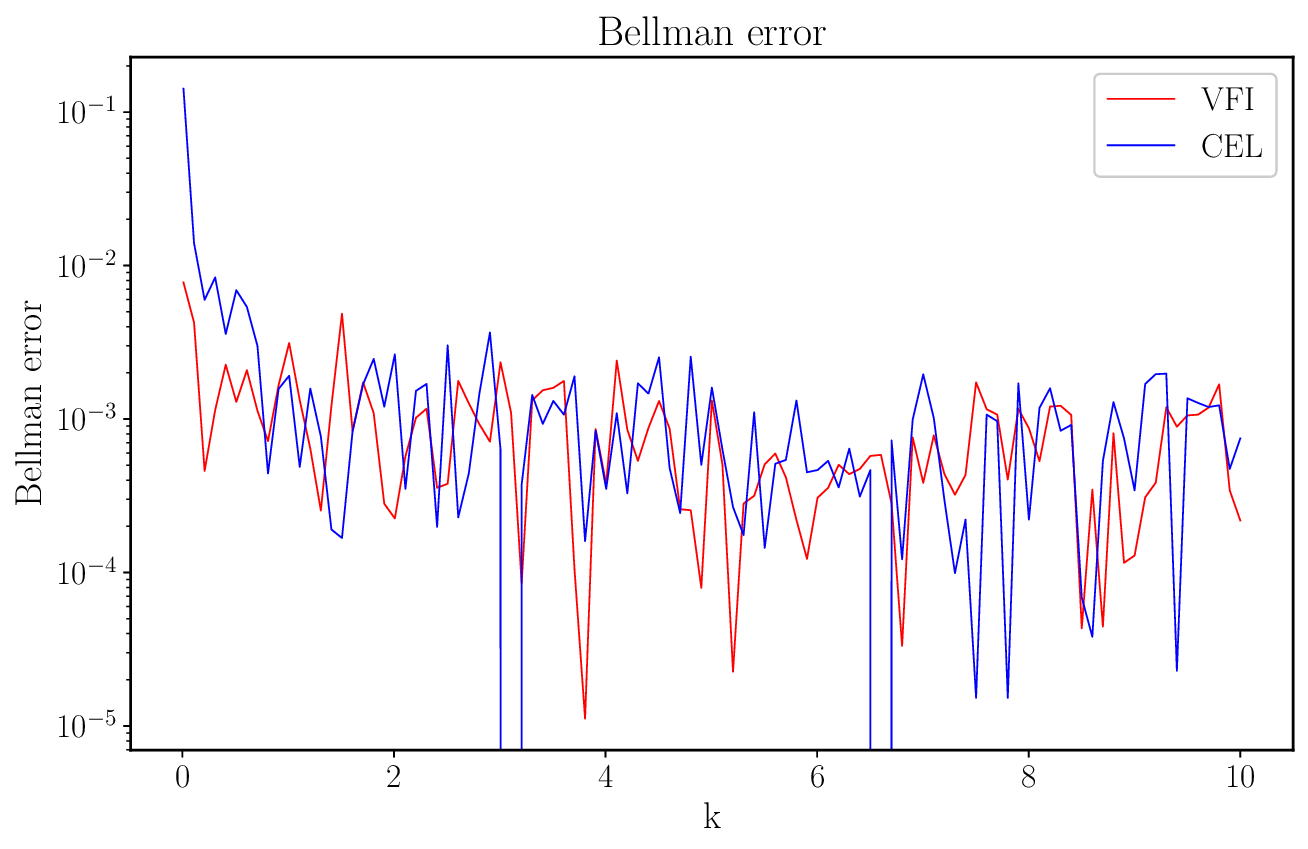}
		\caption{Bellman error for the general nonlinear small-noise model with $\sigma=50$.}
		\label{fig:be_small_noise_50}
	\end{figure}

	Figure~\ref{fig:be_small_noise_50} reports the Bellman error. The Bellman error from the CEL method is also comparable to that of VFI, suggesting that the learned value function satisfies the Bellman equation to a similar degree as the reported grid-based benchmark. Together, the Euler residual and Bellman error indicate that the proposed method remains numerically stable under stronger robustness concerns.
	
	Overall, the general nonlinear small-noise results show that the proposed deep learning method provides value and policy functions that are close to the VFI benchmark. The residual-based diagnostics further show that the Euler residuals and Bellman errors of the CEL and VFI solutions are of similar magnitude. Since the general nonlinear case does not admit an analytical solution, these comparisons should be interpreted as validation against a standard numerical benchmark rather than as a direct ranking relative to the true solution.
	
	\FloatBarrier

	\subsubsection{Homothetic Benchmark Case with Analytical Solution}
	\label{subsubsec:smallnoise_homothetic}
	
	We next consider a homothetic benchmark case with an analytical solution. This case is useful because it allows us to compare both the CEL and VFI solutions directly with the analytical benchmark. Instead of describing it simply as a case with a closed-form solution, we refer to it as a homothetic benchmark case because the analytical tractability comes from the special homothetic structure of the model.
	
	A closed-form solution exists for this optimization problem under the following special case:
	\begin{align}
		V(K_{t-1}, a_t) &= \max_{C_t} \left\{ \ln(C_t) - \frac{1}{\sigma} \log E_t\left[\exp \left(-\sigma \beta V(K_t,a_{t+1})\right) \right] \right\}, \\
		C_t + K_t &= e^{P a_t}K_{t-1}^{\alpha}, \\
		a_{t+1} &= \Omega_0 + \sqrt{\epsilon} \Omega_v w_{t+1}, \quad w_{t+1} \sim \mathcal{N}(0,1),
	\end{align}
	where the logarithmic utility and the simplified production and shock structure generate a homothetic solution. This analytical benchmark allows us to evaluate the numerical accuracy of the value function and policy function more directly than in the general nonlinear case.
	
	The model is trained with state $s_t=(K_{t-1},a_t)$, control $c_t\in(0,1)$, discount factor $\beta=0.9$, and robustness parameters $\sigma=50$ and $\sigma=200$. In the error plots below, the CEL solution, the VFI solution, and the analytical solution are evaluated at the VFI capital grid points. Therefore, the reported relative errors compare the numerical methods with the analytical solution at the same evaluation points.
	
	\noindent\textbf{Results for moderate robustness: $\sigma=50$.}
	
\begin{figure}[!htbp]
		\centering
		\includegraphics[width=0.85\textwidth]{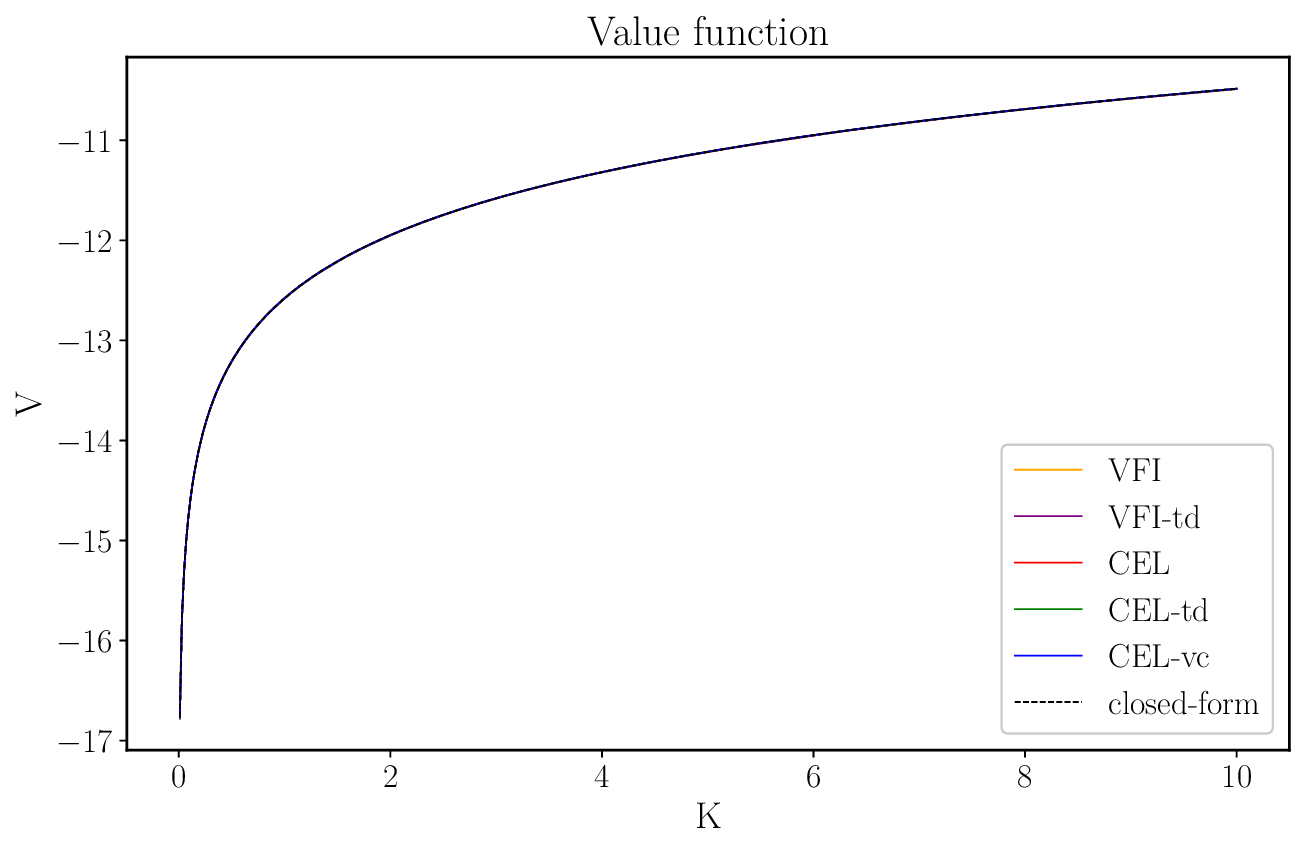}
		\caption{Value function and its one-step expansion from CEL method and VFI when $\sigma=50$.}
		\label{fig:small_noise_vcf50}
	\end{figure}

\begin{figure}[!htbp]
		\centering
		\includegraphics[width=0.85\textwidth]{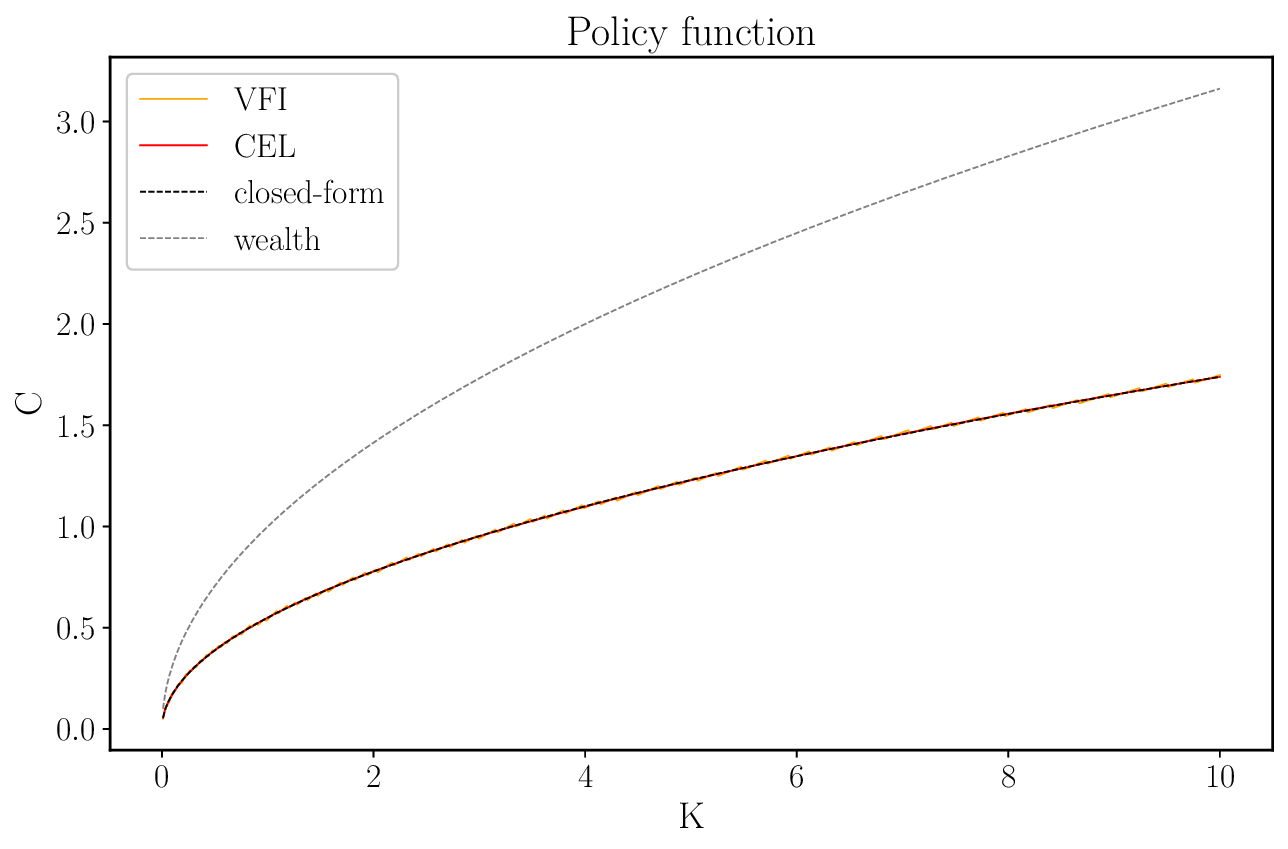}
		\caption{Policy function comparison with the analytical solution when $\sigma=50$.}
		\label{fig:small_noise_ccf50}
	\end{figure}

\begin{figure}[!htbp]
		\centering
		\includegraphics[width=0.85\textwidth]{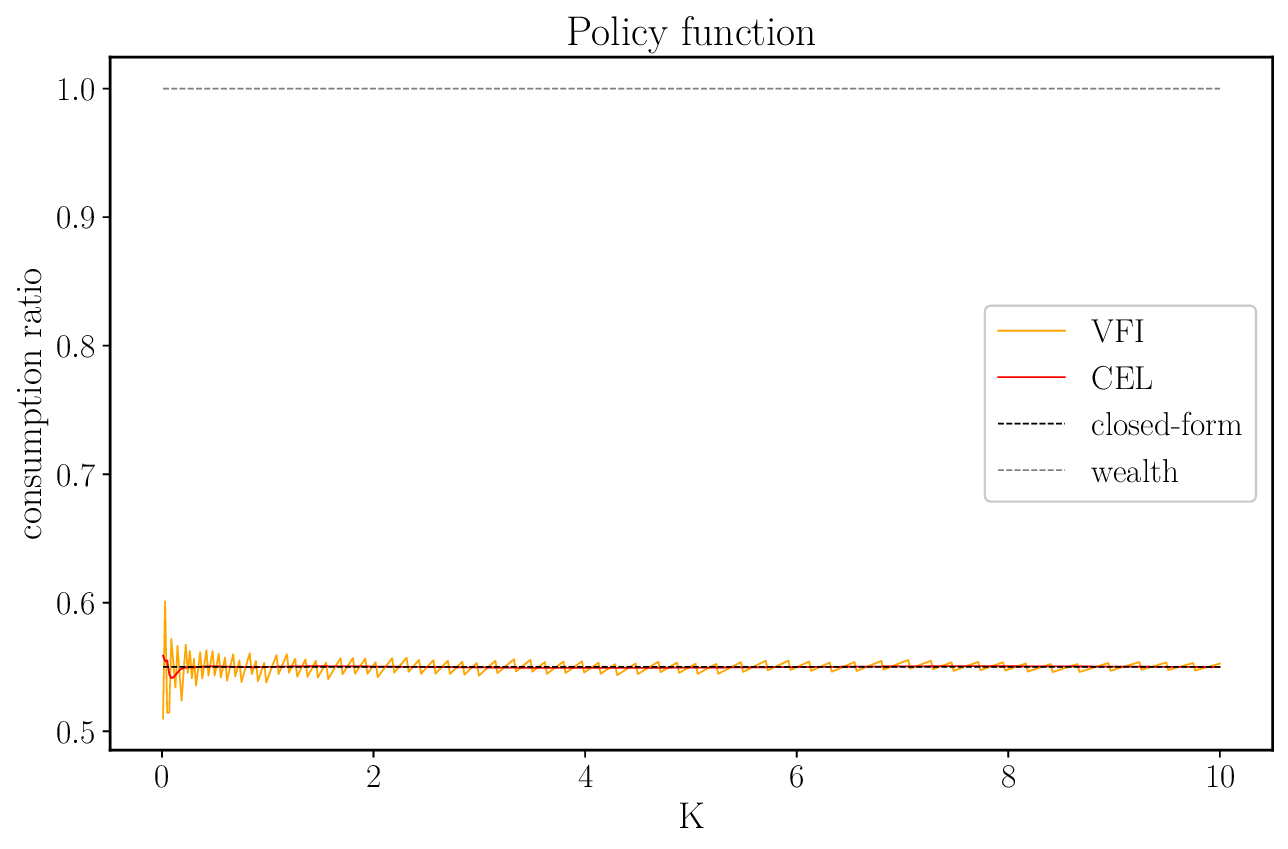}
		\caption{Consumption-to-wealth ratio comparison with the analytical solution when $\sigma=50$.}
		\label{fig:small_noise_crcf50}
	\end{figure}

	Figures~\ref{fig:small_noise_vcf50}, \ref{fig:small_noise_ccf50}, and \ref{fig:small_noise_crcf50} report the value function, policy function, and consumption-to-wealth ratio for the case $\sigma=50$. The learned value function closely follows the analytical solution, and the policy function also tracks the benchmark well.
	
	Figure~\ref{fig:small_noise_vcf50} shows that the CEL value function is very close to the analytical benchmark. The one-step certainty-equivalent expansion is also closely aligned with the direct value network, indicating that the learned value function approximation is internally consistent. Compared with the VFI benchmark, the CEL value function is visibly closer to the analytical solution over most of the evaluation range.
	
	Figures~\ref{fig:small_noise_ccf50} and \ref{fig:small_noise_crcf50} show that both CEL and VFI recover the policy function. The CEL policy is close to the analytical benchmark and appears smoother over the evaluation range. The comparison suggests that the proposed method can accurately recover the consumption-saving decision in this homothetic benchmark case.
	
\begin{figure}[!htbp]
		\centering
		\includegraphics[width=0.85\textwidth]{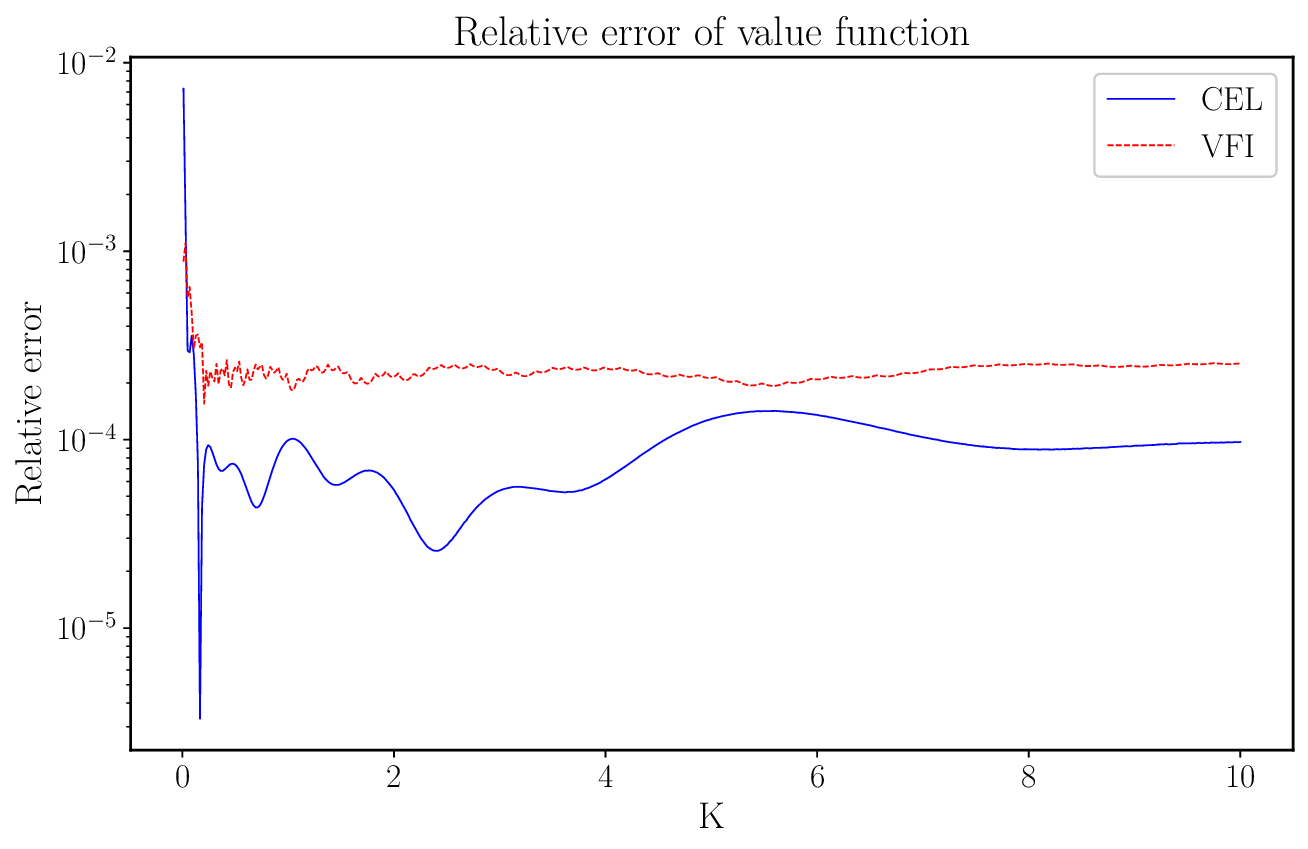}
		\caption{Relative error of value function with respect to the analytical solution when $\sigma=50$. The errors are evaluated at the VFI capital grid points.}
		\label{fig:small_noise_error_value_cf50}
	\end{figure}

\begin{figure}[!htbp]
		\centering
		\includegraphics[width=0.85\textwidth]{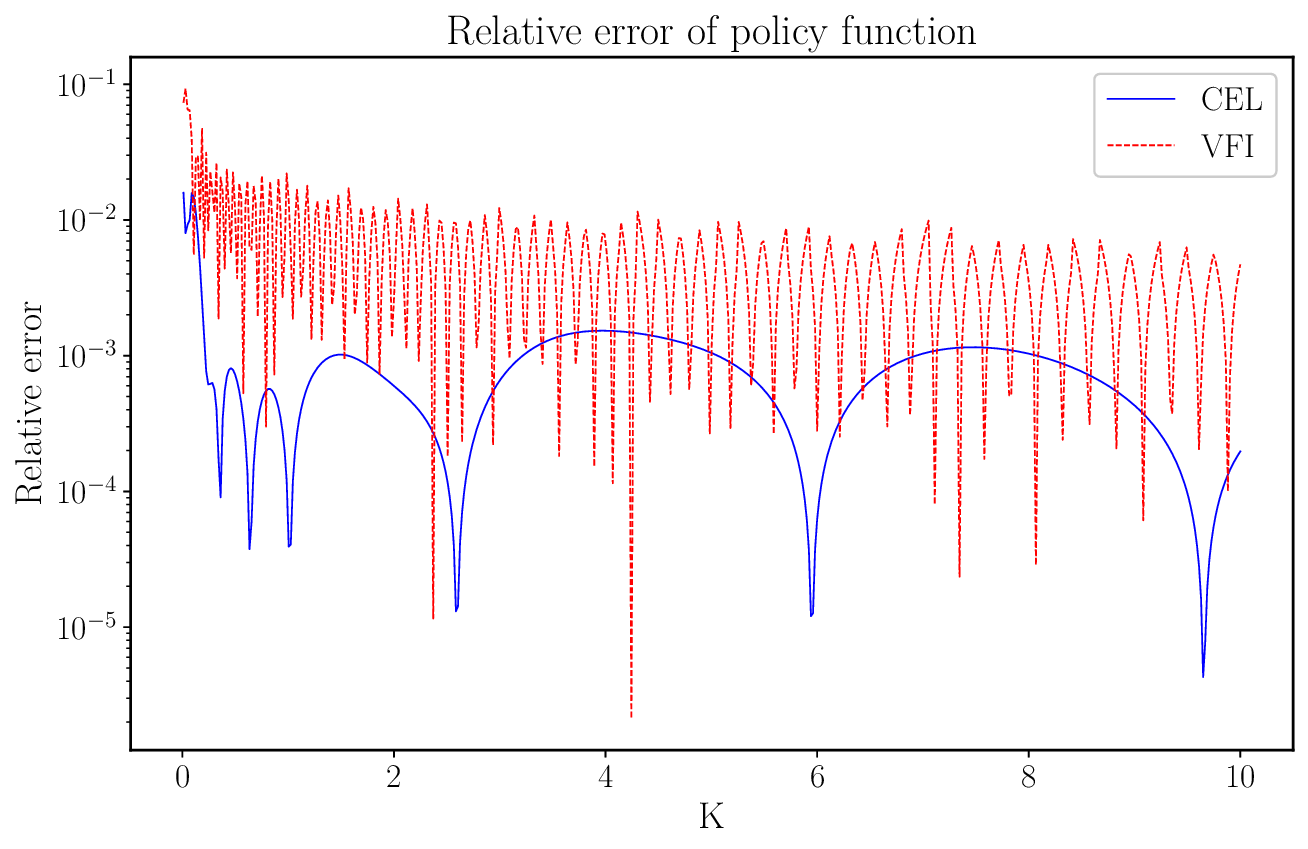}
		\caption{Relative error of policy function with respect to the analytical solution when $\sigma=50$. The errors are evaluated at the VFI capital grid points.}
		\label{fig:small_noise_error_policy_cf50}
	\end{figure}

	Figures~\ref{fig:small_noise_error_value_cf50} and \ref{fig:small_noise_error_policy_cf50} plot the relative errors with respect to the analytical solution for the value function and policy function. The errors are evaluated at the VFI capital grid points, and the vertical axis uses a logarithmic scale to display the magnitude of the errors.
	
	Figure~\ref{fig:small_noise_error_value_cf50} shows that the CEL value-function error is generally lower than that of the VFI benchmark. Over most of the capital grid, the CEL error is around or below $10^{-4}$, whereas the VFI error is typically between $10^{-4}$ and $10^{-3}$. There is a localized deviation near the lower boundary of the grid, but away from this boundary region the CEL value approximation remains closer to the analytical solution.
	
	Figure~\ref{fig:small_noise_error_policy_cf50} shows that the CEL policy-function error is generally lower and smoother than that of the VFI benchmark. The VFI policy error exhibits visible oscillations across the capital grid, while the CEL policy error remains smoother over most of the evaluation range. This pattern suggests that the neural policy approximation can reduce some of the local oscillations associated with grid-based policy approximation.
	
	\FloatBarrier
	
	\noindent\textbf{Results for high robustness: $\sigma=200$.}
	
	We further consider a high-robustness case with $\sigma=200$. A larger value of $\sigma$ increases the strength of the recursive exponential adjustment and makes the problem numerically more challenging. This case therefore provides a useful robustness check for the proposed method.
	
\begin{figure}[!htbp]
		\centering
		\includegraphics[width=0.85\textwidth]{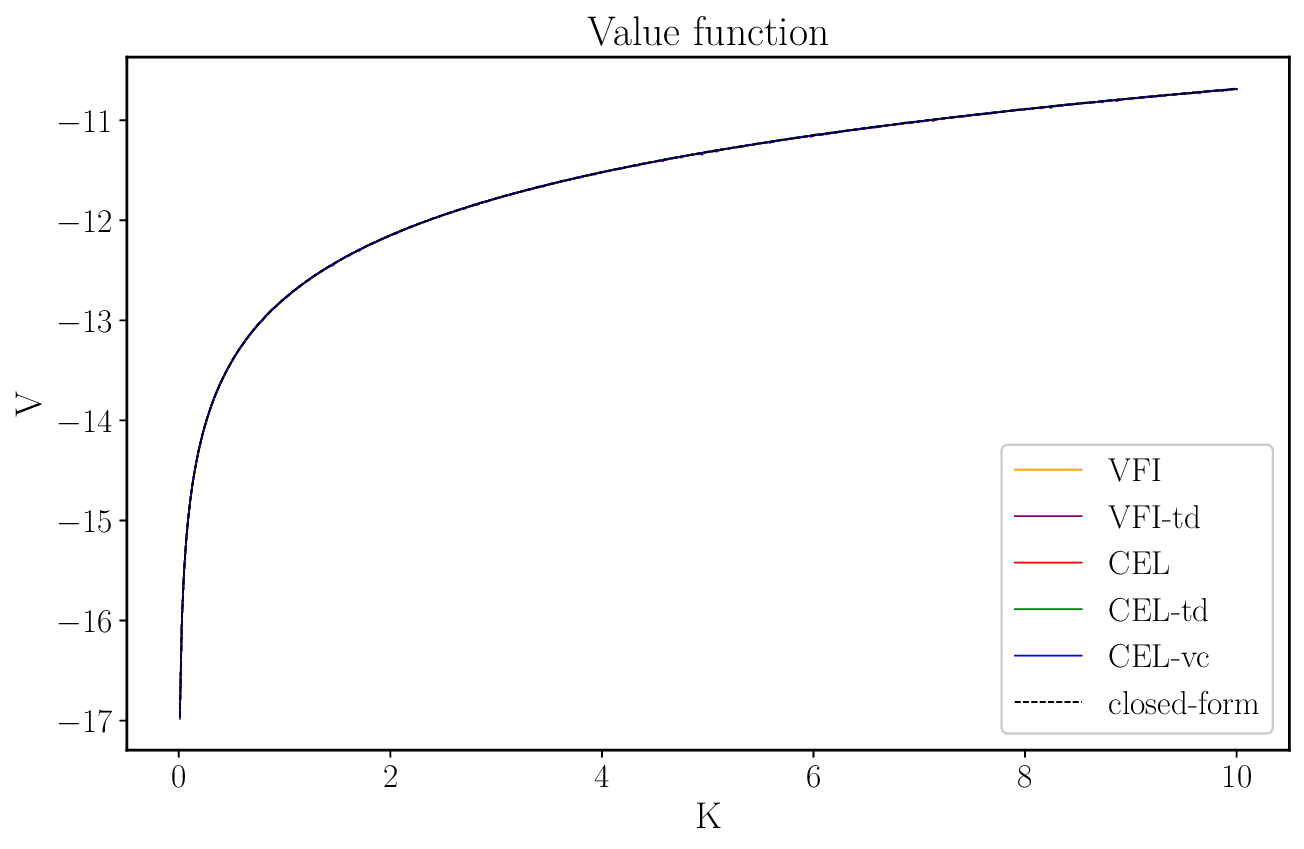}
		\caption{Value function and its one-step expansion from CEL method and VFI when $\sigma=200$.}
		\label{fig:small_noise_vcf200}
	\end{figure}

\begin{figure}[!htbp]
		\centering
		\includegraphics[width=0.85\textwidth]{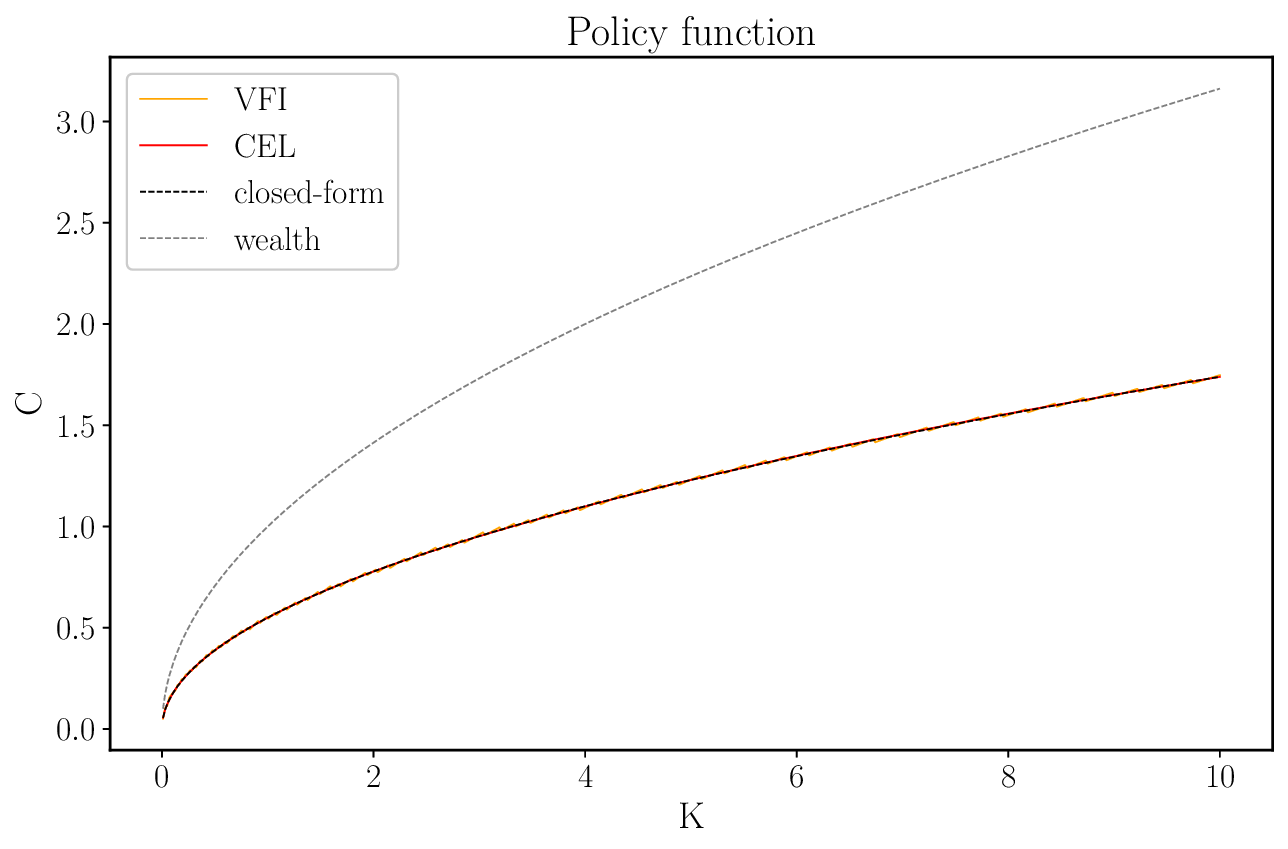}
		\caption{Policy function comparison with the analytical solution when $\sigma=200$.}
		\label{fig:small_noise_ccf200}
	\end{figure}

\begin{figure}[!htbp]
		\centering
		\includegraphics[width=0.85\textwidth]{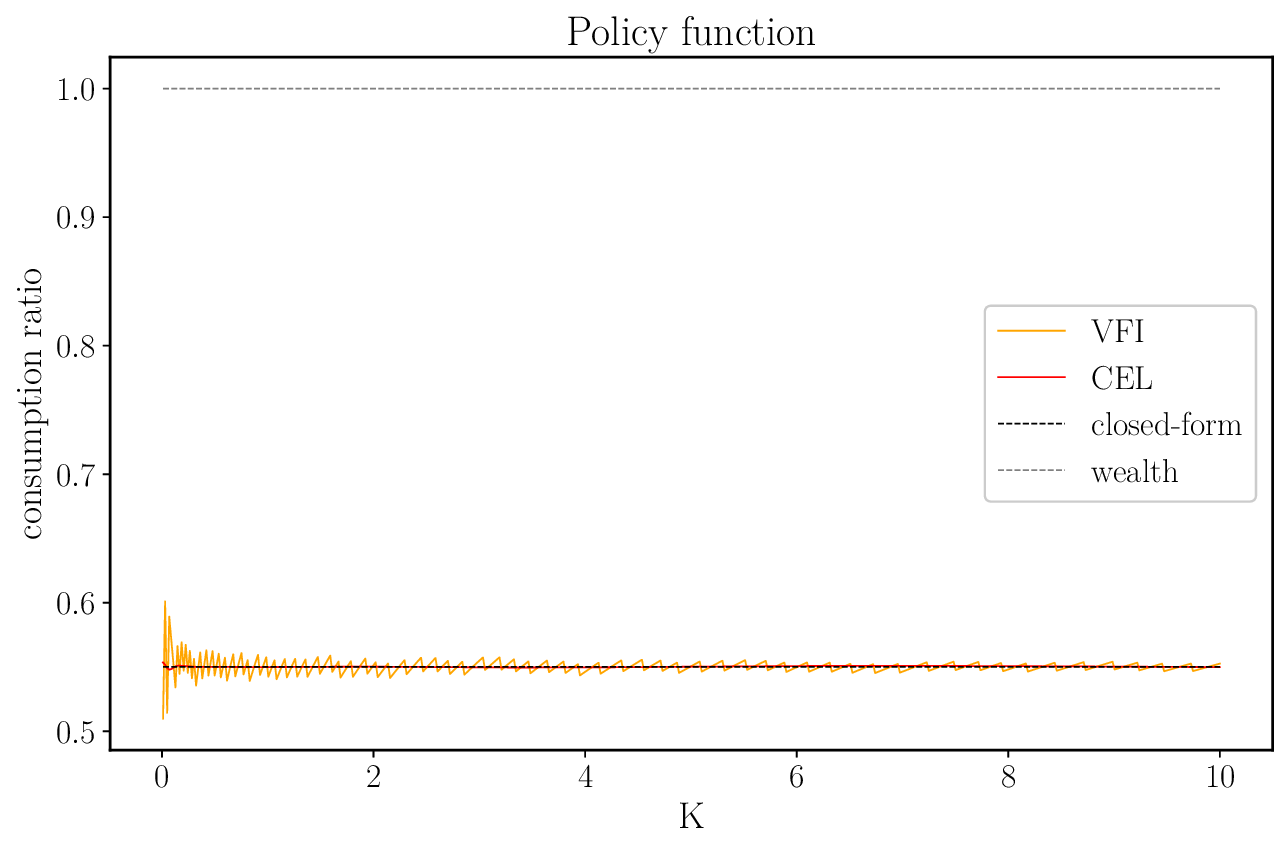}
		\caption{Consumption-to-wealth ratio comparison with the analytical solution when $\sigma=200$.}
		\label{fig:small_noise_crcf200}
	\end{figure}

	Figures~\ref{fig:small_noise_vcf200}, \ref{fig:small_noise_ccf200}, and \ref{fig:small_noise_crcf200} report the value function, policy function, and consumption-to-wealth ratio. The learned value and policy functions remain close to the analytical benchmark, showing that the method is stable even when the robustness parameter is large.
	
	Figure~\ref{fig:small_noise_vcf200} shows that the CEL value function remains close to the analytical benchmark under high robustness. The certainty-equivalent expansion also stays aligned with the direct value network, suggesting that the recursive structure is still well captured when the robustness parameter becomes large.
	
	Figures~\ref{fig:small_noise_ccf200} and \ref{fig:small_noise_crcf200} show that the learned policy remains stable and close to the analytical benchmark. Similar to the $\sigma=50$ case, the CEL policy is smooth and tracks the analytical solution over the capital range. This suggests that the proposed method can recover the main shape of the policy function even under stronger robustness concerns.
	
\begin{figure}[!htbp]
		\centering
		\includegraphics[width=0.85\textwidth]{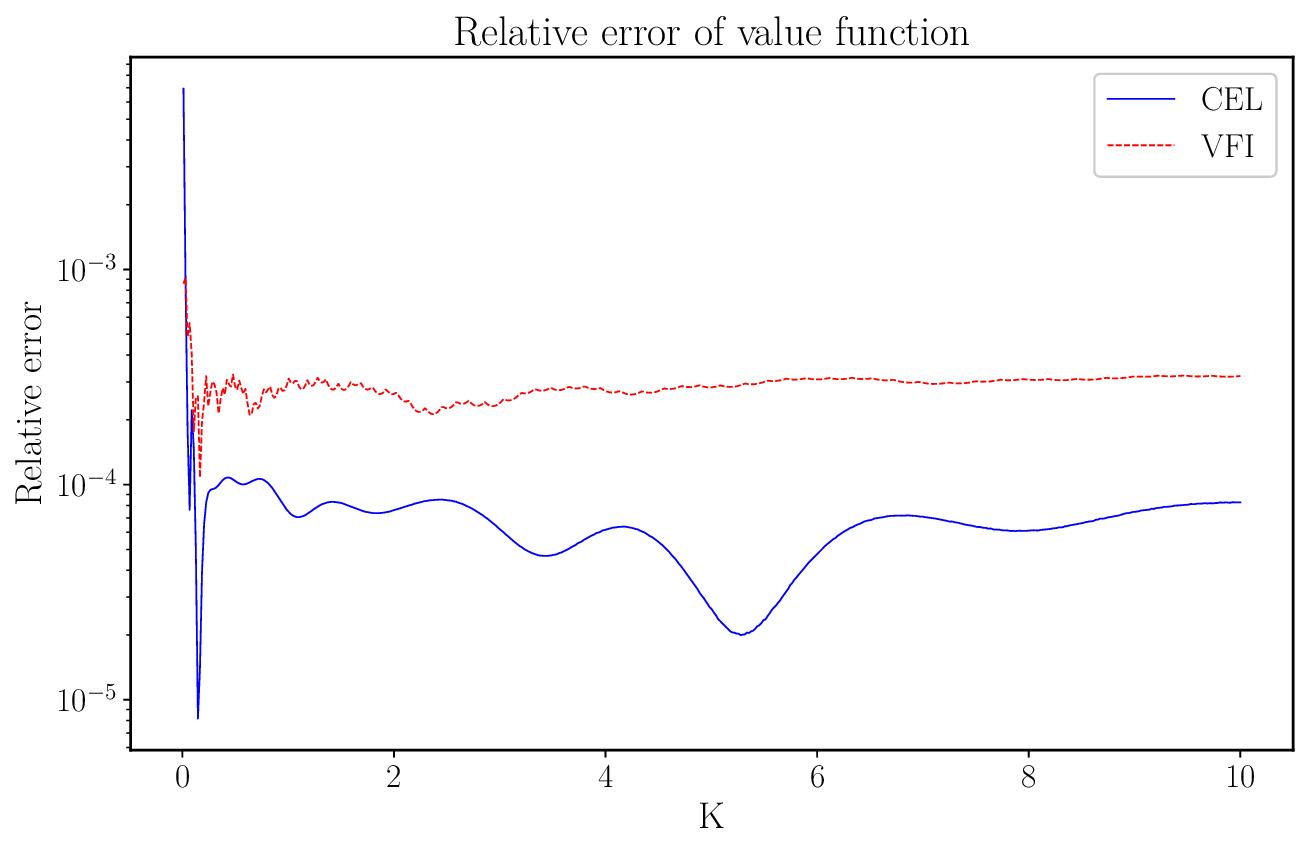}
		\caption{Relative error of value function with respect to the analytical solution when $\sigma=200$. The errors are evaluated at the VFI capital grid points.}
		\label{fig:small_noise_error_value_cf200}
	\end{figure}

\begin{figure}[!htbp]
		\centering
		\includegraphics[width=0.85\textwidth]{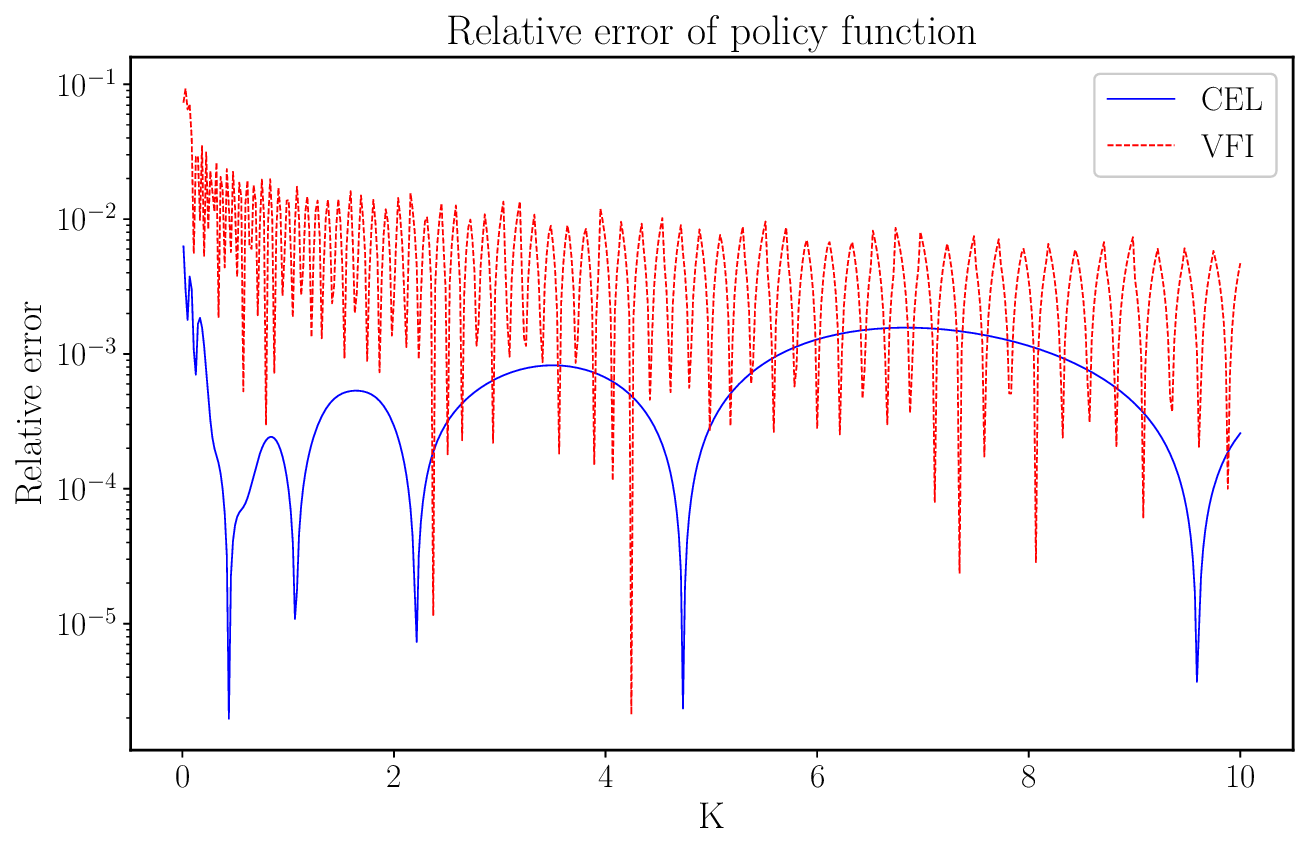}
		\caption{Relative error of policy function with respect to the analytical solution when $\sigma=200$. The errors are evaluated at the VFI capital grid points.}
		\label{fig:small_noise_error_policy_cf200}
	\end{figure}

	Figures~\ref{fig:small_noise_error_value_cf200} and \ref{fig:small_noise_error_policy_cf200} report the relative errors with respect to the analytical solution. As in the $\sigma=50$ case, all relative errors are evaluated at the VFI capital grid points.
	
	Figure~\ref{fig:small_noise_error_value_cf200} shows that the CEL value-function error remains generally lower than that of the VFI benchmark. The CEL error is mostly around or below $10^{-4}$, while the VFI value-function error is typically between $10^{-4}$ and $10^{-3}$. Thus, even in the high-robustness case, the proposed method provides a more accurate value-function approximation in most regions of the evaluation grid.
	
	Figure~\ref{fig:small_noise_error_policy_cf200} shows that the CEL policy-function error is generally lower and smoother than the VFI policy-function error. The oscillatory pattern in the VFI policy error is less pronounced in the CEL solution, indicating that the learned policy network provides a smoother approximation to the analytical policy.
	
	\noindent\textbf{Accuracy comparison.}
	
	The homothetic benchmark results show a consistent pattern across both $\sigma=50$ and $\sigma=200$. Because an analytical solution is available, the reported quantities are true relative errors with respect to the closed-form benchmark, evaluated at the VFI capital grid points. For the value function, the CEL approximation is generally more accurate than the VFI benchmark, with relative errors mostly around or below $10^{-4}$, compared with VFI errors that are typically between $10^{-4}$ and $10^{-3}$. For the policy function, the CEL approximation also delivers lower and smoother relative errors over most of the evaluation range, while the VFI policy error exhibits visible local oscillations.
	
	Overall, the homothetic benchmark case confirms that the proposed deep learning method can accurately solve recursive utility problems when an analytical benchmark is available. In contrast to the general nonlinear case, where VFI is only a numerical benchmark, this closed-form case permits a direct error comparison. The results suggest that CEL achieves accuracy comparable to, and often better than, the VFI benchmark in both value-function and policy-function approximation, especially away from localized boundary regions.
	
	\FloatBarrier

	\subsection{DSGE Model with Recursive Preferences and Stochastic Volatility}
	\label{subsec:dsge_results}
	
	Our third example is adapted from \citet{caldara2012computing}, who study a dynamic stochastic general equilibrium (DSGE) model with recursive preferences and time-varying volatility in total factor productivity (TFP). This framework extends the standard neoclassical growth model by incorporating \citet{epstein1989substitution,epstein1991substitution}-style preferences and a stochastic volatility process, making it particularly suited for studying business cycle fluctuations and long-run asset pricing phenomena. The model's key innovation lies in its ability to separate risk aversion from intertemporal substitution while capturing time-varying economic uncertainty.

	\subsubsection{Problem Setup and Network Representation}
	\label{subsubsec:dsge_setup}

	The representative agent's preferences are specified through the following recursive utility function:
	\begin{equation}
		V(K_t,z_t,\sigma_t) = \max_{C_t, L_t} \left\{
		(1-\beta) \left(C_t^\nu (1-L_t)^{1-\nu}\right)^{\frac{1-\gamma}{\theta}}
		+ \beta \left(\mathbb{E}_t\left[V(K_{t+1},z_{t+1},\sigma_{t+1})^{1-\gamma}\right]\right)^{\frac{1}{\theta}}
		\right\}^{\frac{\theta}{1-\gamma}}, \label{eq:vf_dsge}
	\end{equation}
	where $\nu$ governs the consumption-leisure trade-off, $\gamma$ measures risk aversion, and $\theta = \frac{1-\gamma}{1-1/\psi}$ captures the deviation from standard CRRA preferences, with $\psi$ denoting the elasticity of intertemporal substitution (EIS). This recursive formulation allows for distinct roles of risk aversion and EIS, enabling more flexible modeling of investor behavior under uncertainty.
	
	The production side follows a standard Cobb-Douglas technology:
	\begin{equation}
		C_t + K_{t+1} = e^{z_t}K_t^{\zeta}L_t^{1-\zeta} + (1-\delta)K_t,
	\end{equation}
	where $\zeta$ represents the capital share and $\delta$ is the depreciation rate. The total factor productivity shock $z_t$ evolves as:
	\begin{align}
		z_t &= \lambda z_{t-1} + e^{\sigma_t} \epsilon_t, \quad \epsilon_t \sim \mathcal{N}(0,1), \\
		\sigma_t &= (1-\rho)\bar{\sigma} + \rho \sigma_{t-1} + \eta \omega_t, \quad \omega_t \sim \mathcal{N}(0,1).
	\end{align}
	Here, $\sigma_t$ follows a stationary AR(1) process with persistence $\rho$ and innovation volatility $\eta$, introducing stochastic volatility into the economic system. This specification captures the time-varying uncertainty in productivity shocks, which helps generate realistic asset pricing implications and business cycle dynamics.
	
	For the deep learning implementation, the state is $s_t=(K_t,z_t,\sigma_t)^\top$, and the controls are the consumption-to-resources ratio $c_t$ and labor supply $L_t$. The consumption ratio is
	\begin{equation}
		c_t
		=
		\frac{C_t}{e^{z_t}K_t^{\zeta}L_t^{1-\zeta}+(1-\delta)K_t}.
	\end{equation}
	We impose $c_t,L_t\in(0,1)$ through the final-layer activation functions, which guarantees positive consumption and feasible labor allocations.

	To capture the nested recursive utility structure, we employ a four-network architecture:
	\begin{align*}
		\begin{array}{r@{\;}c@{\;}l@{}l}
			V(s_t)
			& \equiv
			& V(s_t;\phi),
			& \\[0.3em]
			c_t
			& \equiv
			& c(s_t;\theta_c),
			& \smash{\raisebox{-0.65\normalbaselineskip}{$
					\left.
					\begin{array}{c}
						\vphantom{c(s_t;\theta_c)}\\[0.2em]
						\vphantom{L(s_t;\theta_L)}
					\end{array}
					\right\}
					\,\text{policy networks}
					$}} \\[0.1em]
			L_t
			& \equiv
			& L(s_t;\theta_L),
			& \\[0.3em]
			\mathbb{E}\left[V(s_{t+1})\mid s_t,c_t,L_t\right]
			& \equiv
			& V_e(s_t,c_t,L_t;\rho),
			& \\[0.3em]
			\mathbb{E}\left[V(s_{t+1})\right]
			-
			f^{-1}\left(\mathbb{E}\left[f(V(s_{t+1}))\right]\right)
			& \equiv
			& D(s_t,c_t,L_t;\nu).
			&
		\end{array}
	\end{align*}

	The two policy components, $c(s_t;\theta_c)$ and $L(s_t;\theta_L)$, are grouped as the policy-network block in the four-network CEL architecture.
	
	This DSGE framework presents a computationally challenging test case due to its three-dimensional state space and the interaction between recursive preferences and stochastic volatility. The model's solution requires accurately capturing precautionary saving motives and time-varying risk premia, making it an ideal benchmark for evaluating CEL's performance in a rich economic environment.
	
	\subsubsection{Numerical Results}
	\label{subsubsec:dsge_numerical_results}
	
	In this subsection, we evaluate the learned value function, policy functions, and several performance measures for the DSGE model. Let $K_{ss}$ denote the steady-state capital stock. In all figures below, we vary capital $K_t$ over the interval $[0.5K_{ss},1.5K_{ss}]$, while fixing the other state variables at their steady-state values. This allows us to examine how the learned solution changes with capital while keeping productivity and volatility constant.
	
\begin{figure}[!htbp]
		\centering
		\includegraphics[width=0.85\textwidth]{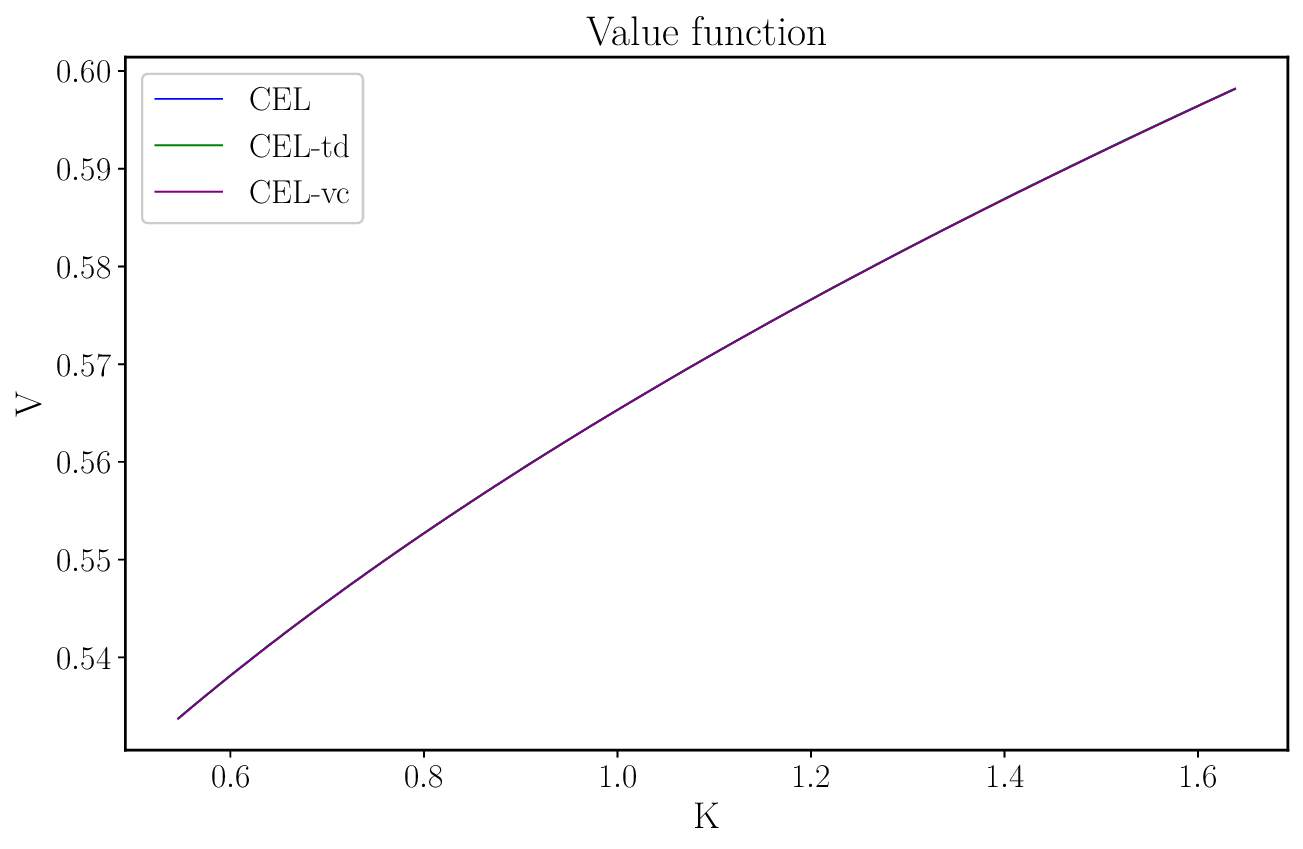}
		\caption{Value function comparison for the DSGE model. Capital $K_t$ varies from $0.5K_{ss}$ to $1.5K_{ss}$, while productivity $z_t$ and volatility $\sigma_t$ are fixed at their steady-state values.}
		\label{fig:dsge_value}
	\end{figure}

	Figure~\ref{fig:dsge_value} reports the learned value function. The figure compares the direct value network, denoted by CEL, with the value estimates obtained from the temporal-difference update and the certainty-equivalent component, denoted by \textit{CEL-td} and \textit{CEL-vc}, respectively. The three curves are very close to each other over the full range of capital. This indicates that the value function has converged well and that the different components of the certainty-equivalent approximation are internally consistent.
	
\begin{figure}[!htbp]
		\centering
		\includegraphics[width=0.85\textwidth]{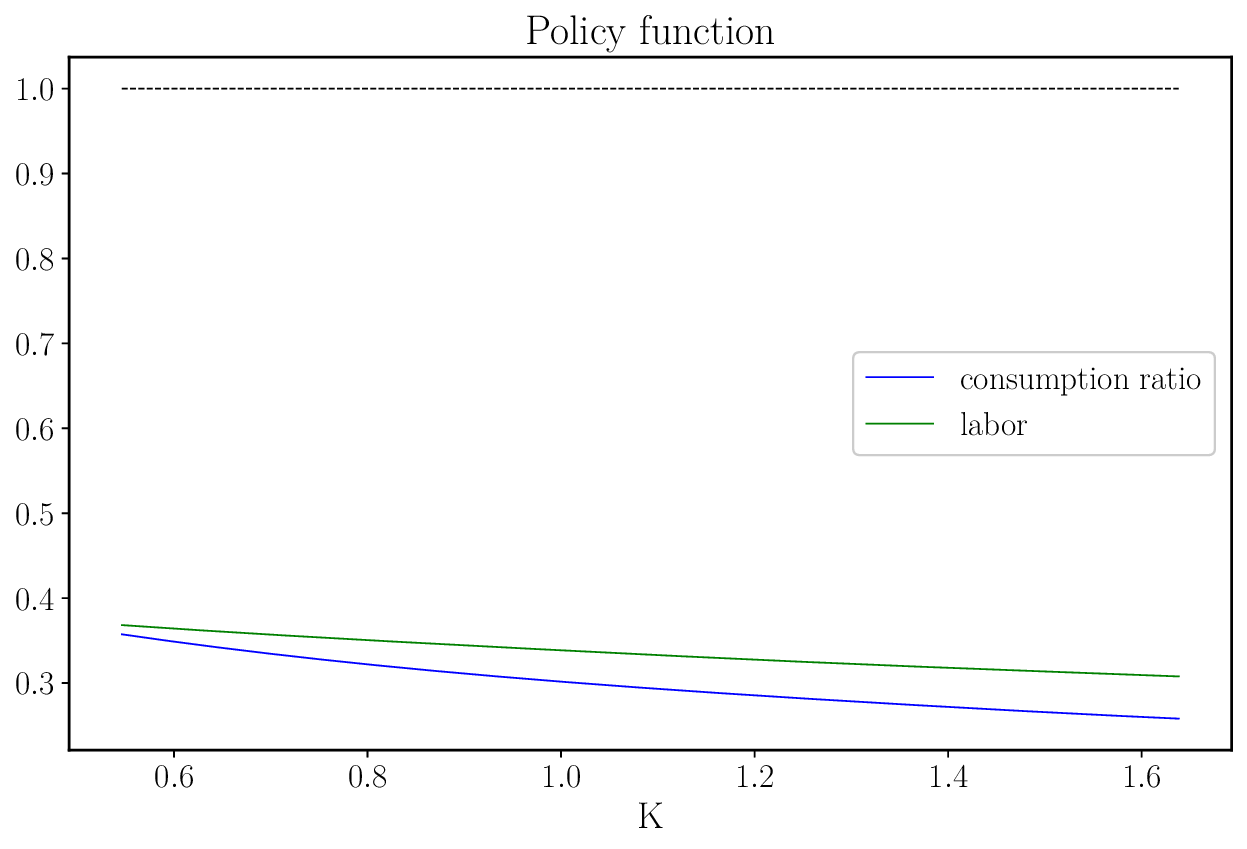}
		\caption{Policy functions for the DSGE model. The figure reports the consumption-to-wealth ratio $c_t$ and labor supply $L_t$ as functions of capital $K_t$, with $z_t$ and $\sigma_t$ fixed at their steady-state values.}
		\label{fig:dsge_policy}
	\end{figure}

	Figure~\ref{fig:dsge_policy} reports the learned policy functions. The two policy components are the consumption-to-wealth ratio $c_t$ and labor supply $L_t$. The consumption-to-wealth ratio describes how much of total available resources is allocated to current consumption, while the labor policy describes the optimal labor supply decision. The smoothness of both policy functions suggests that the neural network learns stable policy functions over the capital range.
	
\begin{figure}[!htbp]
		\centering
		\includegraphics[width=0.85\textwidth]{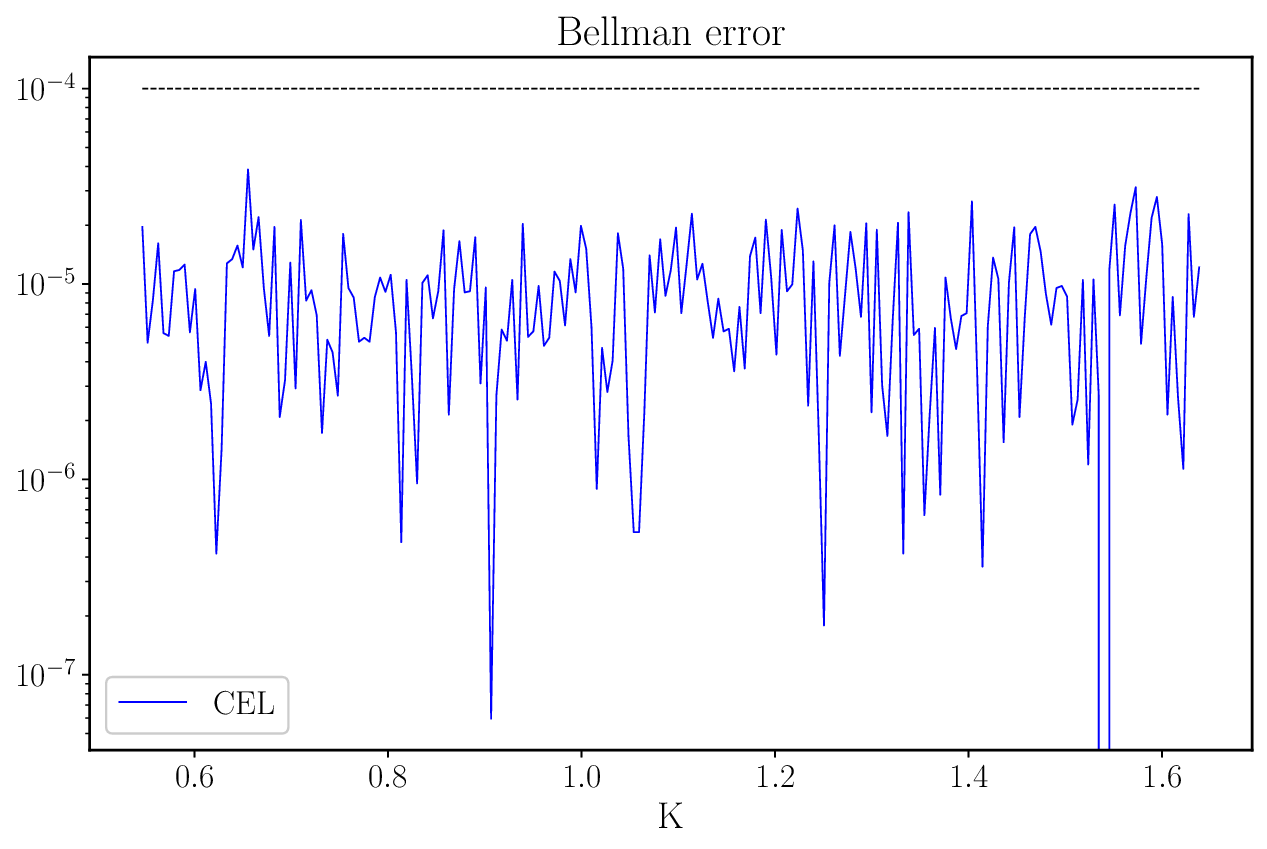}
		\caption{Relative Bellman error for the DSGE model. Capital $K_t$ varies from $0.5K_{ss}$ to $1.5K_{ss}$, while the other state variables are fixed at their steady-state values.}
		\label{fig:dsge_be}
	\end{figure}

	Figure~\ref{fig:dsge_be} reports the relative Bellman error. The relative Bellman error measures whether the learned value function is consistent with the equation after normalizing the Bellman discrepancy by the right-hand side of the Bellman equation. In this experiment, the relative Bellman error remains below $10^{-4}$ across the evaluation range. This provides strong evidence that the learned value function satisfies the Bellman equation with high numerical accuracy.
	
\begin{figure}[!htbp]
		\centering
		\includegraphics[width=0.85\textwidth]{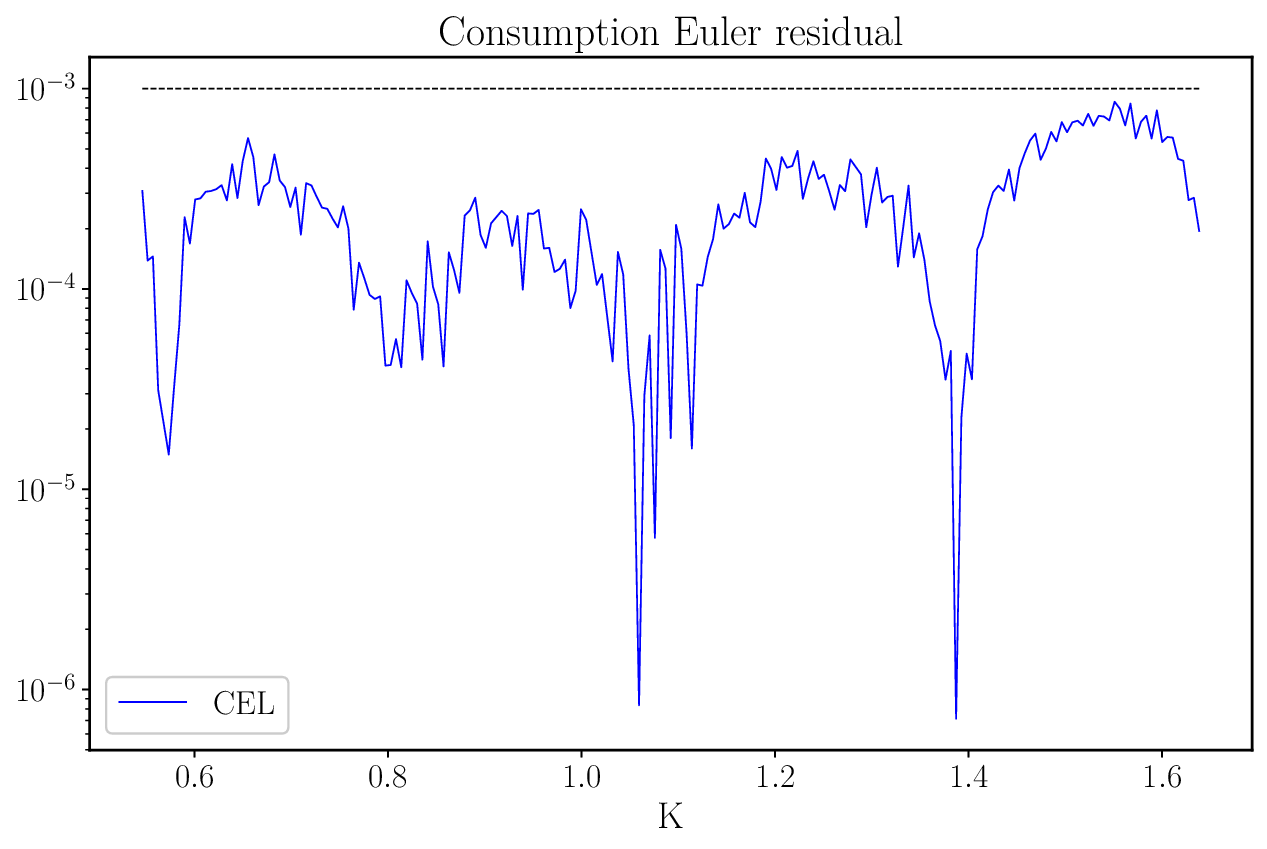}
		\caption{Consumption Euler residual for the DSGE model. Capital $K_t$ varies from $0.5K_{ss}$ to $1.5K_{ss}$, while productivity $z_t$ and volatility $\sigma_t$ are fixed at their steady-state values. The dashed horizontal line marks the $10^{-3}$ threshold.}
		\label{fig:dsge_consumption_ee}
	\end{figure}

\begin{figure}[!htbp]
		\centering
		\includegraphics[width=0.85\textwidth]{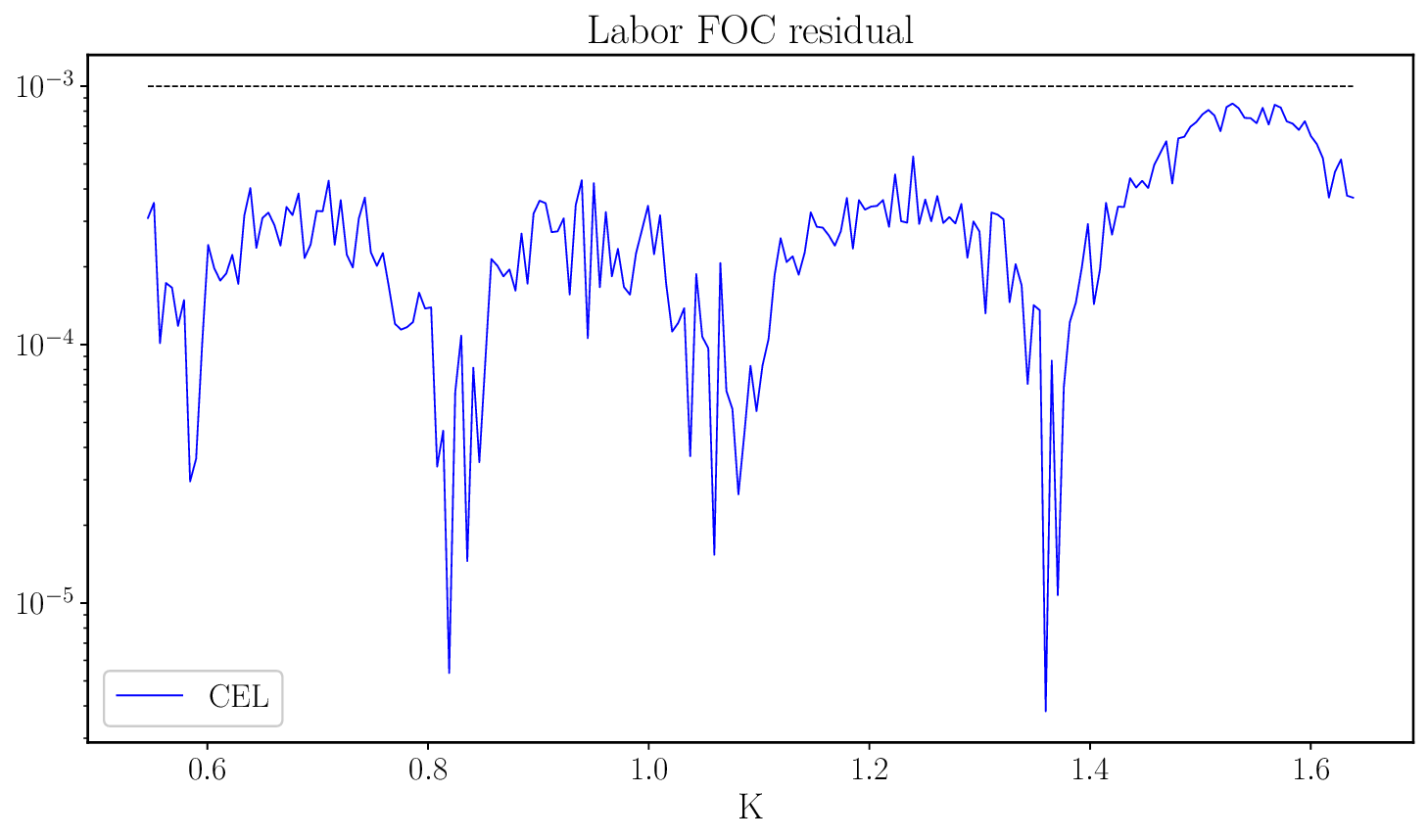}
		\caption{Labor FOC residual for the DSGE model. Capital $K_t$ varies from $0.5K_{ss}$ to $1.5K_{ss}$, while productivity $z_t$ and volatility $\sigma_t$ are fixed at their steady-state values. The dashed horizontal line marks the $10^{-3}$ threshold.}
		\label{fig:dsge_labor_foc}
	\end{figure}

	Figures~\ref{fig:dsge_consumption_ee} and~\ref{fig:dsge_labor_foc} report two optimality-condition diagnostics. The consumption Euler residual evaluates the intertemporal optimality condition for consumption and investment, while the labor FOC residual evaluates the intratemporal first-order condition for labor supply. The formulas used to compute the relative Bellman error and these optimality-condition residuals are reported in Appendix~\ref{app:dsge_bellman_euler_labor_residual}. Both residuals remain below $10^{-3}$ throughout the capital range, indicating that the learned consumption, investment, and labor decisions are close to satisfying the model's first-order optimality conditions.
	
\begin{figure}[!htbp]
		\centering
		\includegraphics[width=0.85\textwidth]{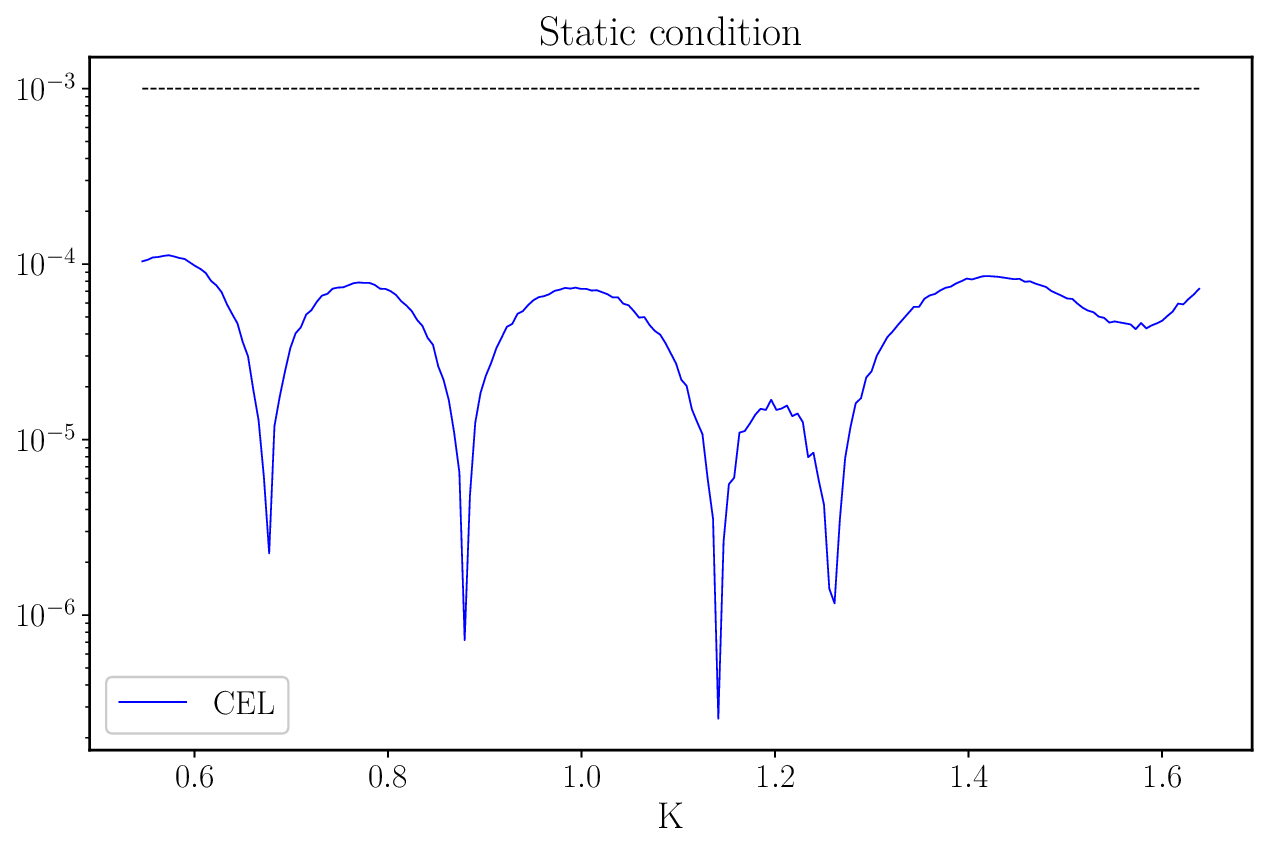}
		\caption{Static condition error for the DSGE model. The condition is derived from the homothetic relationship between the value function and consumption. Capital varies from $0.5K_{ss}$ to $1.5K_{ss}$, while the other state variables are fixed at their steady-state values.}
		\label{fig:dsge_static_condition}
	\end{figure}

	Finally, Figure~\ref{fig:dsge_static_condition} reports the static condition error. This condition is derived from the homothetic structure of the DSGE model, which implies a relationship between the value function and consumption. Therefore, the static condition provides an additional performance measure for checking the internal consistency of the learned solution. The error remains below $10^{-3}$ over the full capital range, indicating that the learned value and policy functions also satisfy this homotheticity-based restriction with high accuracy.
	
	Overall, the DSGE results show that the proposed deep learning method can accurately solve a model with recursive preferences, stochastic volatility, and two controls. The close agreement among the value-function approximations, together with small Bellman errors, consumption Euler residuals, labor FOC residuals, and static condition errors, suggests that the learned solution is both numerically accurate and internally consistent.
	
	\FloatBarrier

	\subsection{Multivariate Strategic Asset Allocation with Predictable Returns}
	\label{subsec:campbell_results}
	
	Our fourth example considers a multivariate strategic asset allocation problem with predictable returns, following the framework of \citet{campbell2003multivariate}. The investor has Epstein--Zin recursive preferences and chooses consumption and portfolio weights in an environment where asset returns are driven by a vector autoregression. 
    This example is useful for evaluating whether the proposed method can solve portfolio-choice problems with recursive preferences, multiple predictive state variables, and multiple controls.
	
	\subsubsection{Problem Setup and Network Representation}
	
	The investor's value function is
	\begin{equation}
		U_t
		=
		\left[
		(1-\delta)C_t^{\varrho}
		+
		\delta
		\left(
		\mathbb{E}_t
		\left[
		U_{t+1}^{1-\gamma}
		\right]
		\right)^{1/\theta}
		\right]^{1/\varrho},
		\qquad
		\varrho \equiv \frac{1-\gamma}{\theta},
		\label{eq:campbell_original_utility}
	\end{equation}
	where $\gamma$ is the coefficient of relative risk aversion, $\delta$ is the discount factor, and $\theta$ controls the elasticity of intertemporal substitution. Wealth evolves according to
	\begin{equation}
		W_{t+1}
		=
		(W_t-C_t)R_{p,t+1}.
		\label{eq:campbell_budget}
	\end{equation}
	The predictive state vector follows a vector autoregression (VAR(1)) process:
	\begin{equation}
		\mathbf{z}_{t+1}
		=
		\Phi_0+\Phi_1\mathbf{z}_t+\mathbf{v}_{t+1},
		\qquad
		\mathbf{v}_{t+1}\sim \mathcal{N}(0,\Sigma_v).
		\label{eq:campbell_var}
	\end{equation}
	The state vector $\mathbf{z}_t$ includes the short-term return, excess returns on risky assets, and predictive variables such as dividend-price ratios and yield spreads.
	
	Given portfolio weights $\boldsymbol{\alpha}_t$ on risky assets, the gross portfolio return is
	\begin{equation}
		R_{p,t+1}
		=
		R_{1,t+1}
		+
		\boldsymbol{\alpha}_t^\top
		\left(
		\mathbf{R}_{t+1}
		-
		R_{1,t+1}\mathbf{1}
		\right),
		\label{eq:campbell_portfolio_return}
	\end{equation}
	where $R_{1,t+1}$ is the gross return on the short-term benchmark asset and $\mathbf{R}_{t+1}$ is the vector of gross returns on risky assets.
	
	The original problem is written in terms of both wealth $W_t$ and the predictive state $\mathbf{z}_t$. However, Epstein--Zin preferences are homothetic, so the value function is homogeneous of degree one in wealth. We can therefore write
	\begin{equation}
		U(W_t,\mathbf{z}_t)
		=
		W_t V(\mathbf{z}_t),
		\label{eq:campbell_homothetic_value}
	\end{equation}
	where $V(\mathbf{z}_t)$ is the normalized value function per unit of wealth. Define the consumption-wealth ratio
	\begin{equation}
		c_t
		\equiv
		\frac{C_t}{W_t}.
	\end{equation}
	Then the normalized wealth transition is
	\begin{equation}
		\frac{W_{t+1}}{W_t}
		=
		(1-c_t)R_{p,t+1}.
		\label{eq:campbell_normalized_wealth}
	\end{equation}
	Substituting \eqref{eq:campbell_homothetic_value} and \eqref{eq:campbell_normalized_wealth} into the Bellman equation and dividing both sides by $W_t$ removes wealth from the state space. The normalized stationary problem becomes
	\begin{equation}
		V(\mathbf{z}_t)
		=
		\max_{c_t,\boldsymbol{\alpha}_t}
		\left[
		(1-\delta)c_t^{\varrho}
		+
		\delta
		\left(
		\mathbb{E}_t
		\left[
		\left(
		(1-c_t)R_{p,t+1}
		V(\mathbf{z}_{t+1})
		\right)^{1-\gamma}
		\right]
		\right)^{1/\theta}
		\right]^{1/\varrho},
		\label{eq:campbell_normalized_problem}
	\end{equation}
	subject to the VAR law of motion \eqref{eq:campbell_var} and the feasibility constraints
	\begin{equation}
		0<c_t<1,
		\qquad
		\boldsymbol{\alpha}_t\in\mathcal{A}.
	\end{equation}
	Equivalently, since $c_t$ is chosen at time $t$, the term $(1-c_t)$ can be taken outside the conditional expectation:
	\begin{equation}
		V(\mathbf{z}_t)
		=
		\max_{c_t,\boldsymbol{\alpha}_t}
		\left[
		(1-\delta)c_t^{\varrho}
		+
		\delta
		(1-c_t)^{\varrho}
		\left(
		\mathbb{E}_t
		\left[
		\left(
		R_{p,t+1}
		V(\mathbf{z}_{t+1})
		\right)^{1-\gamma}
		\right]
		\right)^{1/\theta}
		\right]^{1/\varrho}.
		\label{eq:campbell_normalized_problem_simplified}
	\end{equation}
	This is the formulation used in the numerical implementation. The key simplification is that the state variable is now only $\mathbf{z}_t$, rather than $(W_t,\mathbf{z}_t)$. The policy functions are also normalized: the algorithm learns the consumption-wealth ratio $c(\mathbf{z}_t)$ and the portfolio policy function $\boldsymbol{\alpha}(\mathbf{z}_t)$.
	
	To approximate the solution, we use a four-network architecture:
	\begin{align*}
		\begin{array}{r@{\;}c@{\;}l@{}l}
			V(\mathbf{z}_t)
			& \equiv
			& V(\mathbf{z}_t;\phi),
			&
			\\[0.3em]
			c_t
			& \equiv
			& c(\mathbf{z}_t;\theta_c),
			& \smash{\raisebox{-0.65\normalbaselineskip}{$
					\left.
					\begin{array}{c}
						\vphantom{c(\mathbf{z}_t;\theta_c)}\\[0.2em]
						\vphantom{\boldsymbol{\alpha}(\mathbf{z}_t;\theta_{\alpha})}
					\end{array}
					\right\}
					\,\text{policy networks}
					$}}
			\\[0.1em]
			\boldsymbol{\alpha}_t
			& \equiv
			& \boldsymbol{\alpha}(\mathbf{z}_t;\theta_{\alpha}),
			&
			\\[0.3em]
			\mathbb{E}
			\left[
			V(\mathbf{z}_{t+1})
			\mid
			\mathbf{z}_t,c_t,\boldsymbol{\alpha}_t
			\right]
			& \equiv
			& V_e(\mathbf{z}_t,c_t,\boldsymbol{\alpha}_t;\rho),
			&
			\\[0.3em]
			\mathbb{E}
			\left[
			V(\mathbf{z}_{t+1})
			\right]
			-
			f^{-1}
			\left(
			\mathbb{E}
			\left[
			f(V(\mathbf{z}_{t+1}))
			\right]
			\right)
			& \equiv
			& D(\mathbf{z}_t,c_t,\boldsymbol{\alpha}_t;\nu).
			&
		\end{array}
	\end{align*}

	The consumption policy function $c(\mathbf{z}_t;\theta_c)$ and the portfolio policy function $\boldsymbol{\alpha}(\mathbf{z}_t;\theta_{\alpha})$ are grouped as the policy-network block in the four-network CEL architecture.

	\subsubsection{Numerical Results}
	\label{subsubsec:campbell_numerical_results}
	
	In this subsection, we report the numerical results for the multivariate strategic asset allocation problem. The results are computed under the calibration with risk aversion $\gamma=20$. All figures are presented in a $2\times 3$ subplot format. In each panel, one state variable is varied over a range of $\pm 3$ stationary standard deviations around its steady-state mean, while all other state variables are fixed at their means. This design allows us to isolate the dependence of the value function, consumption policy, and portfolio policy on each component of the predictive state vector.

\begin{figure}[!htbp]
		\centering
		\includegraphics[width=0.9\textwidth]{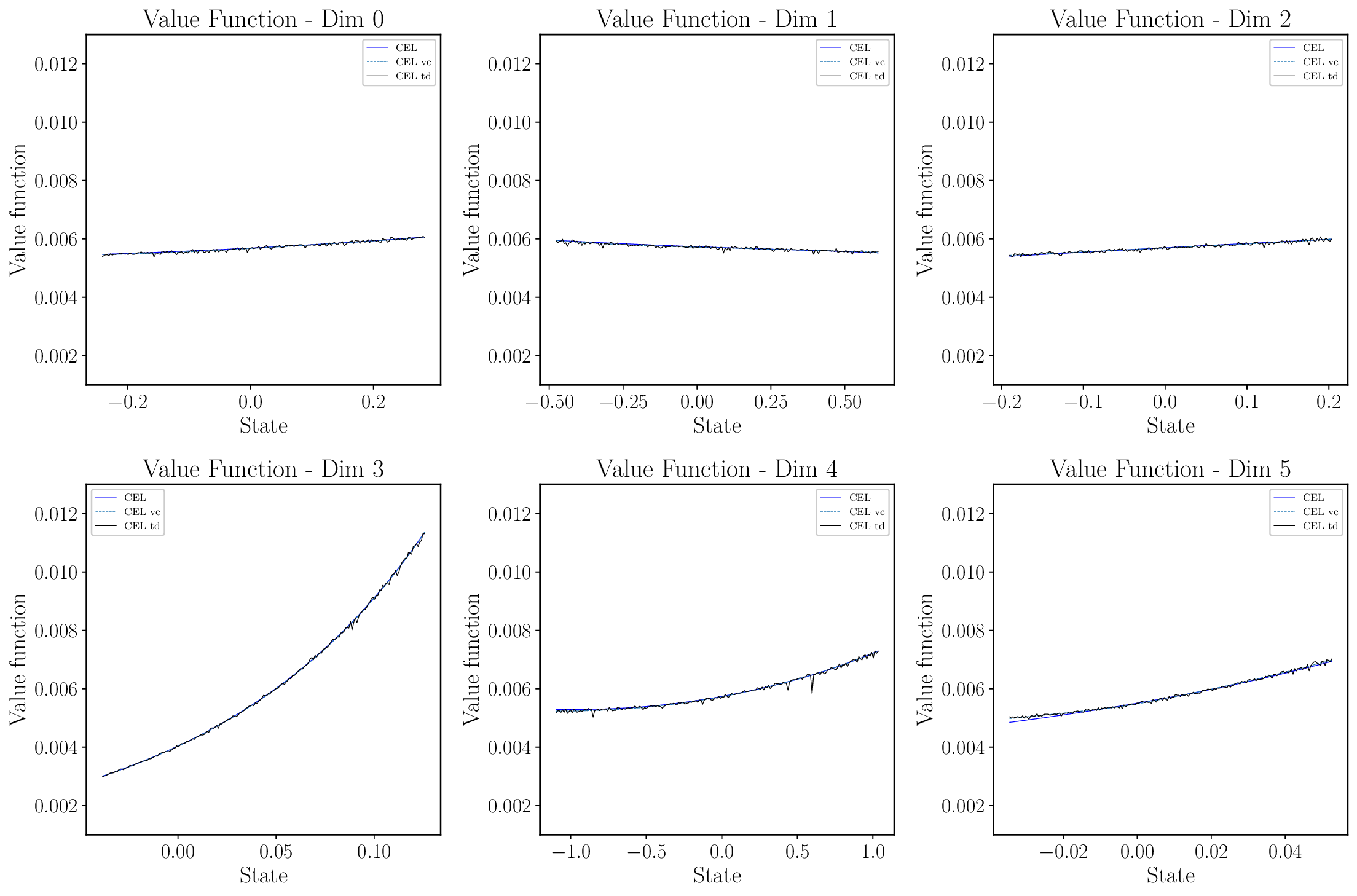}
		\caption{Normalized value function for the multivariate strategic asset allocation model. The figure is shown in a $2\times 3$ subplot format. In each panel, one state variable is varied over $\pm 3$ stationary standard deviations around its steady-state mean, while the remaining state variables are fixed at their means.}
		\label{fig:campbell_value}
	\end{figure}

	\noindent\textbf{Value function approximation.}
	Figure~\ref{fig:campbell_value} reports the normalized value function. Each of the six panels varies one dimension of the state vector while keeping the remaining state variables fixed at their means. The value function is smooth across all dimensions, indicating that the value network captures the dependence of continuation utility on the predictive state variables in a stable manner. The shape of the value function also reflects the effect of time-varying investment opportunities encoded in the VAR state vector.
	
	Overall, Figure~\ref{fig:campbell_value} shows that the learned normalized value function behaves regularly across the state space and exhibits a stable response to changes in the predictive variables. This suggests that the proposed deep learning method is able to approximate the value function accurately in a multivariate asset allocation environment.
	
\begin{figure}[!htbp]
		\centering
		\includegraphics[width=0.9\textwidth]{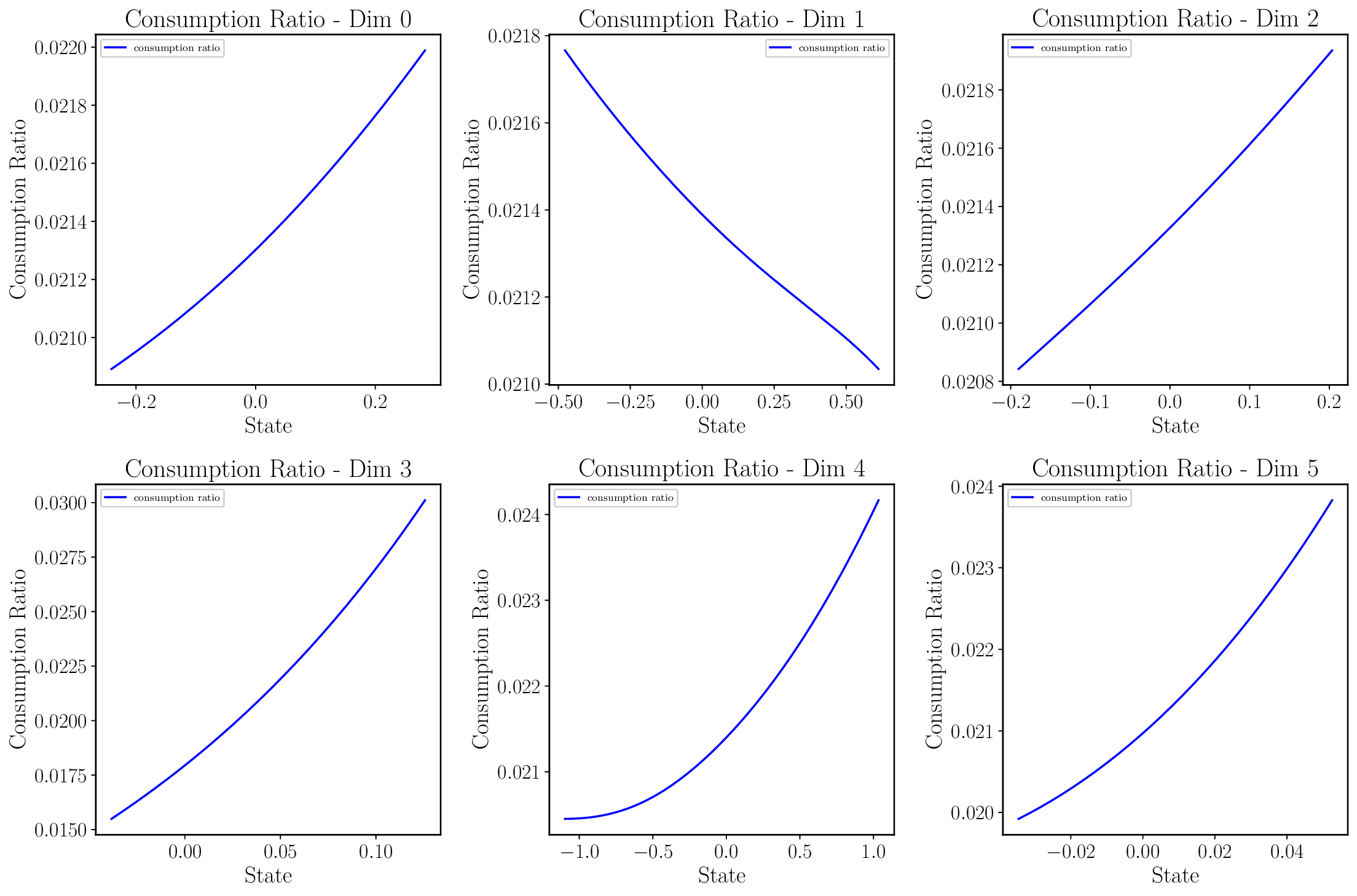}
		\caption{Consumption-wealth ratio policy function for the multivariate strategic asset allocation model. The figure is shown in a $2\times 3$ subplot format. In each panel, one state variable is varied over $\pm 3$ stationary standard deviations around its steady-state mean, while the remaining state variables are fixed at their means.}
		\label{fig:campbell_policy_c}
	\end{figure}

	\noindent\textbf{Policy function approximation: consumption ratio.}
	Figure~\ref{fig:campbell_policy_c} reports the policy function for the consumption-wealth ratio $c_t$. As in the value-function figure, each panel varies one state variable while fixing the remaining state variables at their means. The learned consumption policy function is smooth across all dimensions, which indicates that the policy network produces stable consumption decisions over the relevant region of the state space.
	
	The smooth variation of the consumption-wealth ratio across the six dimensions suggests that the model successfully captures how intertemporal consumption decisions respond to changes in expected returns and other predictive state variables. Since the consumption policy function is a key object in the Epstein--Zin problem, this result provides additional support for the accuracy of the learned policy function.
	
\begin{figure}[!htbp]
		\centering
		\includegraphics[width=0.9\textwidth]{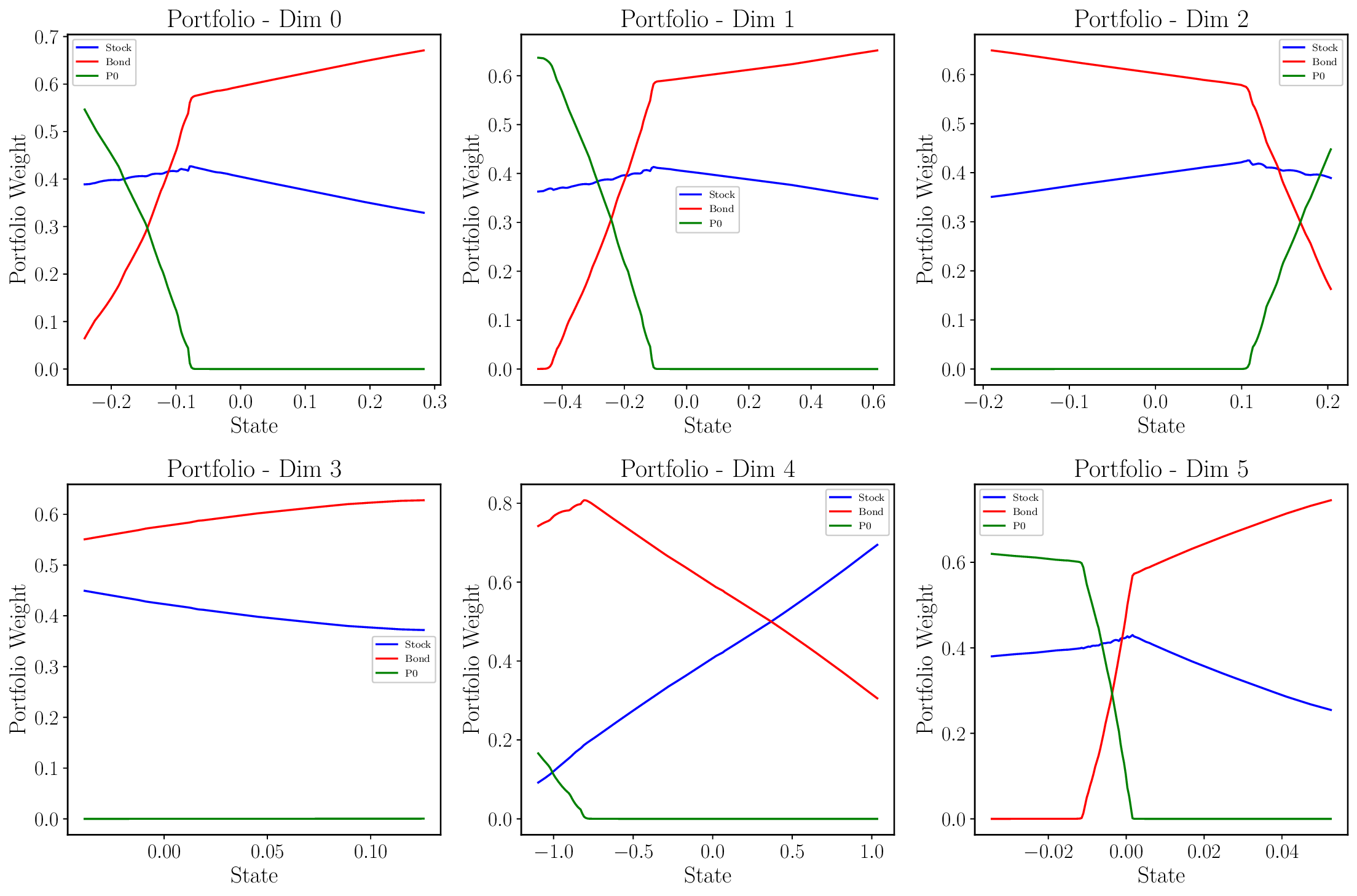}
		\caption{Portfolio-weight policy function for the multivariate strategic asset allocation model. The figure is shown in a $2\times 3$ subplot format. In each panel, one state variable is varied over $\pm 3$ stationary standard deviations around its steady-state mean, while the remaining state variables are fixed at their means.}
		\label{fig:campbell_policy_weight}
	\end{figure}

	\noindent\textbf{Policy function approximation: portfolio weights.}
	Figure~\ref{fig:campbell_policy_weight} reports the portfolio-weight policy function. Each panel shows how the optimal asset allocation responds to changes in one state variable, while all other state variables are held fixed at their means. The resulting portfolio policy functions are state-dependent and respond systematically to changes in the predictive variables, reflecting the fact that the investor adjusts the asset allocation in response to time-varying expected returns and risks.
	
	Figure~\ref{fig:campbell_policy_weight} shows that the learned portfolio policy functions respond systematically to the predictive state vector. This is economically important because the main feature of strategic asset allocation is precisely that portfolio weights vary with the investment opportunity set. The systematic variation of the learned policy suggests that the neural network approximation captures economically meaningful state dependence in portfolio choice.
	
\begin{figure}[!htbp]
		\centering
		\includegraphics[width=0.9\textwidth]{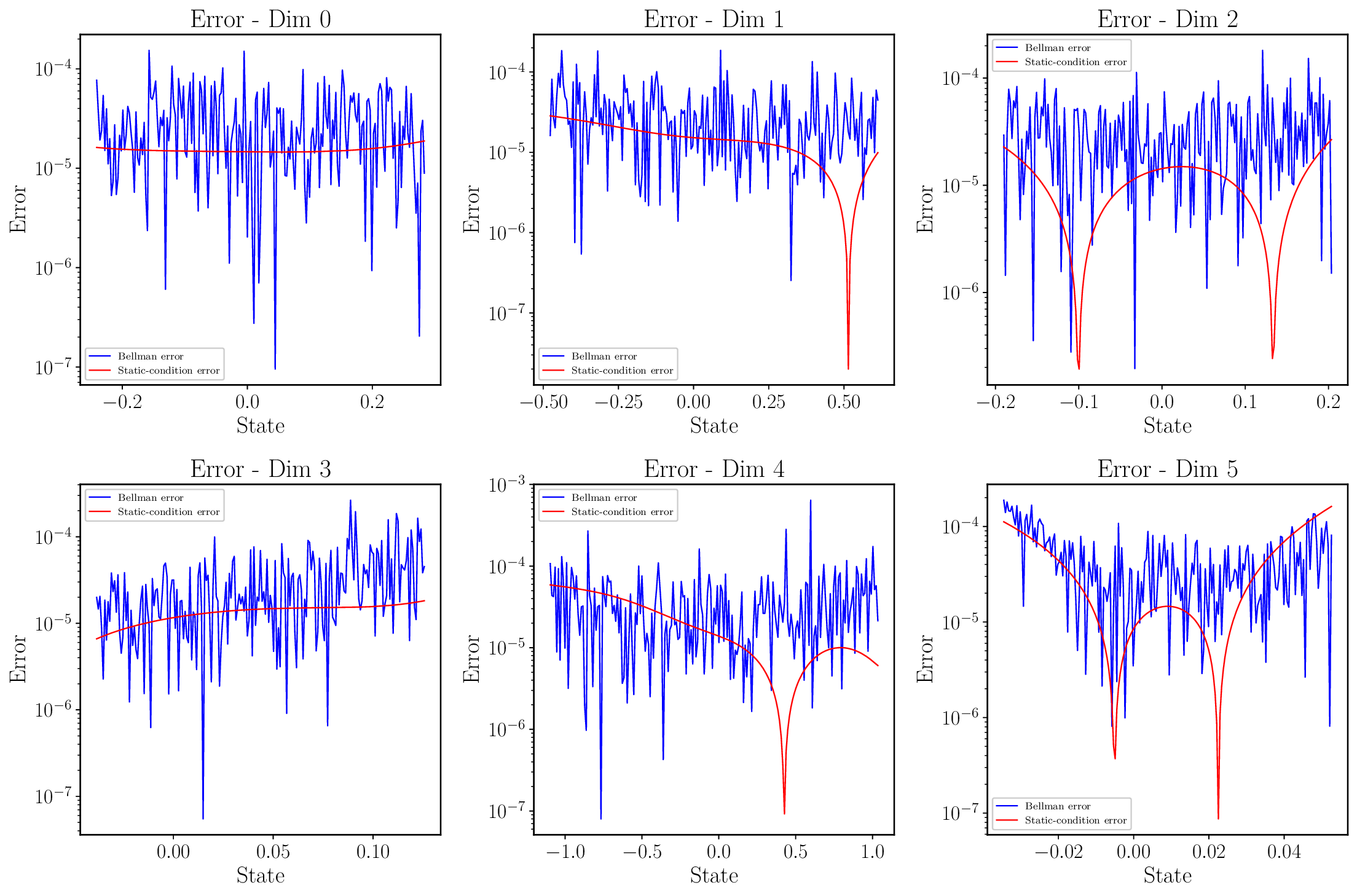}
		\caption{Bellman error and static condition error for the multivariate strategic asset allocation model. The figure is shown in a $2\times 3$ subplot format. In each panel, one state variable is varied over $\pm 3$ stationary standard deviations around its steady-state mean, while the remaining state variables are fixed at their means.}
		\label{fig:campbell_error}
	\end{figure}

	\noindent\textbf{Performance measures.}
	To further assess the numerical quality of the solution, Figure~\ref{fig:campbell_error} reports two performance measures: the Bellman error and the static condition error. The Bellman error evaluates how well the learned value function satisfies the Bellman equation. The static condition error provides an additional internal-consistency check for the policy functions and the value-function approximation. Since the value function is small in magnitude in this calibration, the Bellman error is reported in absolute terms.
	
	Both performance measures are very small over most of the evaluation range, with the majority of the reported values below $10^{-4}$. This indicates that the networks are well trained and that the learned solution is highly consistent with the underlying problem with recursive utility. In particular, the small Bellman errors suggest that the value function approximation is accurate, while the small static condition errors indicate that the policy functions are also internally coherent.

    Overall, the results for the multivariate strategic asset allocation problem show that the proposed deep learning framework can successfully solve a high-dimensional portfolio-choice problem with recursive preferences and predictable returns. The learned value function, consumption policy function, and portfolio weights all respond regularly to the predictive state vector, and the small Bellman and static condition errors provide strong evidence of numerical accuracy.
	
	\FloatBarrier

	\section{Conclusion}
	\label{sec:conclusion}
	
	This paper proposes the first Certainty Equivalent Learning (CEL) algorithm, a deep learning framework for solving high-dimensional discrete-time dynamic programming problems with recursive utility. The central idea is to approximate the certainty-equivalent value directly as a separate state-control function that enters the Bellman equation. Instead of treating the nonlinear certainty-equivalent term as an object that must be recomputed repeatedly during policy and value updates, CEL represents the relevant value functions, policy functions, and certainty-equivalent values with neural networks. This provides a mesh-free and simulation-based approach for nonlinear, stochastic, and high-dimensional decision problems with recursive utility.
	
	The proposed framework includes several architectural variants. The three-network version learns the value function, policy function, and certainty-equivalent value. The four-network version further decomposes the certainty-equivalent value into an expected continuation component and a nonlinear difference component, which is useful when the certainty-equivalent term involves strong nonlinear transformations. The two-network version removes the explicit value network and defines the current value implicitly through the policy and certainty-equivalent networks. These variants should be viewed as different decompositions of the same problem. The algorithm can also be combined with target-network updates for certainty-equivalent components and randomized exploration to stabilize bootstrapped training and policy search. Importantly, CEL does not rely on Euler equations, first-order conditions, perturbation around a steady state, or problem-specific projection bases.
	
	The numerical results lead to several main conclusions. First, in the Gaussian linear-quadratic benchmark, the proposed method accurately recovers both the value function and the policy function. In the benchmark case with $n_s=8$ and $n_c=4$, the learned value function closely matches the closed-form solution, and the relative value-function error remains small across the evaluation range. The learned policy also tracks the analytical benchmark well, indicating that the policy network successfully recovers the optimal-control structure. In the high-dimensional case with $n_s=100$ and $n_c=50$, the method continues to perform well: the value function remains close to the closed-form benchmark, the relative error stays below $10^{-4}$, and the learned policy remains accurate. These results demonstrate both the accuracy and scalability of the proposed algorithm.
	
	Second, in the general nonlinear small-noise robust control problem, where no analytical solution is available, the learned value and policy functions are close to the VFI benchmark. Since VFI is itself a numerical approximation, the comparison is interpreted as a relative difference rather than as an error against the true solution. For both the baseline case with $\sigma=1$ and the high-robustness case with $\sigma=50$, the CEL solution remains close to the VFI benchmark in value and policy functions. The Euler residuals and Bellman errors are also of a similar magnitude to those obtained from the VFI benchmark. These results suggest that the proposed method remains stable as robustness concerns become stronger and can recover solutions consistent with a standard low-dimensional numerical benchmark.
	
	Third, in the homothetic small-noise benchmark case with an analytical solution, both the CEL and VFI solutions can be evaluated directly against the closed-form benchmark. The relative errors are computed at the VFI capital grid points, so the two numerical methods are compared with the analytical solution at the same evaluation points. Across both $\sigma=50$ and $\sigma=200$, the CEL value-function approximation is generally more accurate than the VFI benchmark, with relative errors mostly around or below $10^{-4}$, compared with VFI errors that are typically between $10^{-4}$ and $10^{-3}$. The CEL policy-function approximation also displays lower and smoother relative errors over most of the evaluation range, while the VFI policy error exhibits visible local oscillations. These results indicate that, in the analytical homothetic benchmark, CEL achieves accuracy comparable to, and often better than, the VFI benchmark, especially away from localized boundary regions.
	
	Fourth, in the DSGE model with recursive preferences and stochastic volatility, the proposed method successfully handles a richer macroeconomic environment with a three-dimensional state vector and two controls. The learned value function is internally consistent across the direct value network, the temporal-difference update, and the certainty-equivalent approximation. The learned consumption and labor policies are smooth over the capital range. Moreover, the Bellman error remains below $10^{-4}$, while the consumption Euler residual and the labor FOC residual remain below $10^{-3}$. The former evaluates the intertemporal optimality condition for consumption and investment, whereas the latter evaluates the intratemporal first-order condition for labor supply. The static condition error also remains below $10^{-3}$. These results suggest that the learned solution satisfies the Bellman equation, the intertemporal and intratemporal optimality conditions, and the homotheticity-based static restriction with high numerical accuracy.

    Finally, in the multivariate strategic asset allocation problem with predictable returns, the proposed framework is also able to solve a high-dimensional portfolio-choice problem with recursive preferences. The learned normalized value function, consumption-wealth ratio, and portfolio weights respond systematically to the predictive state variables. The Bellman error and static condition error are mostly below $10^{-4}$, indicating that the learned value function and policy functions are internally consistent with the Bellman equation and the model-specific optimality restrictions.
	
    Overall, these results show that the proposed deep learning framework can solve a wide range of high-dimensional discrete-time dynamic programming problems with recursive utility and nonlinear certainty-equivalent values. The method accurately approximates both value and policy functions, remains stable under stronger robustness concerns, scales to high-dimensional state and control spaces, and performs well in richer DSGE and portfolio-choice environments where traditional grid-based methods become computationally expensive or infeasible. In particular, the small Bellman errors, consumption Euler residuals, labor FOC residuals, and static condition errors provide evidence that the learned solutions are not only regular in shape but also internally consistent with the underlying optimality conditions.
	
    Taken together, the findings suggest that CEL provides a practical computational tool for high-dimensional discrete-time dynamic programming with recursive utility in macroeconomics and finance. By learning certainty-equivalent functions directly, the method addresses a key numerical difficulty created by nonlinear certainty-equivalent terms. CEL's mesh-free neural-network structure allows the method to scale to high-dimensional state spaces and complex stochastic environments where conventional value function iteration and projection methods can become computationally expensive or sensitive to interpolation choices.
	
   The CEL algorithm can also be applied for dynamic programming problems with expected utility, as expected utility is a special case of recursive utility.

	\appendix
	\section{Appendix}
	
	This appendix collects supplementary material that supports the model, algorithm, and numerical analysis in the main text. It first summarizes the neural-network notation and parameter choices, then provides analytical derivations for the benchmark models, and finally defines the diagnostic measures used to evaluate Bellman errors and optimality residuals in the numerical experiments.

	\subsection{Notation for Neural-Network Objects}
	\label{app:notation_table}
	
	Table~\ref{tab:notation_networks} summarizes the main neural-network objects and parameter symbols used in the CEL framework. In applications with multiple control components, the policy-network entry represents the policy block, which may contain several policy subnetworks.
	
	\begin{table}[H]
		\centering
		\caption{Notation for the main CEL networks and parameters.}
		\label{tab:notation_networks}
		\begin{tabular}{lll}
			\toprule
			Symbol & Meaning & Parameter \\
			\midrule
			$V(s_t;\phi)$ & Value network & $\phi$ \\
			$c(s_t;\theta)$ & Policy network & $\theta$ \\
			$V_c(s_t,c_t;\xi)$ & Certainty-equivalent network & $\xi$ \\
			$V_e(s_t,c_t;\rho)$ & Expected continuation network & $\rho$ \\
			$D(s_t,c_t;\nu)$ & Nonlinear difference network in the decomposition $V_c=V_e-D$ & $\nu$ \\
			$\bar V_c$, $\bar V_e$, $\bar D$ & Target-network versions used for stabilized updates & $\bar\xi,\bar\rho,\bar\nu$ \\
			$\widehat V(s_t;\theta,\xi)$ & Implicit value in the two-network architecture & $(\theta,\xi)$ \\
			\bottomrule
		\end{tabular}
	\end{table}
	
	\subsection{Parameters}
	\label{app:high_dim_parameters}
	
	This appendix reports the full numerical specification used in the high-dimensional experiment in Section~\ref{subsec:gaussian_results}. The state dimension is $n_s=8$, the control dimension is $n_c=4$, and the noise covariance matrix is $\Sigma = I_8$.
	
	The state transition matrix is
	\begin{equation}
		A =
		\begin{pmatrix}
			1.0204 & 0.0027 & 0.0149 & 0.0027 & -0.0086 & 0.0018 & -0.0071 & -0.0088 \\
			0.0027 & 1.0143 & -0.0067 & 0.0078 & 0.0188 & -0.0021 & -0.0005 & -0.0047 \\
			0.0149 & -0.0067 & 0.9942 & 0.0084 & 0.0054 & -0.0008 & 0.0106 & 0.0078 \\
			0.0027 & 0.0078 & 0.0084 & 0.9848 & 0.0175 & 0.0007 & -0.0043 & 0.0262 \\
			-0.0086 & 0.0188 & 0.0054 & 0.0175 & 1.0016 & 0.0071 & 0.0086 & 0.0004 \\
			0.0018 & -0.0021 & -0.0008 & 0.0007 & 0.0071 & 0.9943 & 0.0133 & -0.0139 \\
			-0.0071 & -0.0005 & 0.0106 & -0.0043 & 0.0086 & 0.0133 & 1.0271 & -0.0086 \\
			-0.0088 & -0.0047 & 0.0078 & 0.0262 & 0.0004 & -0.0139 & -0.0086 & 0.9996
		\end{pmatrix}.
	\end{equation}
	
	The control matrix is
	\begin{equation}
		B =
		\begin{pmatrix}
			-0.9870 & -0.0152 & -0.0361 & -0.0271 \\
			0.0079 & -0.9409 & -0.0287 & 0.0232 \\
			0.0178 & -0.0166 & -1.0259 & -0.0015 \\
			-0.0165 & 0.0048 & -0.0255 & -1.0064 \\
			-0.9770 & 0.0116 & 0.0378 & -0.0060 \\
			0.0187 & -1.0003 & 0.0115 & -0.0297 \\
			-0.0191 & -0.0403 & -0.9769 & -0.0135 \\
			-0.0203 & -0.0480 & 0.0050 & -1.0436
		\end{pmatrix}.
	\end{equation}
	
	The covariance-related matrix is
	\begin{equation}
		C =
		\begin{pmatrix}
			0.1007 & -0.0034 & -0.0105 & -0.0068 & -0.0095 & 0.0021 & -0.0061 & 0.0104 \\
			-0.0034 & 0.0923 & 0.0041 & -0.0086 & 0.0007 & 0.0109 & -0.0044 & -0.0053 \\
			-0.0105 & 0.0041 & 0.0998 & -0.0078 & 0.0067 & -0.0097 & 0.0006 & 0.0022 \\
			-0.0068 & -0.0086 & -0.0078 & 0.1031 & -0.0011 & 0.0007 & -0.0001 & 0.0091 \\
			-0.0095 & 0.0007 & 0.0067 & -0.0011 & 0.0935 & -0.0001 & 0.0000 & -0.0023 \\
			0.0021 & 0.0109 & -0.0097 & 0.0007 & -0.0001 & 0.0755 & 0.0044 & 0.0095 \\
			-0.0061 & -0.0044 & 0.0006 & -0.0001 & 0.0000 & 0.0044 & 0.1039 & 0.0003 \\
			0.0104 & -0.0053 & 0.0022 & 0.0091 & -0.0023 & 0.0095 & 0.0003 & 0.0979
		\end{pmatrix}.
	\end{equation}
	
	The state cost matrix is
	\begin{equation}
		R =
		\begin{pmatrix}
			1.0100 & 0.0014 & 0.0096 & -0.0087 & 0.0075 & 0.0038 & 0.0022 & 0.0039 \\
			0.0014 & 0.9834 & 0.0072 & -0.0097 & -0.0019 & -0.0222 & 0.0019 & 0.0242 \\
			0.0096 & 0.0072 & 0.9824 & 0.0079 & -0.0089 & 0.0002 & 0.0161 & -0.0016 \\
			-0.0087 & -0.0097 & 0.0079 & 0.9893 & -0.0030 & -0.0026 & 0.0209 & -0.0075 \\
			0.0075 & -0.0019 & -0.0089 & -0.0030 & 1.0158 & -0.0063 & 0.0018 & 0.0082 \\
			0.0038 & -0.0222 & 0.0002 & -0.0026 & -0.0063 & 0.9807 & 0.0021 & 0.0185 \\
			0.0022 & 0.0019 & 0.0161 & 0.0209 & 0.0018 & 0.0021 & 1.0121 & 0.0122 \\
			0.0039 & 0.0242 & -0.0016 & -0.0075 & 0.0082 & 0.0185 & 0.0122 & 1.0010
		\end{pmatrix}.
	\end{equation}
	
	The control cost matrix is
	\begin{equation}
		Q =
		\begin{pmatrix}
			1.0002 & 0.0146 & -0.0016 & 0.0007 \\
			0.0146 & 1.0680 & -0.0142 & 0.0090 \\
			-0.0016 & -0.0142 & 0.9751 & -0.0266 \\
			0.0007 & 0.0090 & -0.0266 & 1.0300
		\end{pmatrix}.
	\end{equation}

	\subsection{Euler Residual and Relative Bellman Error for the Small-Noise Robust Control Model}
	\label{app:smallnoise_euler_bellman_residual}
	
	This appendix defines the diagnostic residuals used to evaluate the learned solution of the small-noise robust control model. These residuals are used only as ex-post accuracy measures and are not used as training objectives in CEL.
	
	Let the state be \(s_t=(K_{t-1},a_t)\), where \(K_{t-1}\) is predetermined capital and \(a_t\) is the current technology state. The production technology and resource constraint are given by
	\begin{equation}
		W_t
		=
		\exp(Pa_t)K_{t-1}^{\alpha}
		+
		(1-\delta)K_{t-1},
		\qquad
		C_t+K_t=W_t.
		\label{eq:app_smallnoise_resource}
	\end{equation}
	In the numerical implementation, the policy network outputs the consumption-to-resources ratio \(c_t\in(0,1)\). Hence,
	\begin{equation}
		C_t=c_tW_t,
		\qquad
		K_t=(1-c_t)W_t.
		\label{eq:app_smallnoise_policy_mapping}
	\end{equation}
	The technology state follows
	\begin{equation}
		a_{t+1}
		=
		\Omega_0+\Omega_a a_t+\sqrt{\epsilon}\Omega_v w_{t+1},
		\qquad
		w_{t+1}\sim\mathcal N(0,1).
		\label{eq:app_smallnoise_shock}
	\end{equation}
	
	The recursive utility specification is
	\begin{equation}
		V(K_{t-1},a_t)
		=
		\frac{C_t^{1-\gamma}}{1-\gamma}
		-
		\frac{1}{\sigma}
		\log
		\mathbb E_t
		\left[
		\exp\left(
		-\sigma\beta V(K_t,a_{t+1})
		\right)
		\right].
		\label{eq:app_smallnoise_recursive_utility}
	\end{equation}
	Define the risk-sensitive change-of-measure term by
	\begin{equation}
		\mathcal M_{t+1}
		\equiv
		\frac{
			\exp\left(
			-\sigma\beta V(K_t,a_{t+1})
			\right)
		}{
			\mathbb E_t
			\left[
			\exp\left(
			-\sigma\beta V(K_t,a_{t+1})
			\right)
			\right]
		}.
		\label{eq:app_smallnoise_rs_weight}
	\end{equation}
	The gross marginal return on capital is
	\begin{equation}
		R^K_{t+1}
		\equiv
		\alpha\exp(Pa_{t+1})K_t^{\alpha-1}
		+
		1-\delta.
		\label{eq:app_smallnoise_capital_return}
	\end{equation}
	
	The standard Euler equation for the small-noise robust control model can be written as
	\begin{equation}
		1
		=
		\beta
		\mathbb E_t
		\left[
		\mathcal M_{t+1}
		\left(
		\frac{C_{t+1}}{C_t}
		\right)^{-\gamma}
		R^K_{t+1}
		\right].
		\label{eq:app_smallnoise_euler_equation}
	\end{equation}
	Accordingly, we define the Euler residual as
	\begin{equation}
		\mathcal R^E(s_t)
		\equiv
		1
		-
		\beta
		\mathbb E_t
		\left[
		\mathcal M_{t+1}
		\left(
		\frac{C_{t+1}}{C_t}
		\right)^{-\gamma}
		R^K_{t+1}
		\right].
		\label{eq:app_smallnoise_euler_residual}
	\end{equation}
	At the exact solution, \(\mathcal R^E(s_t)=0\) for all states.
	
	We also define the relative Bellman error. Given the learned value function and policy function, define the Bellman right-hand side as
	\begin{equation}
		\mathcal B(s_t)
		\equiv
		\frac{C_t^{1-\gamma}}{1-\gamma}
		-
		\frac{1}{\sigma}
		\log
		\mathbb E_t
		\left[
		\exp\left(
		-\sigma\beta V(K_t,a_{t+1})
		\right)
		\right].
		\label{eq:app_smallnoise_bellman_rhs}
	\end{equation}
	The relative Bellman error is then defined as the Bellman discrepancy normalized by the right-hand side of the Bellman equation:
	\begin{equation}
		\mathcal R^B(s_t)
		\equiv
		\frac{
			V(K_{t-1},a_t)-\mathcal B(s_t)
		}{
			\mathcal B(s_t)
		}.
		\label{eq:app_smallnoise_relative_bellman_error}
	\end{equation}
	At the exact solution, \(\mathcal R^B(s_t)=0\) for all states.
	
	In the numerical implementation, the conditional expectations in \eqref{eq:app_smallnoise_rs_weight}, \eqref{eq:app_smallnoise_euler_residual}, and \eqref{eq:app_smallnoise_bellman_rhs} are approximated by sample means over simulated next-period shocks. We report the absolute residuals \(|\mathcal R^E(s_t)|\) and \(|\mathcal R^B(s_t)|\), or their logarithmic transformations, as diagnostic measures of solution accuracy.

	\subsection{Relative Bellman Error, Euler Residual, and Labor Optimality Residual for the DSGE Model}
	\label{app:dsge_bellman_euler_labor_residual}
	
	This appendix defines the diagnostic residuals used to evaluate the learned solution of the DSGE model with recursive utility. These residuals are used only as ex-post accuracy measures and are not used as training objectives in CEL.
	
	Let the state be \(s_t=(K_t,z_t,\sigma_t)\), and let \(C_t\) and \(L_t\) denote consumption and labor implied by the learned policy functions. The production technology and resource constraint are given by
	\begin{equation}
		Y_t=\exp(z_t)K_t^{\zeta}L_t^{1-\zeta},
		\qquad
		C_t+K_{t+1}=Y_t+(1-\delta)K_t.
		\label{eq:app_dsge_resource}
	\end{equation}
	The one-period utility composite is
	\begin{equation}
		u_t
		=
		u(C_t,L_t)
		=
		C_t^{\nu}(1-L_t)^{1-\nu}.
		\label{eq:app_dsge_period_utility}
	\end{equation}
	We write the Epstein--Zin aggregator as
	\begin{equation}
		V_t
		=
		\left[
		(1-\beta)u_t^{1-\rho}
		+
		\beta
		\left(
		\mathbb E_t
		\left[
		V_{t+1}^{1-\gamma}
		\right]
		\right)^{1/\theta}
		\right]^{1/(1-\rho)},
		\qquad
		\theta
		\equiv
		\frac{1-\gamma}{1-\rho}.
		\label{eq:app_dsge_ez_aggregator}
	\end{equation}
	Define
	\begin{equation}
		M_t
		\equiv
		\mathbb E_t
		\left[
		V_{t+1}^{1-\gamma}
		\right].
		\label{eq:app_dsge_Mt}
	\end{equation}
	Given the learned policy functions and value function, define the Bellman right-hand side by
	\begin{equation}
		\mathcal B(s_t)
		\equiv
		\left[
		(1-\beta)u_t^{1-\rho}
		+
		\beta
		M_t^{1/\theta}
		\right]^{1/(1-\rho)}.
		\label{eq:app_dsge_bellman_rhs}
	\end{equation}
	The relative Bellman error is the Bellman discrepancy normalized by the right-hand side of the Bellman equation:
	\begin{equation}
		\mathcal R^B(s_t)
		\equiv
		\frac{
			V_t-\mathcal B(s_t)
		}{
			\mathcal B(s_t)
		}.
		\label{eq:app_dsge_relative_bellman_error}
	\end{equation}
	At the exact solution, \(\mathcal R^B(s_t)=0\) for all states.
	
	Define the current marginal utility term by
	\begin{equation}
		\Lambda_t
		\equiv
		(1-\beta)u_t^{-\rho}u_C(C_t,L_t),
		\label{eq:app_dsge_marginal_utility}
	\end{equation}
	where
	\begin{equation}
		u_C(C_t,L_t)
		=
		\nu C_t^{\nu-1}(1-L_t)^{1-\nu}.
	\end{equation}
	The gross marginal return on capital is
	\begin{equation}
		R^K_{t+1}
		\equiv
		\zeta \exp(z_{t+1})K_{t+1}^{\zeta-1}L_{t+1}^{1-\zeta}
		+
		1-\delta.
		\label{eq:app_dsge_capital_return}
	\end{equation}
	
	The standard Euler equation for capital with Epstein--Zin preferences can be written as
	\begin{equation}
		\Lambda_t
		=
		\beta
		M_t^{1/\theta-1}
		\mathbb E_t
		\left[
		V_{t+1}^{\rho-\gamma}
		\Lambda_{t+1}
		R^K_{t+1}
		\right].
		\label{eq:app_dsge_euler_equation}
	\end{equation}
	Accordingly, we define the capital Euler residual as
	\begin{equation}
		\mathcal R^K(s_t)
		\equiv
		1
		-
		\frac{
			\beta
			M_t^{1/\theta-1}
			\mathbb E_t
			\left[
			V_{t+1}^{\rho-\gamma}
			\Lambda_{t+1}
			R^K_{t+1}
			\right]
		}{
			\Lambda_t
		}.
		\label{eq:app_dsge_euler_residual}
	\end{equation}
	At the exact solution, \(\mathcal R^K(s_t)=0\) for all states.
	
	We also evaluate the intratemporal labor optimality condition. The marginal product of labor is
	\begin{equation}
		w_t
		\equiv
		(1-\zeta)\exp(z_t)K_t^{\zeta}L_t^{-\zeta}.
		\label{eq:app_dsge_wage}
	\end{equation}
	The labor first-order condition equates the marginal product of labor to the marginal rate of substitution between leisure and consumption:
	\begin{equation}
		w_t
		=
		\frac{1-\nu}{\nu}
		\frac{C_t}{1-L_t}.
		\label{eq:app_dsge_labor_foc}
	\end{equation}
	Equivalently,
	\begin{equation}
		1
		=
		\frac{
			\nu w_t(1-L_t)
		}{
			(1-\nu)C_t
		}.
	\end{equation}
	We therefore define the labor FOC error, or labor optimality residual, as
	\begin{equation}
		\mathcal R^L(s_t)
		\equiv
		1
		-
		\frac{
			\nu w_t(1-L_t)
		}{
			(1-\nu)C_t
		}.
		\label{eq:app_dsge_labor_residual}
	\end{equation}
	At the exact solution, \(\mathcal R^L(s_t)=0\) for all states.
	
	In the numerical implementation, the conditional expectations in \(M_t\), in the relative Bellman error \eqref{eq:app_dsge_relative_bellman_error}, and in the Euler residual \eqref{eq:app_dsge_euler_residual} are approximated by sample means over simulated next-period shocks. We report the absolute residuals \(|\mathcal R^B(s_t)|\), \(|\mathcal R^K(s_t)|\), and \(|\mathcal R^L(s_t)|\), or their logarithmic transformations, as diagnostic measures of solution accuracy.

	\appendix

	\bibliographystyle{dcu}
	\bibliography{reference_final}

\end{document}